\documentclass[12pt]{article}
\usepackage{amssymb}
\usepackage{rotating}
\usepackage{amsmath}
\usepackage{arydshln}
\usepackage{mathrsfs}
\usepackage{bm}
\usepackage{array}
\usepackage[labelfont=bf]{caption}
\usepackage{booktabs}
\usepackage{makecell}
\usepackage{longtable}
\usepackage{graphicx}
\usepackage{epstopdf}
\usepackage{seqsplit}
\usepackage{subfig}
\usepackage[colorlinks=true, citecolor=blue, linkcolor=blue, urlcolor=blue]{hyperref}
\allowdisplaybreaks[4]

\renewcommand{\theequation}{\arabic{section}.\arabic{equation}}

\begin{document}



\def\a{\alpha}
\def\b{\beta}
\def\d{\delta}
\def\e{\epsilon}
\def\g{\gamma}
\def\h{\mathfrak{h}}
\def\k{\kappa}
\def\l{\lambda}
\def\o{\omega}
\def\p{\wp}
\def\r{\rho}
\def\t{\tau}
\def\s{\sigma}
\def\z{\zeta}
\def\x{\xi}
\def\V={{{\bf\rm{V}}}}
 \def\A{{\cal{A}}}
 \def\B{{\cal{B}}}
 \def\C{{\cal{C}}}
 \def\D{{\cal{D}}}
\def\G{\Gamma}
\def\K{{\cal{K}}}
\def\O{\Omega}
\def\R{\bar{R}}
\def\T{{\cal{T}}}
\def\L{\Lambda}
\def\f{E_{\tau,\eta}(sl_2)}
\def\E{E_{\tau,\eta}(sl_n)}
\def\Zb{\mathbb{Z}}
\def\Cb{\mathbb{C}}

\def\R{\overline{R}}

\def\beq{\begin{equation}}
\def\eeq{\end{equation}}
\def\bea{\begin{eqnarray}}
\def\eea{\end{eqnarray}}
\def\ba{\begin{array}}
\def\ea{\end{array}}
\def\no{\nonumber}
\def\le{\langle}
\def\re{\rangle}
\def\lt{\left}
\def\rt{\right}

\newtheorem{Theorem}{Theorem}
\newtheorem{Definition}{Definition}
\newtheorem{Proposition}{Proposition}
\newtheorem{Lemma}{Lemma}
\newtheorem{Corollary}{Corollary}
\newcommand{\proof}[1]{{\bf Proof. }
        #1\begin{flushright}$\Box$\end{flushright}}

\baselineskip=20pt

\newfont{\elevenmib}{cmmib10 scaled\magstep1}
\newcommand{\preprint}{
   \begin{flushleft}
   \end{flushleft}\vspace{-1.3cm}
   \begin{flushright}\normalsize
   \end{flushright}}
\newcommand{\Title}[1]{{\baselineskip=26pt
   \begin{center} \Large \bf #1 \\ \ \\ \end{center}}}

\newcommand{\Author}{\begin{center}
	\large \bf
	Zhouzheng Ji${}^{a}$, Pei Sun${}^{a,e,f}$, Xiaotian Xu${}^{a,e,f}\footnote{Corresponding author: xtxu@nwu.edu.cn}$, Yi Qiao${}^{a,e,f}\footnote{Corresponding author: qiaoyi\_joy@foxmail.com}$, Junpeng Cao${}^{b,c,d,e}\footnote{Corresponding author: junpengcao@iphy.ac.cn}$ and Wen-Li Yang${}^{a,e,f}$
\end{center}}

\newcommand{\Address}{\begin{center}
    ${}^a$ Institute of Modern Physics, Northwest University, Xian 710127, China\\
	${}^b$ Beijing National Laboratory for Condensed Matter Physics, Institute of Physics, Chinese Academy of Sciences, Beijing 100190, China\\
    ${}^c$ School of Physical Sciences, University of Chinese Academy of Sciences, Beijing 100049, China\\
	${}^d$ Songshan Lake Materials Laboratory, Dongguan, Guangdong 523808, China\\
	${}^e$ Peng Huanwu Center for Fundamental Theory, Xian 710127, China\\
	${}^f$ Shaanxi Key Laboratory for Theoretical Physics Frontiers, Xian 710127, China
\end{center}}

\Title{Tensor-Network Analysis of Root Patterns in the XXX Model with Open Boundaries}

\Author

\Address
\vspace{1cm}
\begin{abstract}
The string hypothesis of Bethe roots is a cornerstone in the thermodynamic analysis of quantum integrable systems, since it connects root configurations with physical quantities such as the ground-state energy, surface energy and excitation spectra. For integrable models with \(U(1)\) symmetry, this connection is well established. When the \(U(1)\) symmetry is broken by generic non-diagonal boundary fields, however, the off-diagonal Bethe Ansatz leads to an inhomogeneous \(T\text{--}Q\) relation whose Bethe roots have highly nontrivial distributions. This raises two fundamental questions: whether the zero roots and the ODBA Bethe roots still possess regular and classifiable structures in the large-size limit, and whether such structures can be used to extract physical quantities.

In this work, we address these two questions for the isotropic Heisenberg spin chain with non-diagonal open boundaries. By combining tensor-network algorithms with Bethe-Ansatz techniques, we determine the zero-root and Bethe-root configurations associated with the \(\Lambda\text{--}\theta\) relation and the inhomogeneous Bethe Ansatz equations for large system sizes, up to \(N\simeq 60\) and \(100\). We find that, despite the absence of \(U(1)\) symmetry, the roots exhibit well-organized patterns. The zero roots form bulk strings, boundary strings and additional roots, while the ODBA Bethe roots split into four geometric classes: regular roots, line roots, arc roots and paired-line roots.

We further show that these root structures are not merely numerical patterns, but contain direct thermodynamic information. In particular, the geometric structure of the ODBA Bethe roots allows us to derive analytic expressions for the surface energy and elementary excitation energies. This demonstrates that the full ODBA Bethe-root configuration can serve as a thermodynamically meaningful object in a \(U(1)\)-symmetry-broken open spin chain. A comparison with the \(t\text{--}W\) scheme shows that the zero-root description has a simpler topology, while the ODBA Bethe roots provide a direct route to physical quantities. Our results therefore provide a framework for studying root structures and thermodynamic properties in quantum integrable models without \(U(1)\) symmetry.

\vspace{1truecm}


\end{abstract}
\newpage

\section{Introduction}
\label{intro} \setcounter{equation}{0}
Quantum integrable systems provide a rigorous framework for the exact study of one-dimensional many-body problems.
Since Bethe's solution of the spin-$\tfrac{1}{2}$ XXX chain~\cite{Bethe1931}, the Bethe Ansatz has remained a cornerstone of the field, yielding exact results for a broad class of integrable models~\cite{Korepin1993,Takahashi1999,Bax82}.
For quantum integrable systems with boundary fields, integrability can be preserved within the quantum inverse scattering method (QISM)~\cite{Skl80,Fad80}, provided that the boundary reflections satisfy the boundary Yang--Baxter equations~\cite{Sklyanin1988,Cherednik1984,Vega1993}.
When the boundary fields are parallel, the Hamiltonian retains $U(1)$ symmetry and the Bethe Ansatz is well established~\cite{Alcaraz1987,Mezincescu1991}.
In contrast, generic non-diagonal boundary terms break the $U(1)$ symmetry yet the model remains integrable. To address this issue, several techniques have been developed, including gauge transformation \cite{cao03}, fusion-based $T-Q$ relation \cite{yun95,nep021}, $q$-Onsager algebra \cite{bas1,bas2}, separation of variables \cite{nie2-2,nic13-1}, modified algebraic Bethe Ansatz \cite{bel13-1,bel13-2} and off-diagonal Bethe Ansatz \cite{cysw,book}.
The eigenvalues of the transfer matrix of the quantum integrable systems without $U(1)$ symmetry are characterized by the inhomogeneous $T-Q$ relations.
However, the associated Bethe Ansatz equations (BAEs) are inhomogeneous and the corresponding distributions of Bethe roots are very complicated.
Consequently, the thermodynamic Bethe Ansatz \cite{yang69,gau71,tak71} is no longer valid, and computing the exact physical properties in the thermodynamic limit becomes exceedingly difficult.
For a particular range of boundary parameters, non-linear integral equations have recently been derived directly from the inhomogeneous $T-Q$ relation for the XXX spin-$\tfrac{1}{2}$ quantum chain with non-diagonal boundary fields~\cite{Frahm2026}.
Recently, based on off-diagonal Bethe Ansatz, the zero-root method (termed the $t-W$ scheme) has been proposed \cite{qiao20,qiao21}, which effectively tackles the challenges posed by the inhomogeneous $T-Q$ relations.
The key point is to analyze the zero roots of the transfer matrix, not those of the $Q$-operator (i.e., the Bethe roots).

The investigation of zero roots and Bethe roots is essential because their distribution patterns form the basis of the analysis of the thermodynamic limit and thermodynamic properties.
Although the analytic structure of the inhomogeneous $T-Q$ relation and the $t-W$ relation are well understood, solving the associated constrained equations numerically remains challenging. Previous numerical studies of off-diagonal Bethe roots and zero roots have been restricted to small systems, typically $N \leq 12$, and served mainly to check theoretical conjectures~\cite{Nepomechie2013,Cao2013c}.
Particularly in high-rank quantum integrable models, the dimension of the local Hilbert space is large, and existing numerical approaches are insufficient for analyzing the structures of zero roots and Bethe roots.
In contrast, tensor-network algorithms-especially the density matrix renormalization group (DMRG)~\cite{White1992,Schollwock2005,Schollwock2011,Verstraete2004,McCulloch2007,Pollmann2010}, make it possible to obtain high-precision ground state and low-lying spectra for much larger systems, thereby building a practical bridge between functional formulas and numerical studies~\cite{Murg2012}.

In this paper, we take the spin-1/2 XXX chain subject to arbitrary boundary fields as a benchmark to verify the efficacy and broad applicability of the numerical method developed herein.
Within the QISM framework, we first represent the transfer matrix $t(u)$ as a matrix product operator (MPO) and compute its eigenvalue $\Lambda(u)$ for large lattice sizes using DMRG.
The DMRG calculations were performed using the ITensor Library~\cite{ITensor2022}.
Then we solve the zeros of the transfer matrix from the numerically obtained $\Lambda(u)$ and verify them using the maximum modulus principle and the argument principle~\cite{Ahlfors1979};
on this basis, we verify the correctness of the phase regions given in Ref.~\cite{Dong2023} and describe the concrete structural changes when parameters cross between these regions.
Next, we solve the Bethe roots at large $N$ directly from the inhomogeneous $T-Q$ relation by adapting McCoy's method outlined in Ref.~\cite{nep022} together with the strategy of Ref.~\cite{Belliard2018}, and summarize how their overall structural characteristics evolve with the boundary parameters, in particular the gradual restoration of $U(1)$ symmetry as the non-parallel boundary fields are reduced from finite values to zero.
We further analyze the structures of zero roots and Bethe roots for excited states, and derive an analytic expression for the surface energy and excitation energies based on the geometric structure of the Bethe roots.
To the best of our knowledge, this work provides the first systematic large-size characterization of the ODBA Bethe-root structure in a \(U(1)\)-symmetry-broken open XXX chain with generic non-diagonal boundary fields, and establishes its direct thermodynamic use in deriving physical quantities.

The purpose of this work is to clarify whether root configurations in \(U(1)\)-symmetry-broken integrable systems remain structurally regular and thermodynamically useful. More specifically, we address two questions. First, do the zero roots and the ODBA Bethe roots still organize into regular structures analogous to, or generalizing, conventional Bethe strings? Second, if such structures exist, can they be used to extract physical quantities, rather than serving merely as numerical root patterns?

The remainder of the paper is organized as follows.
Sec.~\ref{sec:2} introduces the model under consideration and its integrability, followed by a concise review of the ODBA formalism and the zero-root approach.
Sec.~\ref{sec:3} describes the DMRG/MPO implementation and the computation of \(\Lambda(u)\).
Sec.~\ref{sec:4} presents the ground-state structures of the zero roots, analyzes their changes across different parameter regions, and discusses elementary excitations in the zero-root picture.
Sec.~\ref{sec:5} discusses the ground-state structures of the ODBA Bethe roots and their structural evolution with boundary parameters. It further presents elementary excitations and derives the surface energy and the corresponding excitation energies from the Bethe-root geometry.
Sec.~\ref{sec:6} concludes the work.

Appendix~\ref{app:A} provides the MPO representation of the transfer matrix, while Appendix~\ref{app:B} presents the derivation of the maximum-modulus-principle criterion.
Appendices~\ref{app:C} and~\ref{app:D} provide validation data for the zero roots and the Bethe roots, respectively.
Appendix~\ref{app:E} illustrates the crossover behavior of the Bethe roots as the boundary parameters increase.
Appendix~\ref{app:F} discusses the remaining cases involved in deriving the surface energy from the Bethe-root structure.

\section{Integrability}
\setcounter{equation}{0}
\label{sec:2}
Initially, let $\bm{V}$ be a two-dimensional linear space with an orthonormal basis $\{ |m\rangle, m = 0, 1 \}$. We adopt standard notations: for any matrix $A \in \mathrm{End}(\bm{V})$, $A_j$ denotes the embedding operator acting as $A$ on the $j$-th space and as identity elsewhere in the tensor product $\bm{V}^{\otimes N}$. For $B \in \mathrm{End}(\bm{V} \otimes \bm{V})$, $B_{ij}$ acts nontrivially only on the $i$-th and $j$-th spaces.

We introduce the $R$-matrix $R_{0,j}(u) \in \mathrm{End}(\bm{V}_0 \otimes \bm{V}_j)$ as
\begin{equation}
R_{0,j}(u) =
\begin{pmatrix}
u+\eta & 0 & 0 & 0 \\
0 & u & \eta & 0 \\
0 & \eta & u & 0 \\
0 & 0 & 0 & u+\eta \\
\end{pmatrix}, \label{eq:R-matrix}
\end{equation}
where $u$ is the spectral parameter and $\eta$ is the crossing parameter.

The $R$-matrix satisfies the following properties:
\begin{align}
\text{Initial condition:} \quad & R_{0,j}(0) = \eta P_{0,j}, \nonumber \\
\text{Unitarity:} \quad & R_{0,j}(u) R_{j,0}(-u) = \phi(u) \cdot \mathrm{id}, \nonumber \\
\text{Crossing symmetry:} \quad & R_{0,j}(u) = -\sigma^y_0 R^{t_0}_{0,j}(-u-\eta) \sigma^y_0, \nonumber \\
\text{PT-symmetry:} \quad & R_{0,j}(u) = R_{j,0}(u) = R^{t_0 t_j}_{0,j}(u), \nonumber \\
\text{$\mathbb{Z}_2$-symmetry:} \quad & \sigma_0^\alpha \sigma_j^\alpha R_{0,j}(u) = R_{0,j}(u) \sigma_0^\alpha \sigma_j^\alpha,\quad \alpha = x,y,z, \nonumber \\
\text{Fusion condition:} \quad & R_{0,j}(\pm \eta) = \pm 2\eta P^{(\pm)}_{0,j}, \label{eq:R-properties}
\end{align}
where $\phi(u) = \eta^2 - u^2$, $t_0$ ($t_j$) denotes transposition in space $\bm{V}_0$ ($\bm{V}_j$), and $P_{0,j}$ is the permutation operator.

The $R$-matrix satisfies the quantum Yang-Baxter equation (QYBE):
\begin{equation}
R_{1,2}(u_1 - u_2) R_{1,3}(u_1 - u_3) R_{2,3}(u_2 - u_3) = R_{2,3}(u_2 - u_3) R_{1,3}(u_1 - u_3) R_{1,2}(u_1 - u_2). \label{eq:QYBE}
\end{equation}

Define the monodromy matrices:
\begin{align}
T_0(u) &= R_{0,N}(u - \theta_N) \cdots R_{0,1}(u - \theta_1), \nonumber \\
\hat{T}_0(u) &= R_{1,0}(u + \theta_1) \cdots R_{N,0}(u + \theta_N), \label{eq:monodromy}
\end{align}
where $\bm{V}_0$ is the auxiliary space, $\bm{V}_1 \otimes \cdots \otimes \bm{V}_N$ is the quantum space, and $\{\theta_j\}$ are the inhomogeneity parameters.

The boundary reflection matrices are
\begin{align}
K^-(u) &= \begin{pmatrix} p + u & 0 \\ 0 & p - u \end{pmatrix}, \label{eq:Kminus} \\
K^+(u) &= \begin{pmatrix} q + u + \eta & \xi (u + \eta) \\ \xi (u + \eta) & q - u - \eta \end{pmatrix}, \label{eq:Kplus}
\end{align}
where $p, q, \xi$ are boundary parameters.

$K^-(u)$ satisfies the reflection equation
\begin{align}
  &R_{1,2}(\lambda - u)\,K^-_1(\lambda)\,R_{2,1}(\lambda + u)\,K^-_2(u)\nonumber\\
  =\quad&K^-_2(u)\,R_{1,2}(\lambda + u)\,K^-_1(\lambda)\,R_{2,1}(\lambda - u). \label{eq:RE}
\end{align}
and $K^+(u)$ satisfies the dual reflection equation
\begin{align}
  &R_{1,2}(-\lambda + u)\,K^+_1(\lambda)\,R_{2,1}(-\lambda - u - 2\eta)\,K^+_2(u)\nonumber\\
  =\quad&K^+_2(u)\,R_{1,2}(-\lambda - u - 2\eta)\,K^+_1(\lambda)\,R_{2,1}(-\lambda + u). \label{eq:dualRE}
\end{align}

Define the double-row monodromy matrix as
\begin{equation}
\bm{U}_0(u) = T_0(u) K^-_0(u) \hat{T}_0(u). \label{eq:double-row}
\end{equation}
It satisfies the same reflection equation:
\begin{equation}
R_{1,2}(\lambda - u) \bm{U}_1(\lambda) R_{2,1}(\lambda + u) \bm{U}_2(u) = \bm{U}_2(u) R_{1,2}(\lambda + u) \bm{U}_1(\lambda) R_{2,1}(\lambda - u). \label{eq:U-RE}
\end{equation}

\begin{figure}[htbp]
  \centering
  \includegraphics[width=0.65\textwidth]{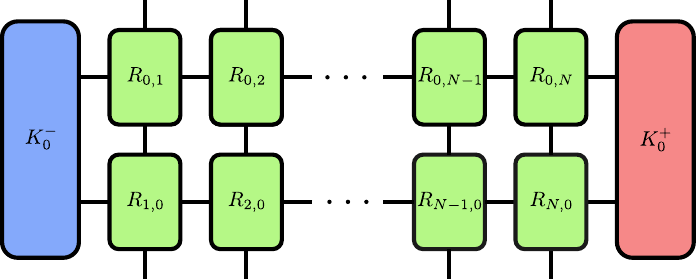}
  \caption{
   Tensor network diagram of the transfer matrix: horizontal bonds denote the auxiliary space labeled 0, and vertical bonds represent the quantum spaces labeled $1$ through $N$.
   }
  \label{fig:tu}
\end{figure}

The transfer matrix is defined by
\begin{equation}
t(u) = \mathrm{tr}_0 \{ K^+_0(u) \bm{U}_0(u) \}. \label{eq:transfer}
\end{equation}
In the tensor network representation, the transfer matrix $t(u)$ is illustrated in Fig.~\ref{fig:tu}.
It also satisfies the crossing symmetry
\begin{equation}
t(u) = t(-u - \eta). \label{eq:t-crossing}
\end{equation}

From the QYBE and reflection equations, the transfer matrices commute:
\begin{equation}
[t(u), t(v)] = 0, \label{eq:commutativity}
\end{equation}
which guarantees the integrability of the system.

The Hamiltonian of the open XXX spin-$\frac{1}{2}$ chain with general boundary fields is
\begin{equation}
H = \sum_{n=1}^{N-1} (\sigma^x_n \sigma^x_{n+1} + \sigma^y_n \sigma^y_{n+1} + \sigma^z_n \sigma^z_{n+1}) + \frac{\eta}{p} \sigma^z_1 + \frac{\eta}{q}(\sigma^z_N + \xi \sigma^x_N).
\label{eq:Hamiltonian}
\end{equation}

This Hamiltonian can be generated by the transfer matrix as
\begin{equation}
H = \eta \left. \frac{\partial \ln t(u)}{\partial u} \right|_{u=0, \{\theta_j = 0\}} - N.
\label{eq:H-from-transfer}
\end{equation}

To ensure the Hermiticity of the Hamiltonian \eqref{eq:Hamiltonian} for real $\eta$, the boundary parameters must satisfy:
\begin{equation}
q^* = q,\quad p^* = p,\quad \xi^* = \xi,\quad \text{when } \eta^* = \eta.
\label{eq:boundary-conditions}
\end{equation}

\subsection{ODBA}
Let us briefly review the ODBA method \cite{book}.
Using the properties of the $R$-matrix \eqref{eq:R-properties}, the eigenvalue $\Lambda(u)$ of $t(u)$ satisfies:
\begin{align}
\Lambda(u) &= \Lambda(-u - \eta), \label{eq:Lambda1} \\
\Lambda(0) &= a(0),  \label{eq:Lambda2}\\
\Lambda(u) &\sim 2u^{2N+2} + \dots,\quad u \to \pm\infty, \label{eq:Lambda3} \\
\Lambda(\theta_j) \Lambda(\theta_j - \eta) &= a(\theta_j) d(\theta_j - \eta),\quad j = 1,\dots,N, \label{eq:Lambda4}
\end{align}
where $a(u)$ and $d(u)$ are defined as
\begin{equation}
\begin{aligned}
a(u) &= \frac{2u + 2\eta}{2u + \eta}(u + p) [ (1 + \xi^2)^{\frac{1}{2}} u + q ] \prod_{j=1}^N (u + \theta_j + \eta)(u - \theta_j + \eta),\\
d(u) &= a(-u - \eta).
\end{aligned}
\end{equation}

These conditions allow us to construct the following inhomogeneous $T-Q$ relation for each eigenvalue $\Lambda(u)$:
\begin{align}
\Lambda(u) &= a(u)\frac{Q(u-\eta)}{Q(u)} + d(u)\frac{Q(u+\eta)}{Q(u)} + 2[1 - (1 + \xi^{2})^{\frac{1}{2}}] u(u + \eta) \nonumber \\
&\quad \times \frac{\prod_{j=1}^{N} (u + \theta_{j})(u - \theta_{j})(u + \theta_{j} + \eta)(u - \theta_{j} + \eta)}{Q(u)}.
\label{eq:T-Q relation}
\end{align}
The function $Q(u)$ is parameterized by $N$ Bethe roots $\{\lambda_{j} \mid j=1,\ldots,N\}$ as follows:
\begin{equation}
Q(u) = \prod_{j=1}^{N} (u - \lambda_{j})(u + \lambda_{j} + \eta).
\label{eq:Q-function}
\end{equation}

Now we can take the homogeneous limit $\theta_{j} \to 0$. In this case, the functional $T-Q$ relation is
\begin{align}
\Lambda(u) &= \frac{2(u+\eta)^{2N+1}}{2u+\eta}(u+p)[(1+\xi^{2})^{\frac{1}{2}} u+q]\frac{Q(u-\eta)}{Q(u)} \nonumber \\
&\quad + \frac{2u^{2N+1}}{2u+\eta}(u-p+\eta)[(1+\xi^{2})^{\frac{1}{2}}(u+\eta)-q]\frac{Q(u+\eta)}{Q(u)} \nonumber \\
&\quad + 2[1-(1+\xi^{2})^{\frac{1}{2}}]\frac{[u(u+\eta)]^{2N+1}}{Q(u)},
\label{eq:inhomogeneous T-Q relation}
\end{align}
and the BAEs read
\begin{align}
&(\frac{\lambda_{j}+\eta}{\lambda_{j}})^{2N+1}\frac{(\lambda_{j}+p)[(1+\xi^{2})^{\frac{1}{2}}\lambda_{j}+q]}{(\lambda_{j}-p+\eta)[(1+\xi^{2})^{\frac{1}{2}}(\lambda_{j}+\eta)-q]} = -\frac{Q(\lambda_{j}+\eta)}{Q(\lambda_{j}-\eta)} \nonumber \\
&\quad - \frac{[1-(1+\xi^{2})^{\frac{1}{2}}](2\lambda_{j}+\eta)(\lambda_{j}+\eta)^{2N+1}}{(\lambda_{j}-p+\eta)[(1+\xi^{2})^{\frac{1}{2}}(\lambda_{j}+\eta)-q] Q(\lambda_{j}-\eta)}, \quad j=1,\ldots, N.
\label{eq:BAEs}
\end{align}

The eigenvalue of the Hamiltonian in \eqref{eq:Hamiltonian} is given by
\begin{equation}
  E = \sum_{j=1}^{N}\frac{2\eta^2}{\lambda_j(\lambda_j+\eta)} + \frac{\eta}{p} + \frac{\eta\sqrt{1+\xi^{2}}}{q} + N - 1.
\label{Eigenvalue1}
\end{equation}

\subsection{Zero-root method}
Now, let us recall the zero-root method \cite{qiao20,qiao21}.
Using \eqref{eq:Lambda1} and \eqref{eq:Lambda3}, together with the fact that $\Lambda(u)$ is a degree-$(2N + 2)$ polynomial, we can simply parameterize it by its zero roots $\{z_{j} \mid j=1,\ldots,N+1\}$ as
\begin{equation}
\Lambda(u) = 2 \prod_{j=1}^{N+1} (u - z_j )(u + z_j + \eta).
\label{eq:lambda}
\end{equation}

Thus, the constrained equations (\ref{eq:Lambda2}) and (\ref{eq:Lambda4}) become
\begin{equation}
  \prod_{j = 1}^{N + 1} (-z_j)(z_j + \eta) = a(0),\label{zbae1}
\end{equation}
and
\begin{equation}
  \prod_{j = 1}^{N + 1} (\theta_j - z_j)(\theta_j + z_j +\eta) = a(\theta_j)a(-\theta_j),\quad j = 1,\ldots,N.\label{zbae2}
\end{equation}
In the homogeneous limit $\{\theta_j=0 \mid j=1,\ldots,N\}$, Eq.~\eqref{zbae1} and \eqref{zbae2} becomes
\begin{equation}
\prod_{j = 1}^{N + 1} (-z_j)(z_j + \eta) = 2pq,
\label{eq:hom-identity1}
\end{equation}
and
\begin{equation}
\big[\Lambda(u)\Lambda(u-\eta)\big]^{(n)}\Big|_{u=0} = \big[a(u)a(-u)\big]^{(n)}\Big|_{u=0},\qquad n=0,1,\ldots,N-1,
\label{eq:hom-identity2}
\end{equation}
where the superscript $(n)$ denotes the $n$-th derivative with respect to $u$.

The eigenvalue of the Hamiltonian \eqref{eq:Hamiltonian} can be expressed in terms of the zero roots
$\{z_j\}$ as follows:
\begin{equation}
  E = -\sum_{j=1}^{N+1}\frac{\eta^{2}}{z_j\bigl(z_j+\eta\bigr)} - N.
\label{eq:Eigenvalue2}
\end{equation}

\section{DMRG}
\label{sec:3}
\setcounter{equation}{0}
The density-matrix renormalization group (DMRG) method was originally a high-precision numerical algorithm for investigating the ground state and low-lying excitations of low-dimensional, strongly-correlated quantum many-body systems. Later, a variety of techniques were developed that extended DMRG from zero temperature to finite temperatures and from the calculation of static quantities to the evaluation of time-dependent or dynamical correlation functions \cite{Xiang2023}. These advances not only broadened the scope of DMRG but also stimulated the emergence of a new generation of algorithms based on so-called tensor-network states, which represent a quantum state as a product of interconnected tensors and provide a new tool for exploring quantum many-body systems.

\begin{figure}[htbp]
  \centering
  \includegraphics[width=0.45\textwidth]{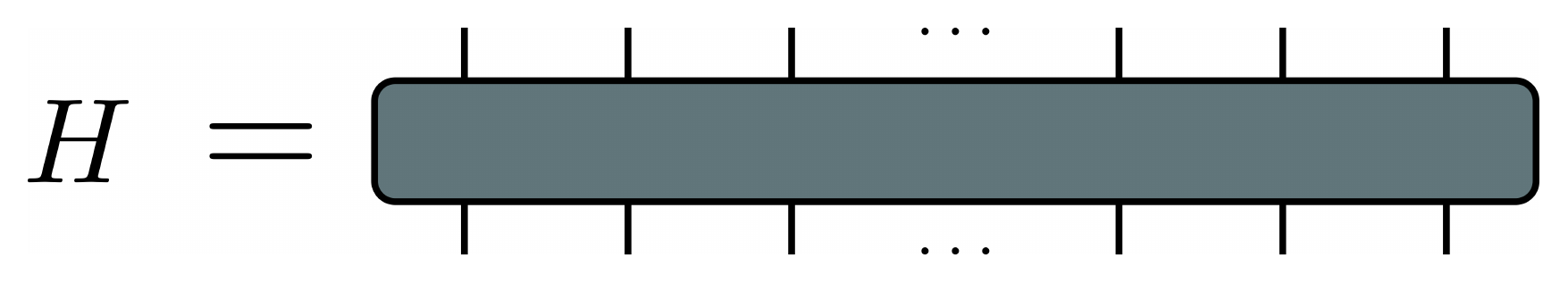}
  \caption{
   Tensor network representation of the Hamiltonian $H$. The network has $N$ input and $N$ output indices, each corresponding to a local Hilbert space of dimension $d$.
   }
  \label{fig:h}
\end{figure}

First, let us explain how DMRG obtains the ground state in the language of tensor networks.
Consider a Hermitian Hamiltonian $H$ acting on a vector space formed by the tensor product of $N$ local spaces, each of dimension $d$.
We depict $H$ as a tensor in Fig.~\ref{fig:h}.

\begin{figure}[htbp]
  \centering
  \includegraphics[width=0.8\textwidth]{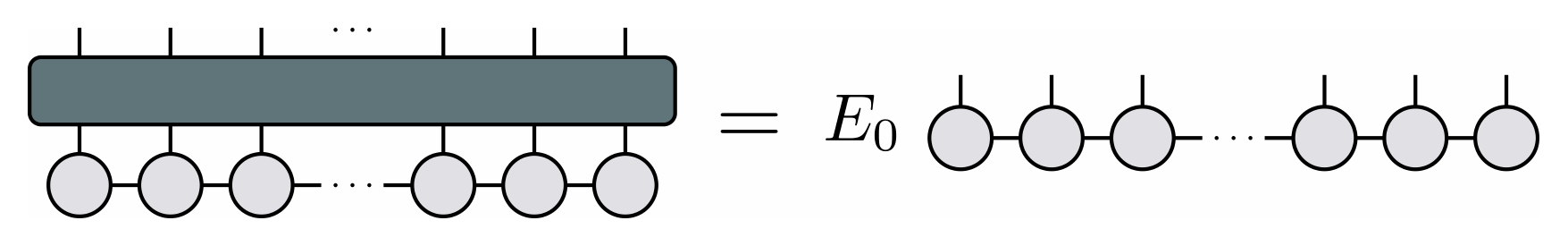}
  \caption{
  Here $E_{0}(\leq E_{1}\leq E_{2} \dots)$ denotes the lowest variational energy obtained by DMRG, which approximates the true ground-state eigenvalue of $H$.
  The shaded region below $H$ indicates the MPS representation of the approximate ground state corresponding to this energy.
  }
  \label{fig:mps}
\end{figure}

The DMRG method approximates the ground state of $H$ by constructing a Matrix Product State (MPS) representation, which is a tensor network that captures the entanglement structure of the system.
Starting from an initial MPS, the algorithm performs $n$ sweeps of energy-minimizing optimization: at each step it updates one or two site tensors and applies a singular-value decomposition (SVD), truncating the singular values according to two criteria---a maximum bond dimension $D$ and a maximum allowed truncation error that limits the discarded weight.
After $n$ sweeps, this variational procedure yields an MPS whose accuracy is controlled jointly by the bond-dimension limit $D$, the truncation-error threshold, and the total number of sweeps.
In our calculations, the truncation error is set to $1 \times 10^{-12}$, the maximum bond dimension is $D = 200$, and the convergence criterion is defined by a minimum of $30$ sweeps with an energy convergence tolerance of $1 \times 10^{-14}$.
The overall goal of the DMRG procedure is illustrated schematically in Fig.~\ref{fig:mps}.

\begin{figure}[htbp]
  \centering
  \includegraphics[width=0.45\textwidth]{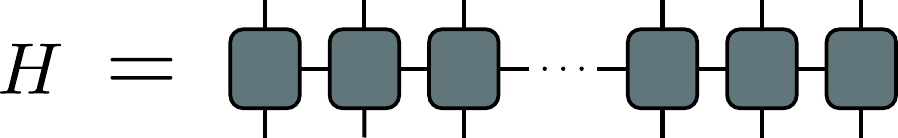}
  \caption{
   Tensor network representation of the Hamiltonian $H$ in MPO form. Each tensor represents a local operator with one physical index on the top and bottom, and virtual indices connecting neighboring sites.
  }
  \label{fig:mpo}
\end{figure}

To make the algorithm efficient, the Hamiltonian $H$ is usually encoded in the form of a Matrix Product Operator (MPO); see Fig.~\ref{fig:mpo}. This MPO representation proves particularly advantageous for numerical implementations, including tensor network contractions.

Meanwhile, Fig.\ref{fig:tu} illustrates that the transfer matrix $t(u)$ can be expressed as a tensor network in MPO form.
The transformation from the double-row transfer matrix to the MPO form is illustrated in Fig.~\ref{fig:transform}. The derivation details are provided in Appendix~\ref{app:A}. Consequently, the transfer matrix is reformulated as an MPO that can be directly handled by standard tensor-network techniques.
\begin{figure}[htbp]
  \centering
  \includegraphics[width=1\textwidth]{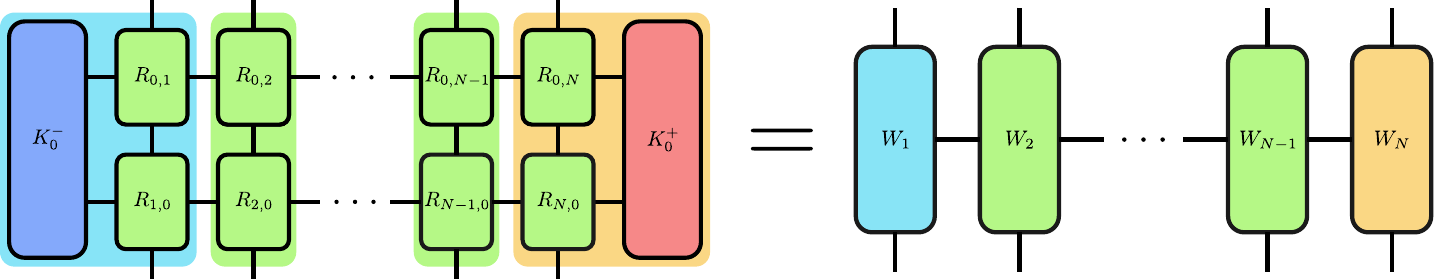}
  \caption{
  Transformation of the double-row transfer matrix into an MPO.
  Each local tensor $W_i$ on the right is obtained by contracting the vertical pair of $R$-matrices at site $i$ in the double-row structure, along with the adjacent boundary tensors at the first and last sites.
  }
  \label{fig:transform}
\end{figure}

It should be noted that, although the Hamiltonian and the transfer matrix commute, their respective ground states are not identical. Our goal is to compute the eigenvalues of the transfer matrix within the ground state of the Hamiltonian given in Eq.~\eqref{eq:Hamiltonian}.
Therefore, we first need to obtain the ground state of the Hamiltonian.
To address this, we employ the ITensor library~\cite{ITensor2022}.
Within ITensor, the MPO representation of the Hamiltonian can be constructed straightforwardly, and its ground state MPS can be efficiently obtained through DMRG. Subsequently, we compute the inner product of the Hamiltonian's ground state MPS with the transfer-matrix MPO, thereby extracting the eigenvalues of the latter.

The errors stem from two sources: first, the numerical error introduced by DMRG in computing $\Lambda(u)$; second, the additional error incurred when extracting the zeros or Bethe roots from the DMRG-supplied $\Lambda(u)$. Using the $U(1)$-symmetric case as a benchmark, we confirm that both contributions are extremely small; in the $U(1)$-symmetry-broken case, a dedicated verification routine further confirms that the second error remains negligible.

\section{Zero roots}
\label{sec:4}
\setcounter{equation}{0}
Using the approach described in the previous section, the eigenvalue $\Lambda(u)$ of the transfer matrix $t(u)$ at any $u$ can be readily computed within the ground state of the Hamiltonian Eq.~\eqref{eq:Hamiltonian}, even for large $N$.
We can use the DMRG-computed $\Lambda(u)$ to locate its zeros. The Bethe Ansatz-DMRG (BA-DMRG) procedure is summarized as follows:
\begin{enumerate}
  \item Obtain the ground state MPS of the Hamiltonian.
  \item Represent the transfer matrix $t(u)$ as an MPO.
  \item Select a set of interpolation points $\{u_j\}$ and compute the corresponding eigenvalues $\Lambda(u_j)$ of the transfer matrix in the ground state.
  \item Reconstruct $\Lambda(u)$ using the Lagrange interpolation method.
  \item Determine the zero roots of $\Lambda(u)$ through numerical root-finding algorithms.
  \item Verify the correctness of the obtained roots (the specific procedures and principles will be presented in the following sections); if the verification fails, return to step~3 and use the current results as new interpolation points.
\end{enumerate}

In step 3, from Eq.~\eqref{eq:lambda}, the eigenvalue polynomial $\Lambda(u)$ can be parameterized in terms of $2N + 2$ points $\{z_j, -z_j-\eta\mid j=1,\dots,N+1\}$.
To improve the accuracy of the interpolation procedure, the interpolation nodes $\{u_j\}_{j=1}^{2N+2}$ are constructed as $\{x_j,-x_j-\eta \mid j=1,\dots,N+1\}$. Specifically, the nodes are chosen as
$x_j=\frac{\eta}{2}+\frac{i}{N}\left(j-\frac{N}{2}\right)^k,$
where $j=1,2,\dots,N$ and $k\in[1,1.1]$, with $i$ denoting the imaginary unit.
These nodes are chosen based on the numerical results obtained for small lattice sizes.
This choice provides a denser sampling near the root accumulation region, thereby improving the accuracy of the interpolation.

In step 4, through a Lagrange polynomial representation, any degree-$(2N + 2)$ polynomial can be uniquely determined by its values at $2N + 3$ arbitrary grid points.
In the present case (see Eq.~\eqref{eq:lambda}), since the leading coefficient of $\Lambda(u)$ is known to be 2, only $2N + 2$ points are required to fully specify the polynomial
\begin{equation}
\Lambda(u) =2\prod_{j=1}^{2N+2} (u-u_j) + \sum_{j=1}^{2N+2} \Lambda(u_j) \prod_{k \ne j} \frac{u - u_k}{u_j - u_k},
\label{lag}
\end{equation}
where the interpolation nodes $\{u_j\}_{j=1}^{2N+2}$ are those defined in step 3.

In step 6, if the root verification fails, the current results are used as updated interpolation nodes, and the zero roots are recalculated iteratively until the desired precision is achieved.

\subsection{Ground state solution with U(1) symmetry}
\label{sec:homogeneous}
At $\xi = 0$, the two boundary fields are parallel, endowing the system with a global $U(1)$ symmetry. Consequently, the BAEs in Eq.~\eqref{eq:BAEs} reduce to the following homogeneous form:
\begin{equation}
(\frac{\lambda_{j}+\eta}{\lambda_{j}})^{2N+1}\frac{(\lambda_{j}+p)(\lambda_{j}+q)}{(\lambda_{j}+\eta-p)(\lambda_{j}+\eta-q)} = -\frac{Q(\lambda_{j}+\eta)}{Q(\lambda_{j}-\eta)}, \quad j=1,\ldots, N.
\label{eq:BAEs-homogeneous}
\end{equation}
In this case, the inhomogeneous term in the $T-Q$ relation vanishes and part of the $N$ Bethe roots may take the value of infinity.
The function $Q(u)$ is thus reduced to
\begin{equation}
Q(u) = \prod_{j=1}^{M} (u - \lambda_j)(u + \lambda_j + \eta), \qquad M = 0, \ldots, N.
\label{eq:Q-function-homogeneous}
\end{equation}
For the homogeneous BAEs \eqref{eq:BAEs-homogeneous}, we recast them into logarithmic form to enhance numerical accuracy. This well-established method is used here to generate reference data for comparison with the BA-DMRG solution, thereby verifying the reliability of the latter.

For convenience, we put $\eta=1$, $\mu_j=i\lambda_j-\tfrac{1}{2}$, $\hat{p}=p-\tfrac{1}{2}$, $\hat{q}=q-\tfrac{1}{2}$.
Eq.~\eqref{eq:BAEs-homogeneous} can be rewritten as
\begin{equation}
\frac{(\mu_j - i \hat{q})(\mu_j - i \hat{p})}{(\mu_j + i \hat{q})(\mu_j + i \hat{p})}
\left( \frac{\mu_j - \tfrac{i}{2}}{\mu_j + \tfrac{i}{2}} \right)^{2N}
= \prod_{\substack{l=1 \\ l \neq j}}^{M}
\frac{(\mu_j - \mu_l - i)(\mu_j + \mu_l - i)}{(\mu_j - \mu_l + i)(\mu_j + \mu_l + i)}.
\label{eq:mu}
\end{equation}
Taking the logarithm of Eq.~\eqref{eq:mu}, we have
\begin{equation}
\theta_{2\hat{p}}(\mu_j) + \theta_{2\hat{q}}(\mu_j) + 2N\theta_{1}(\mu_j)
= 2\pi I_j + \sum_{l=1}^{M} \left[ \theta_{2}(\mu_j - \mu_l) + \theta_{2}(\mu_j + \mu_l) \right] - \theta_{1}(\mu_j),
\label{eq:logrithm}
\end{equation}
where $\theta_n(x) = 2 \arctan(2x / n)$, and $\{I_j\}$ are certain integers (half-odd integers) for $N - M$ odd ($N - M$ even).

Through analytical considerations \cite{book}, it can be concluded that, for the ground state, the ratio $\tfrac{M}{N} = \tfrac{1}{2}$ holds, and the quantum numbers $\{I_j\}$ are continuously distributed.
For even $N$, the quantum numbers can be expressed as $I_j = -\tfrac{M}{2} + j - 1$, where $j = 1, 2, \dots, M$.
By substituting these quantum numbers into Eq.~\eqref{eq:logrithm}, the $M$ Bethe roots $\{\mu_j\}$ corresponding to the ground state can be numerically determined.
After transforming $\{\mu_j\}$ into $\{\lambda_j\}$, $Q(u)$ can be constructed, from which $\Lambda(u)$ is obtained via Eq.~\eqref{eq:inhomogeneous T-Q relation}.
Finally, the zero roots $\{z_j\}$ are extracted using a standard numerical root-finding procedure.

\begin{figure}[htbp]
  \centering
    \subfloat[Global configuration of zero roots]{
    \includegraphics[width=0.47\textwidth]{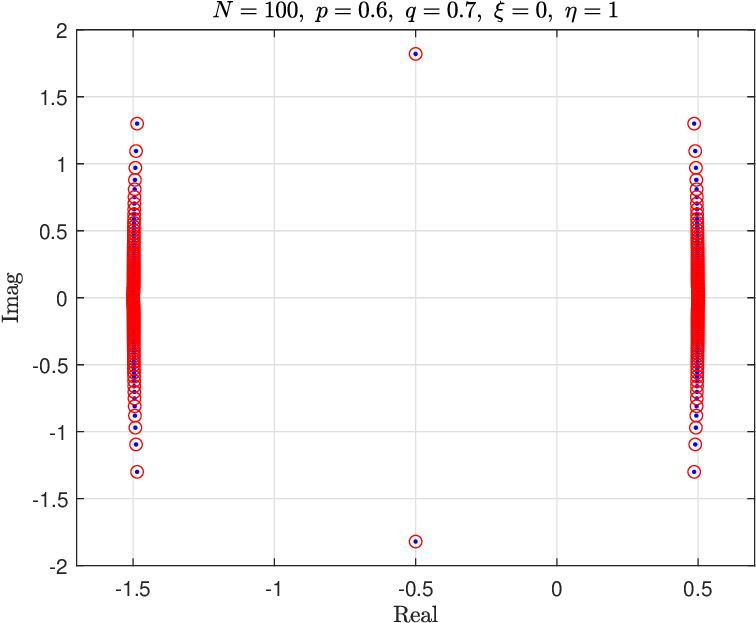}
    }
  \subfloat[Enlarged view of the right cluster]{
    \includegraphics[width=0.4895\textwidth]{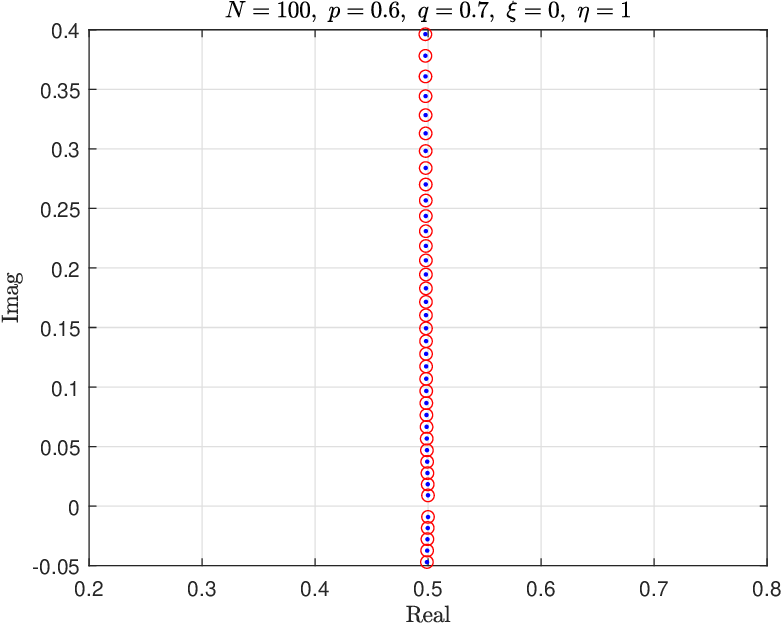}
  }
\caption{
Distribution of zero roots in the complex plane with parameters $N = 100$, $p = 0.7$, $q = 0.6$, $\xi = 0$, and $\eta = 1$.
Red circles represent zero roots obtained from the logarithmic form of the BAEs, and blue dots correspond to those computed using the BA-DMRG method.
Subplot (b) is an enlarged view of the central portion of the right-hand line in subplot (a).}
\label{fig:compare}
\end{figure}

Fig.~\ref{fig:compare} compares the zero roots obtained from the logarithmic BAEs, which are solved with high numerical precision, and those computed using the BA-DMRG method, showing an almost perfect match.
Since the BAEs solutions are taken as the accurate benchmark, the strong agreement between the two sets of results---with the maximum absolute error of $6.72\times10^{-8}$---demonstrates that the BA-DMRG method can reliably and accurately compute the zero roots.

\subsection{Ground state solution without U(1) symmetry}
\begin{figure}[htbp]
  \centering
  \includegraphics[width=0.5\textwidth]{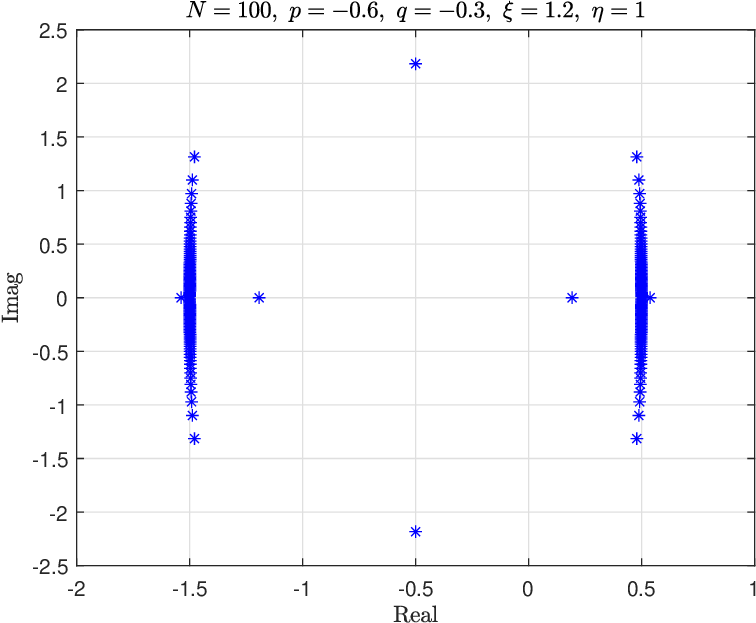}
\caption
{
Distribution of zero roots in the complex plane with parameters $N = 100$, $p = -0.6$, $q = -0.3$, $\xi = 1.2$, and $\eta = 1$.
}
\label{fig:inhomogeneous-roots}
\end{figure}

At $\xi \ne 0$, the two boundary fields are non-parallel, so the system no longer possesses global $U(1)$ symmetry.
The BAEs in Eq.~\eqref{eq:BAEs} are inhomogeneous, so their logarithmic form is unavailable.
Nevertheless, we can still compute the zero roots with the BA-DMRG method.
The results are presented in Fig.~\ref{fig:inhomogeneous-roots}, while the detailed numerical values of the roots and the auxiliary data are provided in Table~\ref{tab:tablec}.

%


For large \(N\), \(\Lambda(u)\) is of high degree, and a poor choice of interpolation nodes can stall convergence. In the homogeneous case, we have verified that $\Lambda(u)$ of the transfer matrix obtained via DMRG is highly accurate (the relative error of $\Lambda(u)$ at $N=100$ computed by DMRG is no worse than $10^{-10}$). Based on this, it suffices to verify the accuracy of the iteration procedure.
Consequently, the zero roots obtained by BA-DMRG demand an independent accuracy check.
At large lattice sizes even sizeable displacements of some roots may leave the energy practically unchanged, so validation via the energy expression \eqref{eq:Eigenvalue2} is no longer sufficiently stringent.
Moreover, directly checking the condition $\Lambda(z_j)=0$ is not a robust
verification criterion in practice. For large $N$, $\Lambda(u)$ is a polynomial
of degree $2N+2$, and the residual $|\Lambda(z_j)|$ can be highly sensitive to
small errors in the root position. Indeed, if $z_j^\ast$ denotes the exact zero
and $z_j$ is a nearby numerical root, then
\[
\Lambda(z_j)\simeq \Lambda'(z_j^\ast)(z_j-z_j^\ast),
\]
where $|\Lambda'(z_j^\ast)|$ may become very large for a high-degree polynomial.
Consequently, even an accurate numerical root can give a sizable absolute
residual. Therefore, the unnormalized value of $|\Lambda(z_j)|$ alone is not a
reliable indicator of root accuracy.

To verify the zero roots obtained by the BA-DMRG method, we instead employ two
independent analytic tests: the maximum-modulus principle and the argument
principle. The former uses the local analytic behavior of $\Lambda(u)$ near a
candidate zero, while the latter directly counts the number of zeros enclosed by
a closed contour.

\subsubsection{Maximum modulus principle}
Let $\{z_j^\prime\}$ denote the computed zero roots, for which $\Lambda(z_j^\prime)$ should ideally vanish.
Owing to the high degree of $\Lambda(u)$ for large system size $N$, the evaluated $\Lambda(z_j^\prime)$ may appear large, so $\{z_j^\prime\}$ might seem not to be true zeros.
In reality, the genuine zeros of $\Lambda(u)$ lie extremely close to the computed $\{z_j^\prime\}$; in other words, $\{z_j^\prime\}$ are in fact highly accurate. This can be rigorously justified using the maximum modulus principle.
The maximum modulus principle is a fundamental result in complex analysis. It states that:

\textit{If a function is analytic and non-constant within a domain, then its maximum modulus cannot occur strictly inside the domain--it must occur only on the boundary.}

This property places strong constraints on the behavior of analytic functions in the complex plane.
Based on this principle, we consider a useful corollary:

\textit{Suppose a function $f(z)$ is analytic on the closed disk $|z| \leq R$, and there exists a constant $a > 0$ such that
\begin{equation}
    |f(z)| > a \quad \text{for all } |z| = R, \quad \text{and} \quad |f(0)| < a,
    \label{eq:corollary-condition}
\end{equation}
then $f(z)$ must have at least one zero in the open disk $|z| < R$.}

The derivation of the corollary is given in Appendix~\ref{app:B}. More generally, if the modulus $|f(z)|$ demonstrates concave behavior around a point $z_j^\prime$, then $f(z)$ must vanish at some point within the enclosed region.
This reflects the fact that analytic functions cannot exhibit a strict local minimum of modulus unless the function is identically zero or vanishes somewhere inside the domain.

\begin{figure}[htbp]
  \centering
  \subfloat[$\mathrm{Im}(u) = 0$]{
    \includegraphics[width=0.45\textwidth]{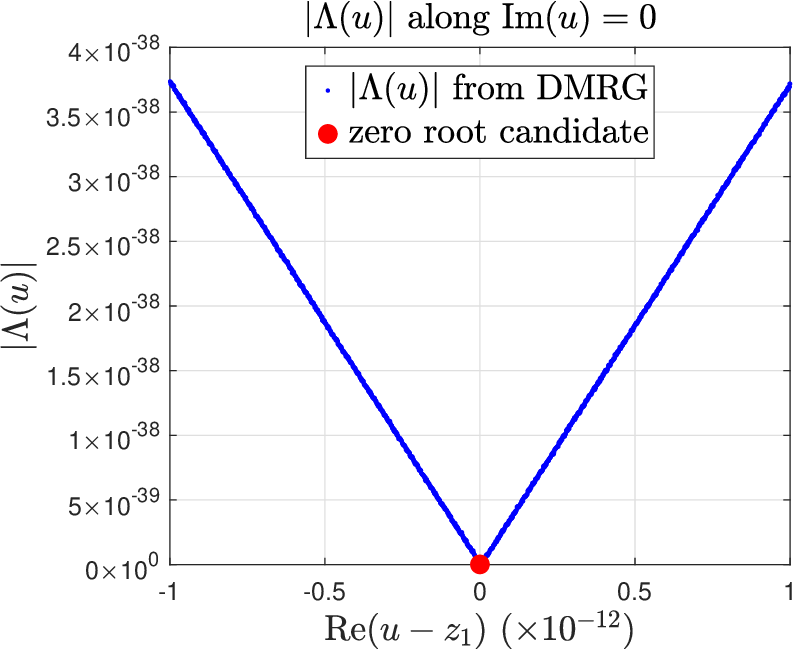}
  }
  \subfloat[$\mathrm{Re}(u) = -0.5$]{
    \includegraphics[width=0.45\textwidth]{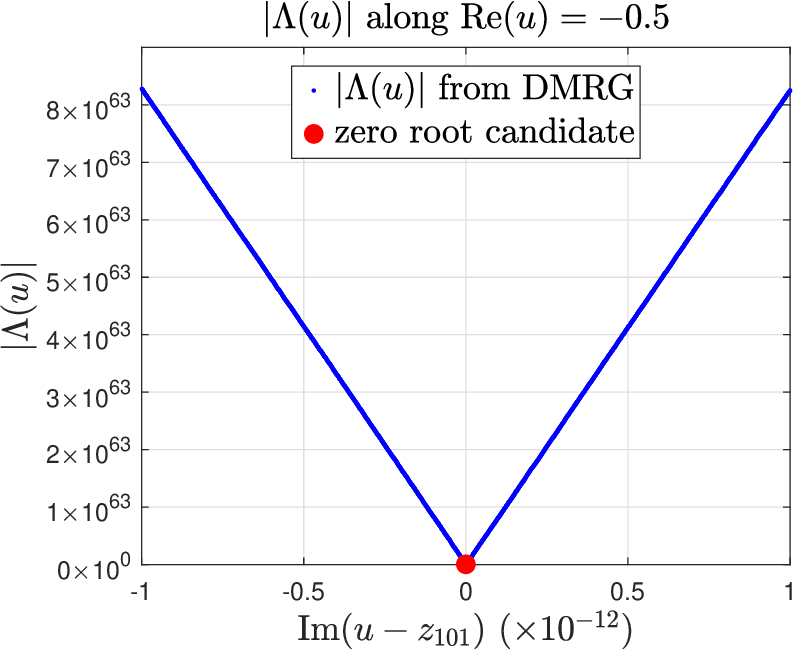}
  }\\
  \subfloat[complex-$u$ plane]{
    \includegraphics[width=0.5\textwidth]{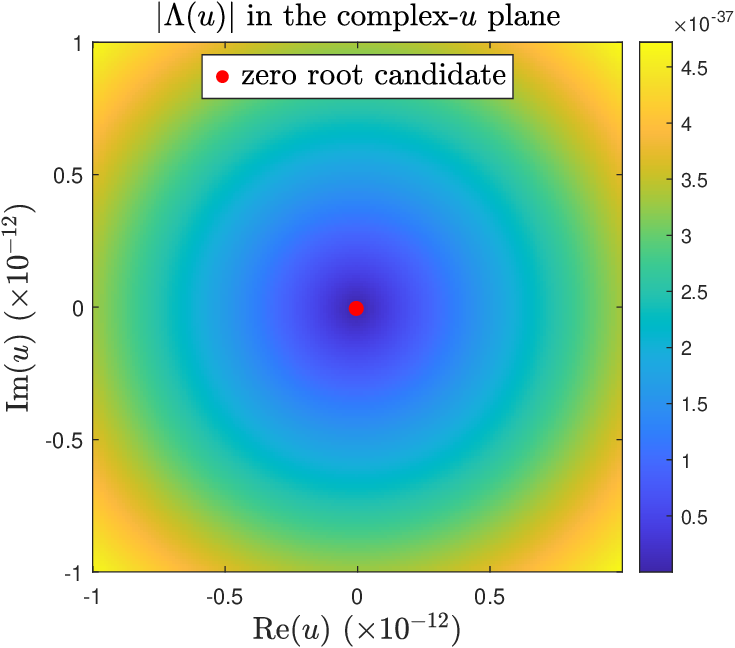}
  }
\caption{
Local behavior of the eigenvalue modulus $|\Lambda(u)|$ around three
representative zero-root candidates (indices 1, 101, and 37 in
Table~\ref{tab:tablec}), with
$z_{1} = -1.537797484404868 + 0.000000000000011i$,
$z_{101} = -0.500000001273242 + 2.182890275358742i$,
and $z_{37} = -1.498541855014614 + 0.190968750516886i$.
All panels are plotted in relative coordinates centered at the respective
zero-root candidates.
(a)~$|\Lambda(u)|$ along the real axis ($\mathrm{Im}(u)=0$).
(b)~$|\Lambda(u)|$ along the vertical line $\mathrm{Re}(u)=-0.5$.
(c)~Two-dimensional modulus distribution of $|\Lambda(u)|$ in the complex-$u$
plane obtained by DMRG.
A uniform coordinate precision of $10^{-12}$ is used throughout.
}
\label{fig:lambda_mod_valleys}
\end{figure}

We apply this result to the eigenvalue function $\Lambda(u)$ obtained from the DMRG calculations.
For each candidate zero root $z_j^\prime$, we numerically evaluate the modulus $|\Lambda(u)|$ over a neighborhood around $z_j^\prime$ in the complex plane.
In all tested cases, the modulus $|\Lambda(u)|$ exhibits a clear valley-shaped behavior: its value at $z_j^\prime$ is significantly smaller than that on the surrounding contour. According to the corollary of the maximum-modulus principle, such a valley-shaped behavior guarantees the existence of a true zero within the neighborhood.
Fig.~\ref{fig:lambda_mod_valleys} shows $|\Lambda(u)|$ sampled around three representative zero-root candidates listed in Table~\ref{tab:tablec}: $z_{1} = \seqsplit{-1.537797484404868 + 0.000000000000011i}$, $z_{101} = \seqsplit{-0.500000001273242 + 2.182890275358742i}$, and $z_{37} = \seqsplit{-1.498541855014614 + 0.190968750516886i}$.
These three points represent the typical structural types of zero roots in the system.
The $|\Lambda(u)|$ computed by DMRG exhibits a pronounced concave dip, thereby confirming the reliability of the identified zero-root candidates.


The maximum modulus principle allows for a visually intuitive identification of zeros via the concave behavior of the modulus. It provides direct and accessible evidence of zero-root locations. This approach is particularly effective when the evaluation points lie along straight lines, where plotting is efficient.
However, verifying zeros across the full complex plane becomes computationally expensive if high visual resolution is required. Therefore, to rigorously validate all candidate points, we further employ the argument principle, which gives a definitive count of zeros within closed contours.

\subsubsection{Argument principle}
The argument principle establishes a fundamental connection between the number of zeros and poles of a meromorphic function inside a closed contour and the total change in the function's argument along that contour:
\begin{equation}
  N_{\text{zeros}} - N_{\text{poles}} = \frac{1}{2\pi i} \oint_C \frac{f'(z)}{f(z)}\, dz = \frac{1}{2\pi} \Delta_C \arg f(z), \label{apcheck}
\end{equation}
where $\Delta_C \arg f(z)$ denotes the net change in the function's argument as the contour $C$ is traversed once in the positive (counterclockwise) direction.

In this work, the eigenvalue function $\Lambda(u)$ is analytic and free of poles.
Hence, the total argument change along any closed contour directly equals the number of enclosed zeros.

\begin{figure}[htbp]
  \centering
  \includegraphics[width=0.25\textwidth]{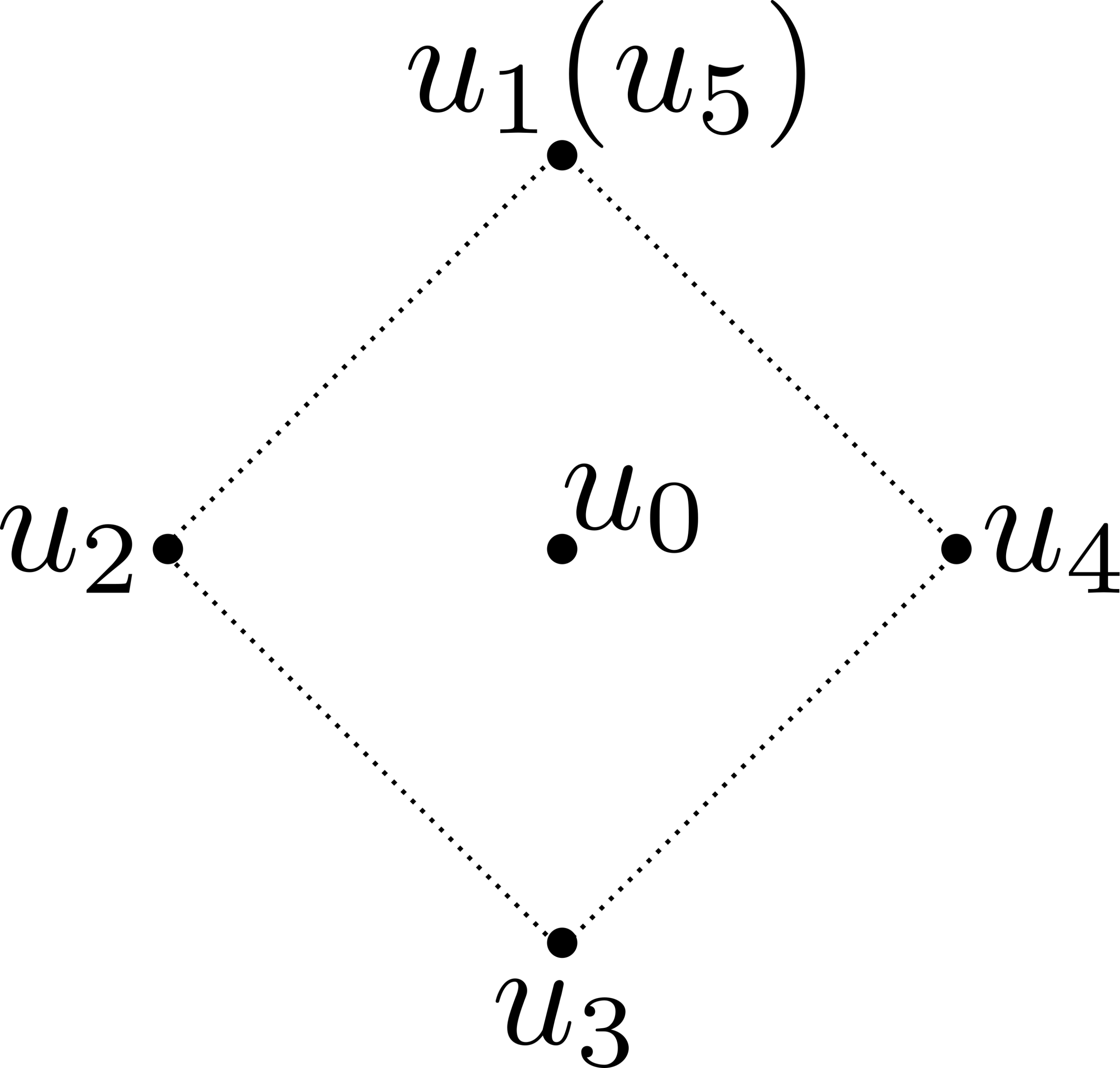}
\caption
{
Illustration of the discrete rectangular contour used in the argument principle test.
The path is centered at the candidate root $u_0$ and consists of five points:
$u_1 = u_0 + i\delta$,
$u_2 = u_0 - \delta$,
$u_3 = u_0 - i\delta$,
$u_4 = u_0 + \delta$,
and $u_5 = u_1$ to close the loop.
The points are symmetrically offset by a small value $\delta$ along the real and imaginary directions. This minimal closed contour enables phase unwrapping to detect whether exactly one zero lies inside.
}
\label{fig:u0}
\end{figure}

To apply this principle numerically, we construct a small discrete closed contour around each candidate zero root $u_0$, composed of five evaluation points arranged as shown in Fig.~\ref{fig:u0}.
These points are symmetrically offset from $u_0$ by a small value $\delta$ along the real and imaginary directions and connected in a counterclockwise order to form a rectangular loop.
The path starts and ends at the same point to ensure formal closure.
Importantly, the parameter $\delta$ is chosen to be much smaller than the minimal distance between any two zero roots. This guarantees that each contour encloses at most one zero.
With this contour, $\Delta_C \arg f(z) = \arg f(u_5)-\arg f(u_1)$.

The rationale for using these five points lies in the nature of argument evaluation: when computing $\arg f(z)$ numerically, the returned values lie within the principal branch $(-\pi, \pi]$. As such, the raw argument at the starting $\arg f(u_1)$ and ending points $\arg f(u_5)$ is always the same, even if the function’s phase has changed by a multiple of $2\pi$ along the path.
To detect any such hidden change, intermediate points are needed. These enable the use of phase unwrapping, a technique that corrects artificial phase jumps by adding or subtracting $2\pi$ whenever adjacent phase differences exceed $\pi$.

With this setup, if the path encloses a zero, the unwrapped argument will exhibit a net change close to $2\pi$, consistent with the theoretical prediction. Conversely, if no zero is enclosed, the total unwrapped change remains zero. This makes the verification result \eqref{apcheck} strictly binary--either 0 or 1.

Table~\ref{tab:phase_unwrap_example} illustrates this approach using the candidate zero root indexed 202 in Table~\ref{tab:tablec}, showing how the unwrapped total argument change depends on the choice of $\delta$ and confirming the detection of a single zero.
The zero is successfully detected when $\delta=10^{-14}$, but it is missed as $\delta$ is reduced to $10^{-15}$, indicating that $z_{202} = \seqsplit{0.537797484404868 - 0.000000000000011i}$ is located with an accuracy of the order of $10^{-14}$.
All candidate roots obtained by the BA-DMRG method and their verifications are summarized in Table~\ref{tab:tablec}.
For the 202 zero roots at \(N=100\), the accuracy is better than \(1 \times 10^{-12}\), providing strong numerical evidence for the correctness of the roots.
\begin{table}[htbp]
\centering
\caption
{
Effect of the unwrapping threshold~$\delta$ on the net argument change and the resulting zero count $\Delta = \frac{1}{2\pi} \Delta_C \arg f(z)$.
The five sampling points $u_1,\dots,u_5$ (locations as shown in Fig.~\ref{fig:u0}) encircle the candidate root $z_{202} = 0.537797484404868 - 0.000000000000011i$.
With $\arg_j\equiv\arg f(u_j)$, the unwrapped $\arg_j$ are listed for each~$\delta$; values in parentheses give the raw principal values.
}
\label{tab:phase_unwrap_example}
\begin{tabular}{ccccccc}
\toprule
$\delta$ & $\arg_1$ & $\arg_2$ & $\arg_3$ & $\arg_4$ & $\arg_5$  & $\Delta$ \\
\midrule
$10^{-13}$ & 1.557 (1.557) &  3.142 (-3.142) & 4.726 (-1.557) & 6.283 (0) & 7.84 (1.557) & 1.0 \\
\midrule
$10^{-14}$ & 1.436 (1.436) & 3.142 (-3.142) & 4.847 ( -1.436) & 6.283 (0) & 7.719 (1.436) & 1.0 \\
\midrule
$10^{-15}$ & 0.634 (0.634) & 0 (0) & -0.634 (-0.634) & 0 (0) & 0.634 (0.634) & 0.0 \\
\bottomrule
\end{tabular}
\end{table}

This approach achieves high efficiency by evaluating the function at only a few strategically chosen points per candidate.
It eliminates the need for dense sampling, making it particularly advantageous for large-scale zero detection tasks that demand rigorous numerical accuracy.
We note that the dominant source of error is the numerical inaccuracy introduced by DMRG when computing $\Lambda(u)$.

\subsection{Structure of ground state}
\label{sec:z_g}
\begin{figure}[htbp]
  \centering
  \includegraphics[width=0.5\textwidth]{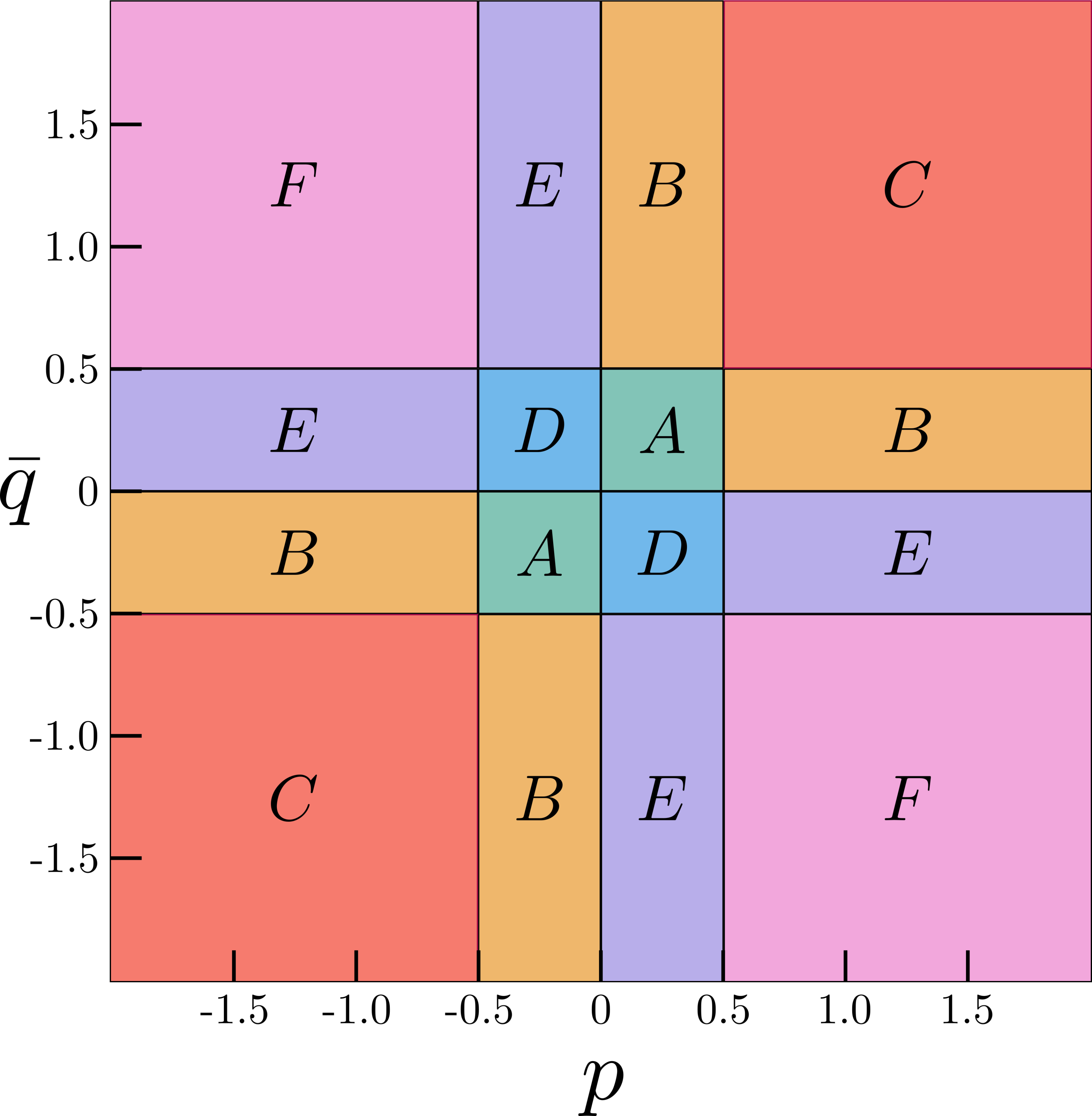}
\caption
{
Phase diagram of zero-root configurations for the ground state in the ($p, \bar{q}$) plane of boundary parameters at $\eta = 1$.
}
\label{fig:phase}
\end{figure}

After confirming the accuracy of the zero roots, we proceed to analyze their structural properties as functions of the boundary parameters.
The phase diagram and accompanying description of root structures, first presented in Ref.~\cite{Dong2023}, are recast into the notation used here in Fig.~\ref{fig:phase} and Table~\ref{tab:phase}.
It should be noted that we have revised the root classification given in Ref.~\cite{Dong2023}.
Unless otherwise specified, all results presented in this paper are for even $N$.
For convenience, we define $\bar{q} = \tfrac{q}{\sqrt{1+\xi^2}}$.
While these structural characteristics have been extensively studied in Ref.~\cite{Dong2023}, the corresponding phase diagram for the off-diagonal case was established using a relatively small system size ($N \sim 10$). In this work, we verify these conclusions for large system size (\(N \sim 100\)) using the BA-DMRG method. Furthermore, we provide a detailed analysis of the root structures as the parameters cross phase boundaries.
\begin{table}[htbp]
\centering
\caption{The patterns of zero roots for the ground state in different regions.
Roots related to $\tilde{z}_j$ form the bulk strings; those tied to $|p|$ or $|\bar{q}|$ constitute boundary strings; roots related to $\tilde{z}_x$ are the additional roots, while those related to $\tilde{z}_0$ are the helical half-spinon.
}
\label{tab:phase}
\begin{tabular}{c c}
\toprule
Region & $\{z_j, -z_j-\eta\}$ \\
\midrule
A & \makecell[c]{
  $\pm(\tilde{z}_j + \eta) - \tfrac{\eta}{2} \;|\; j=1,\ldots,N-2,$ \\
  $\pm \tilde{z}_0 - \tfrac{\eta}{2} ,\; \pm(|p| + \tfrac{\eta}{2})- \tfrac{\eta}{2} ,\; \pm(|\bar{q}| + \tfrac{\eta}{2}) - \tfrac{\eta}{2}\, $
} \\
\midrule
B & \makecell[c]{
  $\pm(\tilde{z}_j + \eta) - \tfrac{\eta}{2}   \;|\; j=1,\ldots,N-2,$ \\
  $\pm \tilde{z}_0 - \tfrac{\eta}{2} ,\; \pm(\min(|p|,|\bar{q}|) + \tfrac{\eta}{2}) - \tfrac{\eta}{2} ,\; \pm(\tilde{z}_x + \tfrac{\eta}{2}) -\tfrac{\eta}{2} \, $
} \\
\midrule
C & \makecell[c]{
  $\pm \tilde{z}_0 - \tfrac{\eta}{2} ,\; \pm(\tilde{z}_j + \eta) - \tfrac{\eta}{2}  \;|\; j=1,\ldots,N$
} \\
\midrule
D & \makecell[c]{
  $\pm(\tilde{z}_j + \eta) - \tfrac{\eta}{2} \;|\; j=1,\ldots,N-2,$ \\
  $\pm(\tilde{z}_x + \tfrac{\eta}{2}) -\tfrac{\eta}{2}\, ,\; \pm(|p| + \tfrac{\eta}{2})- \tfrac{\eta}{2} ,\; \pm(|\bar{q}| + \tfrac{\eta}{2}) - \tfrac{\eta}{2}\, $
} \\
\midrule
E & \makecell[c]{
  $\pm(\tilde{z}_j + \eta) - \tfrac{\eta}{2}  \;|\; j=1,\ldots,N,$ \\
  $\pm(\min(|p|,|\bar{q}|) + \tfrac{\eta}{2}) - \tfrac{\eta}{2}$
} \\
\midrule
F & \makecell[c]{
  $\pm(\tilde{z}_x + \tfrac{\eta}{2}) -\tfrac{\eta}{2}\, ,\; \pm(\tilde{z}_j + \eta) - \tfrac{\eta}{2} \;|\; j=1,\ldots,N$
} \\
\bottomrule
\end{tabular}
\end{table}

From Fig.~\ref{fig:phase} and Table~\ref{tab:phase} we conclude that, for the ground state, the zero roots fall into three classes:
\begin{enumerate}
\item \textbf{Bulk strings:}
\[
\pm(\tilde{z}_j +\eta)-\frac{\eta}{2}.
\]
All $\tilde{z}_j $ are purely imaginary; these roots are the most numerous and form two straight lines parallel to the imaginary axis.

\item \textbf{Boundary strings:}
\begin{itemize}
\item[$\bullet$] \emph{$p$-boundary string:} \quad $\pm\!\left(|p|+\dfrac{\eta}{2}\right)-\dfrac{\eta}{2}$. \\[-3mm]
\item[$\bullet$] \emph{$\bar{q}$-boundary string:} \quad $\pm\!\left(|\bar q|+\dfrac{\eta}{2}\right)-\dfrac{\eta}{2}$.
\end{itemize}
Their positions are completely fixed by the boundary parameters.


\item \textbf{Additional roots:}
\[
\pm z_x=\pm(\tilde{z}_x + \tfrac{\eta}{2})- \tfrac{\eta}{2}.
\]
Here $z_x$ is introduced for convenience.
$\tilde{z_x}$ is a real number greater than $\frac{\eta}{2}$.

\item \textbf{Helical half-spinon:}
\[
\pm z_0=\pm \tilde{z}_0 - \tfrac{\eta}{2}.
\]
Here $z_0$ is introduced for convenience.
$\tilde{z}_0$ is a purely imaginary number whose modulus is larger than the maximum of $\{|\tilde{z}_j|\}$.

The root $z_0$ has a direct physical interpretation: it is closely related to
the low-energy excitations and can be regarded as one half of a helical spinon
excitation, namely a helical half-spinon. In the excited states, $\tilde z_0$ may enter the distribution
range of $\{\tilde z_j\}$, thereby forming an excitation branch. The presence
or absence of $z_0$ is jointly determined by the relative orientation of the
boundary fields and the parity of the lattice size $N$: for even $N$, $z_0$
appears when the $z$ components of the two boundary fields are antiparallel
and disappears when they are parallel, while for odd $N$ the situation is
reversed. Consequently, as shown in Table~\ref{tab:phase}, for even $N$ the
root $z_0$ appears in regions A--C and disappears in regions D--F, whereas
for odd $N$ the opposite occurs.

\end{enumerate}

In fact, the zero-root structures shown in Fig.~\ref{fig:phase} and Table~\ref{tab:phase} are strictly valid only in the thermodynamic limit $N \to \infty$.
For finite system sizes $N$, the zero-root structures exhibit small deviations.
Numerical results indicate that the phase boundaries are typically slightly smaller than $\tfrac{\eta}{2}$, approaching $\tfrac{\eta}{2}$ as $N \to \infty$.
Nevertheless, these deviations do not alter the overall structural characteristics of the zero roots, and the essential features can still be clearly observed even at finite lattice sizes.

In the following, we systematically analyze how zero-root structures evolve as boundary parameters are varied across the phase boundaries in the off-diagonal case.
As the exact transition boundary is uncertain at finite $N$, we provide several representative figures to illustrate the entire crossing process:
The first and last figures correspond to the structures in two adjacent phase regions in Fig.~\ref{fig:phase}, while the others depict structures close to the boundary.

We first discuss the transitions from Region A to Region D and from Region B to Region E.
Exactly at $p = \bar{q} = 0$ the Hamiltonian energy diverges, and the root configuration changes discontinuously:
$A \rightarrow D$ is realized by the conversion of the helical half-spinon $z_0$ into the additional roots $z_x$, whereas $B \rightarrow E$ corresponds to the annihilation of the special and additional roots together.
Away from this singular point the zero-root pattern still adheres to the regularity listed in Table~\ref{tab:phase} without any special deviation; hence no further elaboration is required.

\begin{figure}[htbp]
  \centering
  \subfloat[Region A]{\includegraphics[width=0.4\linewidth]{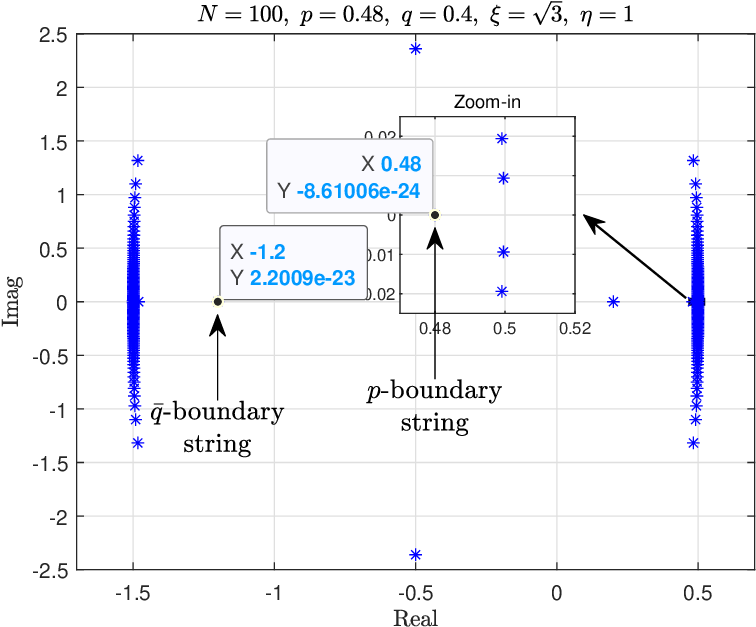}}
  \subfloat[Region A (near the boundary)]{\includegraphics[width=0.4\linewidth]{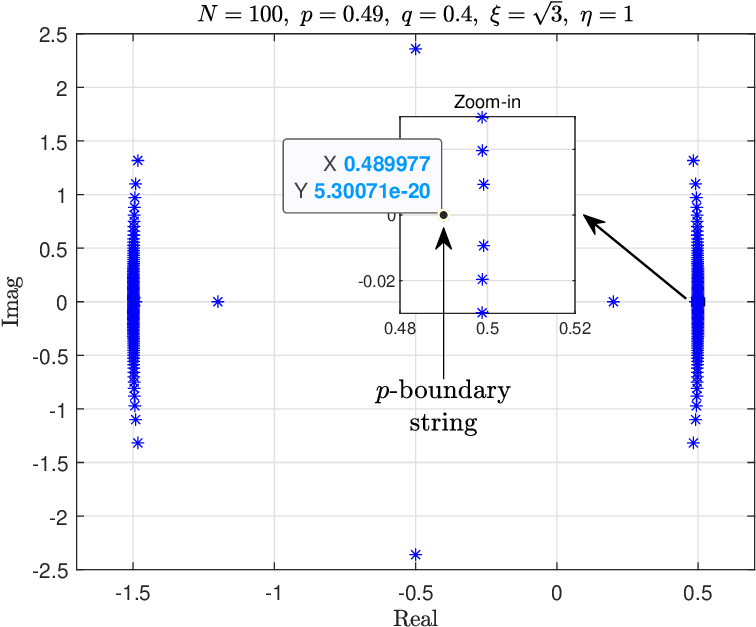}} \\
  \subfloat[Region B (near the boundary)]{\includegraphics[width=0.4\linewidth]{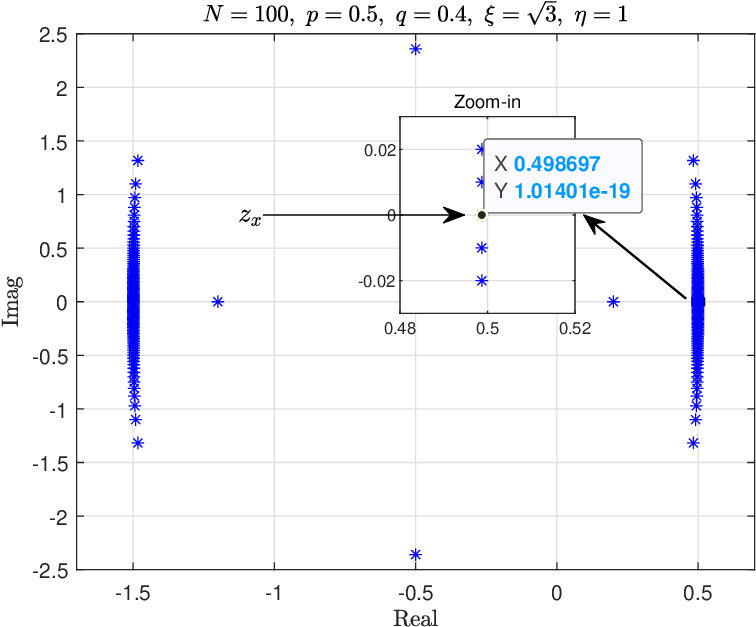}}
  \subfloat[Region B]{\includegraphics[width=0.4\linewidth]{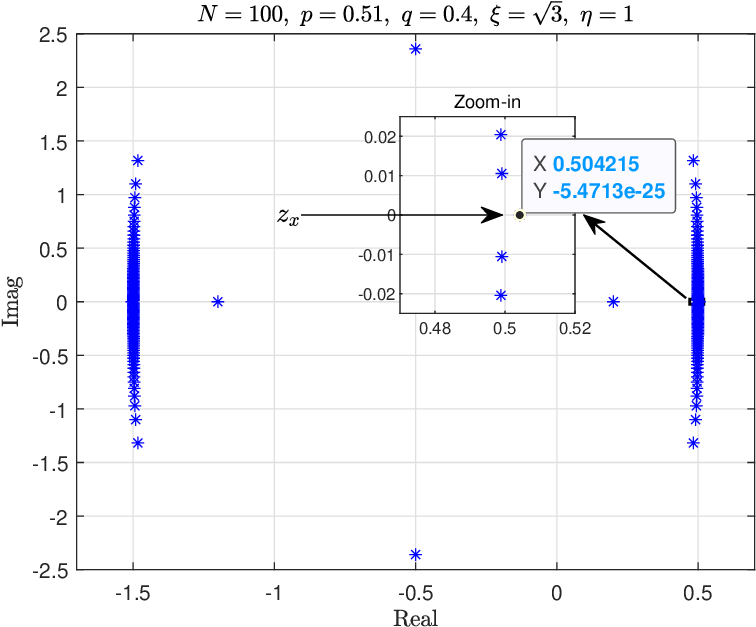}}
  \caption{
  Zero-root configurations as the boundary parameter crosses the A--B phase boundary.
  The parameters are fixed at $N = 100$, $q = 0.4$, $\xi = \sqrt{3}$, and $\eta = 1$,
  while $p$ varies across the four panels:
  (a) $p = 0.48$ (Region A),
  (b) $p = 0.49$ (Region A, near the boundary),
  (c) $p = 0.5$ (Region B, near the boundary),
  (d) $p = 0.51$ (Region B).
  }
  \label{fig:ab}
\end{figure}

Figure~\ref{fig:ab} illustrates how the root pattern evolves from Region~A to Region~B.
In Region~A the two characteristic structures identified in the phase diagram---the $\bar q$-boundary string and the $p$-boundary string---are clearly resolved Fig.~\ref{fig:ab}(a).
As $|p|$ increases toward the phase boundary, the two $p$-boundary strings separate horizontally and each coalesces with an additional root $z_x$ (Fig.~\ref{fig:ab}(b,c)).
Once $|p|$ crosses the boundary, the original $p$-string roots are entirely replaced by the new pair $z_x$ (Fig.~\ref{fig:ab}(d)).

\begin{figure}[htbp]
  \centering
  \subfloat[Region B]{\includegraphics[width=0.4\linewidth]{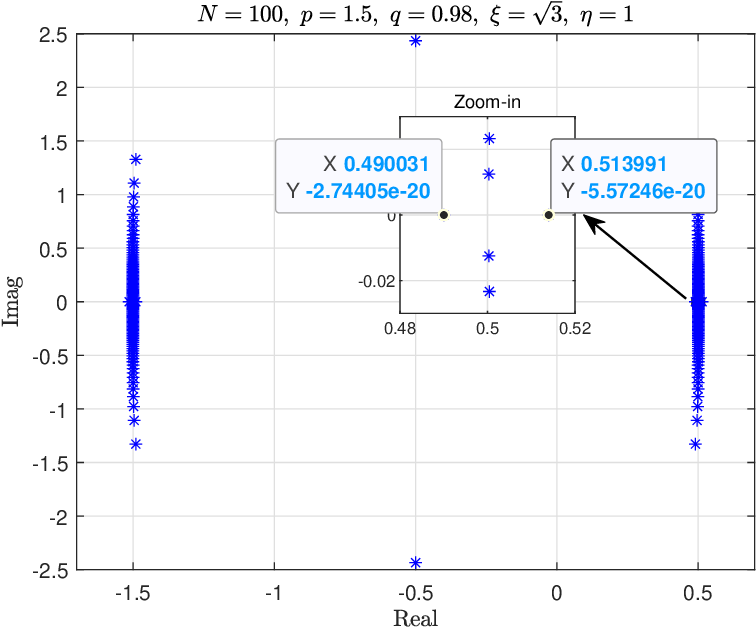}}
  \subfloat[Region B (near the boundary)]{\includegraphics[width=0.4\linewidth]{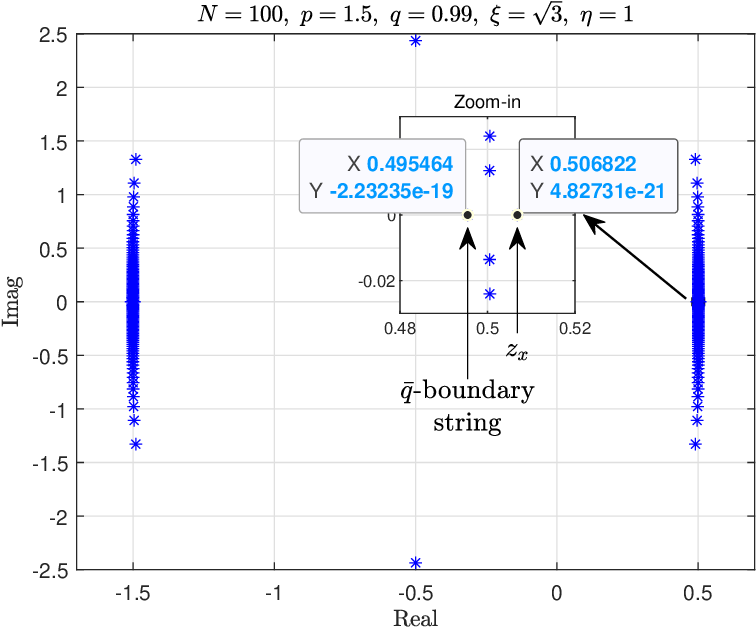}} \\
  \subfloat[Region C]{\includegraphics[width=0.4\linewidth]{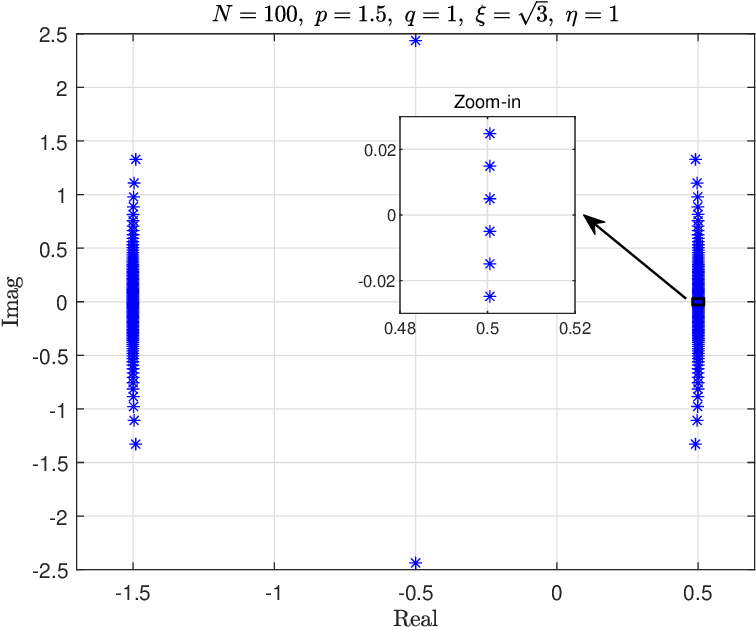}}
  \caption
  {
  Zero-root configurations as the boundary parameter crosses the B--C phase boundary.
  The parameters are fixed as $N = 100$, $p = 1.5$, $\xi = \sqrt{3}$, and $\eta = 1$,
  while $q$ varies across the three panels:
  (a) $q = 0.98$ (Region B),
  (b) $q = 0.99$ (Region B, near the boundary),
  (c) $q = 1$ (Region C).
  }
  \label{fig:bc}
\end{figure}

Fig.~\ref{fig:bc} illustrates the evolution of the zero roots during the transition from Region B to Region C.
In Region B, the $\bar{q}$-boundary string can be clearly identified.
As $|\bar{q}|$ increases, the $\bar{q}$-boundary string and additional root $z_x$ gradually converge; upon entering Region C, the two roots merge into bulk strings.

\begin{figure}[htbp]
  \centering
  \subfloat[Region D]{\includegraphics[width=0.4\linewidth]{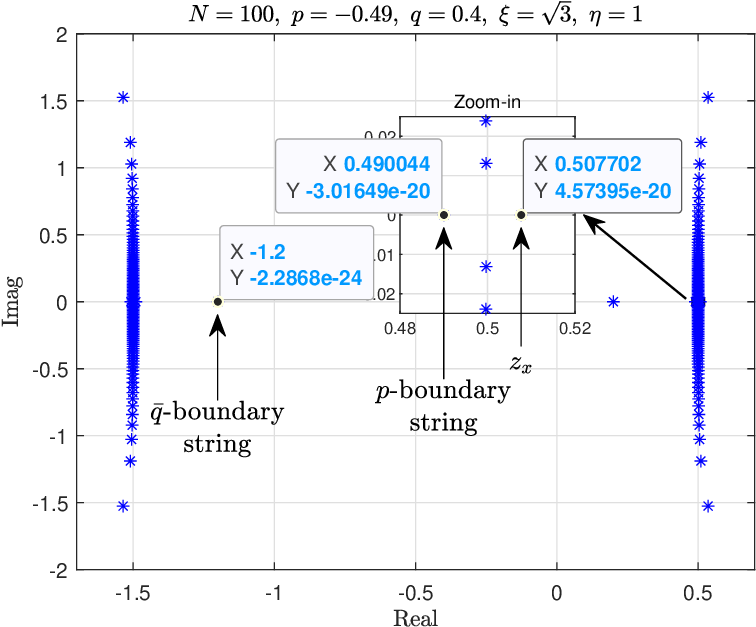}}
  \subfloat[Region D (near the boundary)]{\includegraphics[width=0.4\linewidth]{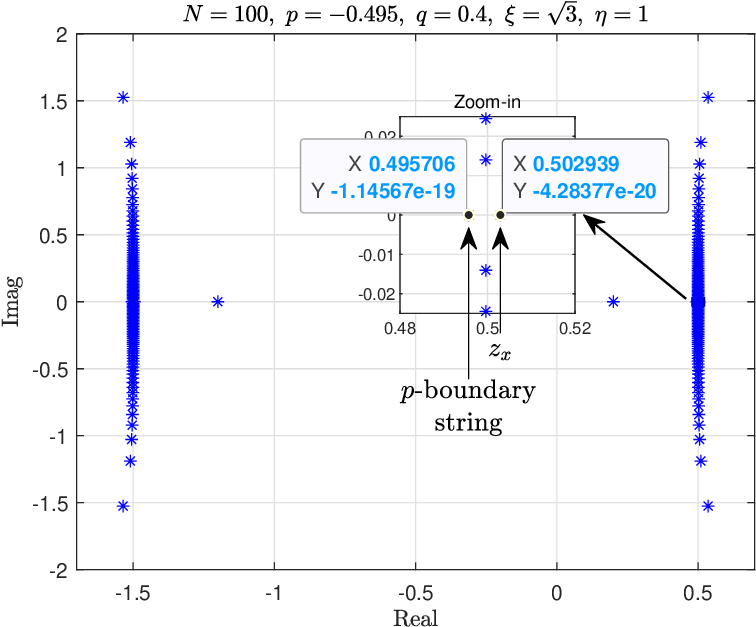}} \\
  \subfloat[Region E]{\includegraphics[width=0.4\linewidth]{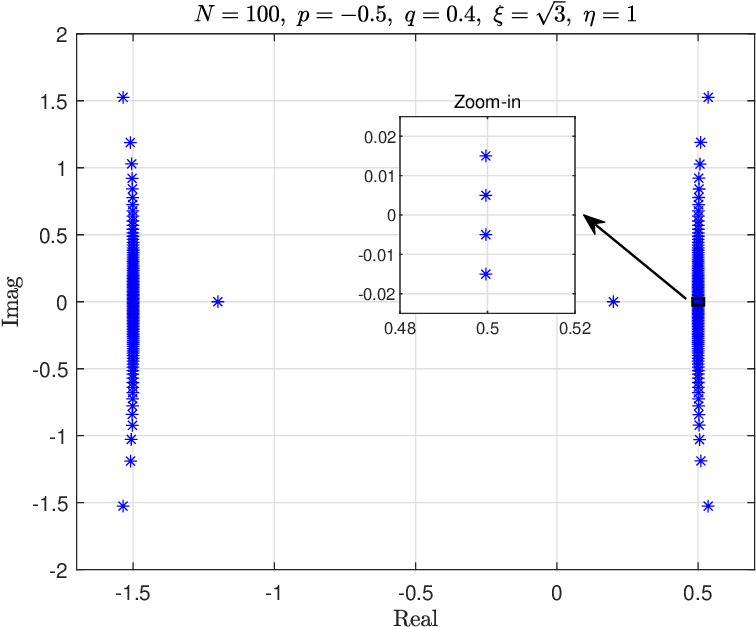}}
  \caption
  {
  Zero-root configurations as the boundary parameter crosses the D--E phase boundary.
  The parameters are fixed as $N = 100$, $q = 0.4$, $\xi = \sqrt{3}$, and $\eta = 1$,
  while $p$ varies across the three panels:
  (a) $p = -0.49$ (Region D),
  (b) $p = -0.495$ (Region D, near the boundary),
  (c) $p = -0.5$ (Region E).
  }
  \label{fig:de}
\end{figure}

Fig.~\ref{fig:de} displays the evolution of zero roots as parameters change from Region D to Region E.
This behavior closely resembles the transition observed between Regions B and C.
In Region D, three characteristic roots are highlighted: $p$-boundary string, $\bar{q}$-boundary string, and the additional root $z_x$.
As the parameter $|p|$ increases, $p$-boundary string and $z_x$ gradually move closer together.
At the limiting position, these two roots merge into bulk strings, while the $\bar{q}$-boundary string is preserved.

\begin{figure}[htbp]
  \centering
  \subfloat[Region E]{\includegraphics[width=0.4\linewidth]{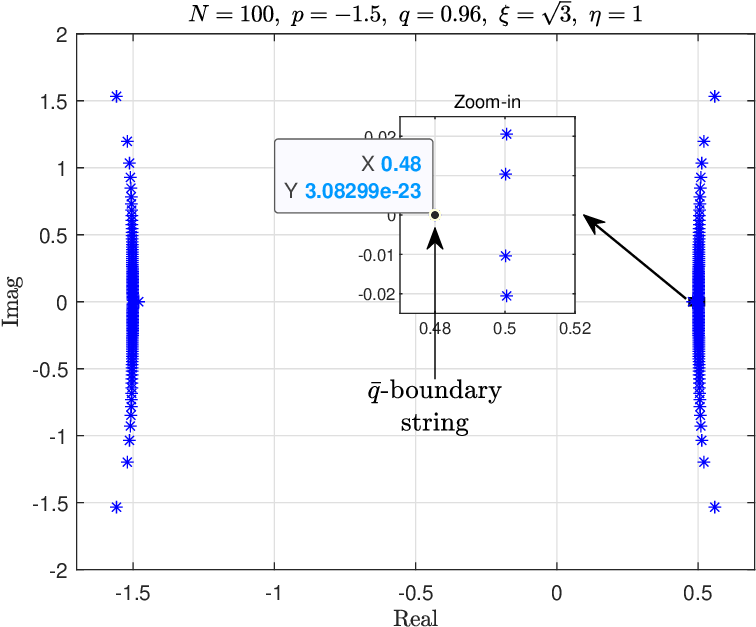}}
  \subfloat[Region E (near the boundary)]{\includegraphics[width=0.4\linewidth]{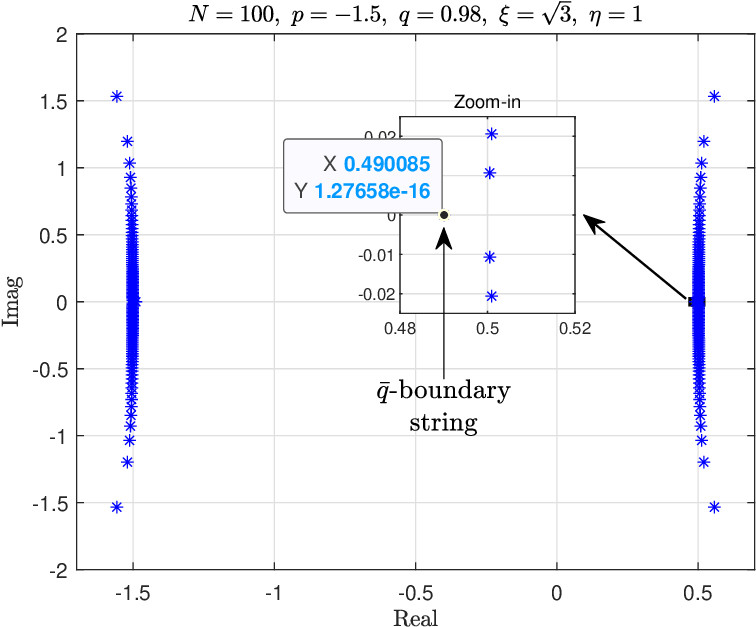}}\\
  \subfloat[Region F (near the boundary)]{\includegraphics[width=0.4\linewidth]{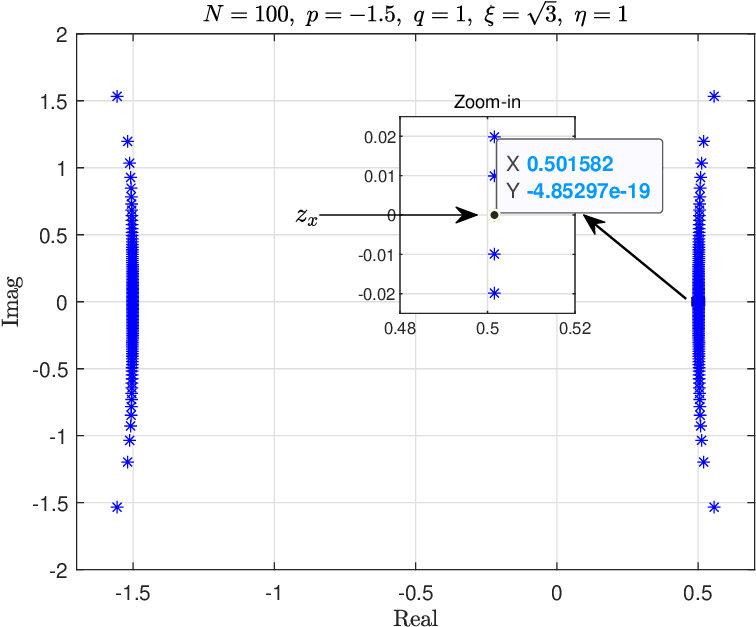}}
  \subfloat[Region F]{\includegraphics[width=0.4\linewidth]{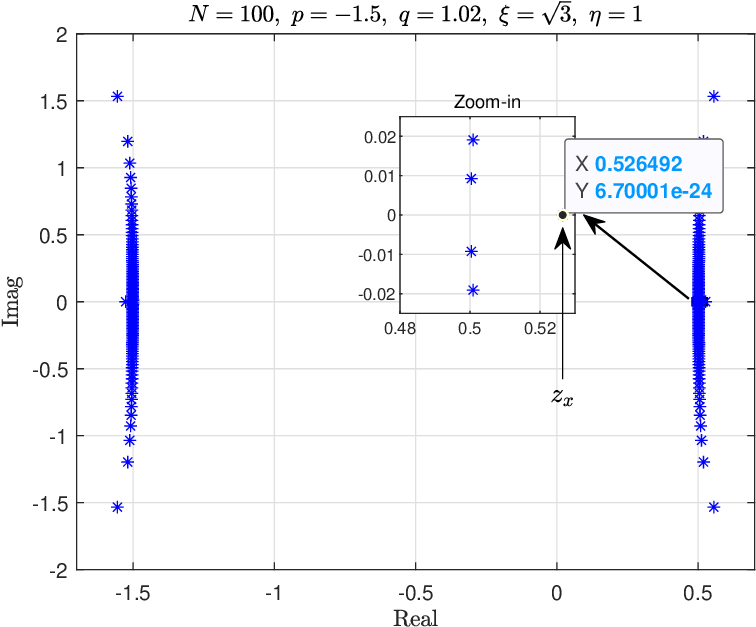}}
  \caption
  {
  Zero-root configurations as the boundary parameter crosses the E--F phase boundary.
  The parameters are fixed as $N = 100$, $p = -1.5$, $\xi = \sqrt{3}$, and $\eta = 1$,
  while $q$ varies across the four panels:
  (a) $q = 0.96$ (Region E),
  (b) $q = 0.98$ (Region E, near the boundary),
  (c) $q = 1$ (Region F, near the boundary),
  (d) $q = 1.02$ (Region F).
  }
  \label{fig:ef}
\end{figure}

As shown in Fig.~\ref{fig:ef}, the evolution from Region E to Region F is more similar to the case between Regions A and B.
In Region E, the $\bar{q}$-boundary string is clearly visible (Fig.~\ref{fig:ef}(a)).
With increasing $|\bar{q}|$, this root shifts rightward and gradually transforms into an additional root $z_x$ (Fig.~\ref{fig:ef}(b, c)).
Once $\bar{q}$ passes the phase boundary, the $\bar{q}$-boundary string is completely replaced by $z_x$ (Fig.~\ref{fig:ef}(d)).

\subsection{Structure of excited states}
\label{sec:zero-excited}
Since the presence or absence of \(U(1)\) symmetry does not essentially affect the structure of the zero roots, in this section we consider the \(U(1)\)-symmetry-broken case to illustrate the root distributions associated with elementary excitations.
In practice, within the DMRG framework one can introduce weighted penalty terms into the Hamiltonian to shift the energies of the already obtained low-energy states.
These weight terms raise the energies of the previously computed eigenstates, thereby excluding them from the variational optimization space.
As a result, a higher-energy eigenstate of the original Hamiltonian becomes the effective ground state of the modified Hamiltonian, which allows the corresponding excited state to be obtained through a ground-state DMRG calculation.
The same strategy also applies in the degenerate case: DMRG typically converges to one state in a degenerate subspace, and by penalizing the obtained state one can successively access the remaining degenerate states or higher excited states.

\begin{figure}[htbp]
  \centering
  \includegraphics[width=0.5\textwidth]{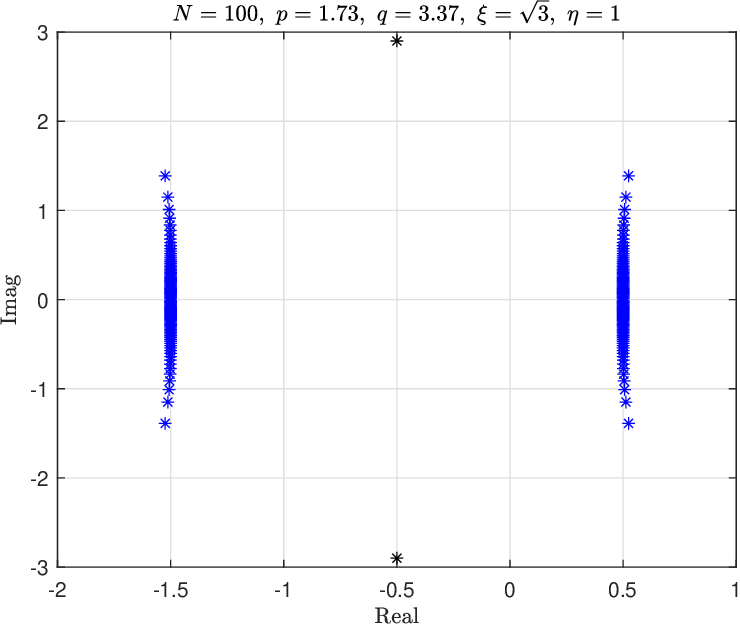}
\caption
{
Ground-state distribution of zero roots in the complex plane with parameters $N = 100$, $p = 1.73$, $q = 3.37$, $\xi = \sqrt{3}$, and $\eta = 1$.
The black points denote the helical half-spinon.
}
\label{fig:ground}
\end{figure}

Within this scheme, we find two representative types of elementary excitations, namely \textbf{helical spinon excitation} and \textbf{string excitation}, whose root distributions exhibit clearly distinguishable patterns.
These two types should be understood as representative rather than exhaustive, as other excitation patterns may also exist within the spectrum.
In this context, following Ref.~\cite{Zhakenov2025}, we regard the elementary excitation 2 described in Ref.~\cite{Dong2023} as a base state, namely, the lowest state within a given excitation tower.
Since excited-state DMRG calculations are more computationally demanding than ground-state calculations, we choose specific parameters such that both types of excitations lie in a relatively low-energy region, facilitating numerical computation and allowing a more stable identification of their root structures.
The corresponding ground-state configuration is shown in Fig.~\ref{fig:ground}.

\begin{figure}[htbp]
  \centering
  \includegraphics[width=0.5\textwidth]{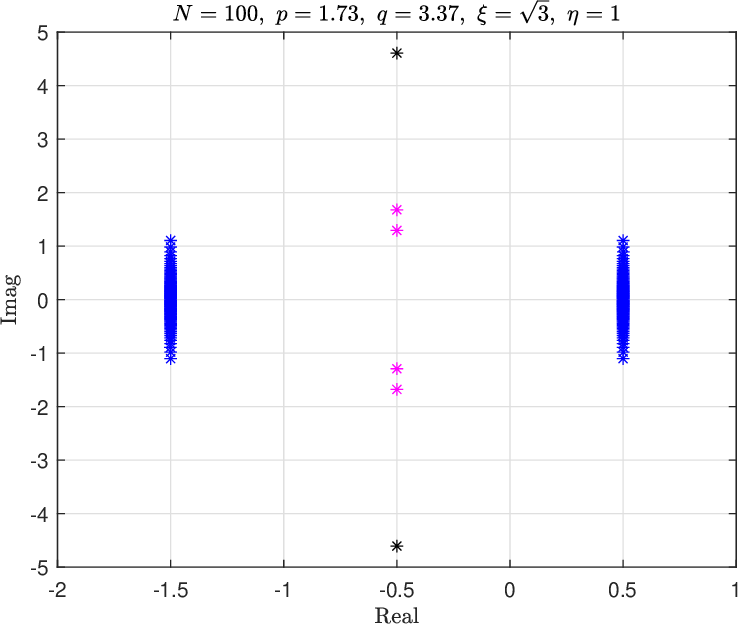}
\caption
{
Third excited-state distribution of zero roots in the complex plane with parameters $N = 100$, $p = 1.73$, $q = 3.37$, $\xi = \sqrt{3}$, and $\eta = 1$.
The black points denote the helical half-spinon, while the magenta points represent the roots associated with the helical spinon excitation.
}
\label{fig:spinon}
\end{figure}

\paragraph{Helical spinon excitation.}
In this case, four roots from the bulk strings shift to the vertical line $\mathrm{Re}(u)=-\tfrac{\eta}{2}$ and become
$\{\pm \tilde{z}_1-\tfrac{\eta}{2},\ \pm \tilde{z}_2-\tfrac{\eta}{2}\}$,
as shown in Fig.~\ref{fig:spinon}. This excitation corresponds to the third excited state.

It is worth noting that the helical half-spinon already present in the ground
state is not removed by the elementary excitation. Instead, it provides a
background contribution of one half-spinon to the excited-state spectrum. As a
result, the total spinon number in this sector is shifted by one half and takes
the half-integer sequence
\[
\frac12,\ \frac32,\ \frac52,\ \ldots .
\]
In contrast, when the $z$ components of the two boundary fields are
antiparallel, or when the parity of the lattice size $N$ is reversed, this
ground-state half-spinon background is absent. The corresponding spinon number
then follows the usual integer sequence
\[
1,\ 2,\ 3,\ \ldots .
\]

\begin{figure}[htbp]
  \centering
  \includegraphics[width=0.5\textwidth]{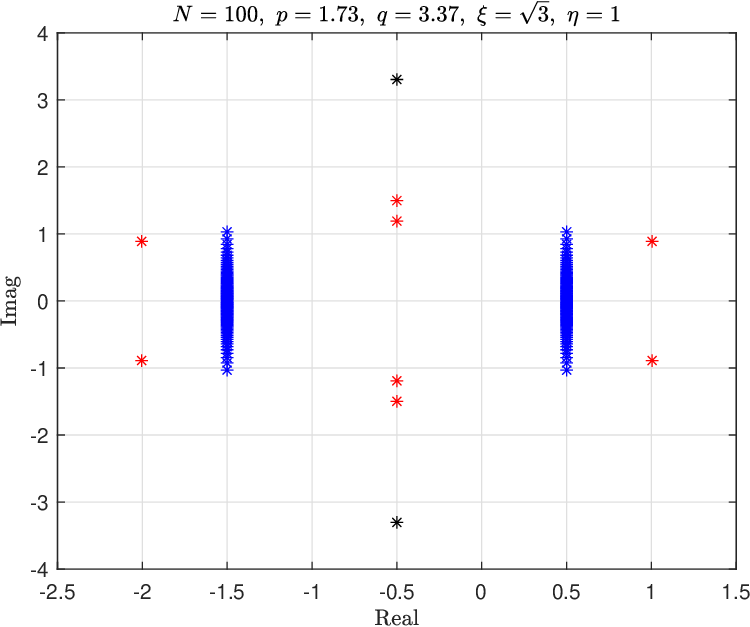}
\caption
{
Sixth excited-state distribution of zero roots in the complex plane with parameters $N = 100$, $p = 1.73$, $q = 3.37$, $\xi = \sqrt{3}$, and $\eta = 1$.
The black points denote the helical half-spinon, while the the red points denote the roots corresponding to the string excitation.
}
\label{fig:string}
\end{figure}

\paragraph{String excitation.}
Compared with the ground-state configuration, eight roots from the bulk strings
are rearranged. Among them, four roots move to higher 2-strings and take the
form
\[
\left\{\pm\left(\tilde z_1+\frac{k\eta}{2}\right)-\frac{\eta}{2},
\ \pm\left(\tilde z_2+\frac{k\eta}{2}\right)-\frac{\eta}{2}\right\},
\]
where $k>2$ is an integer, while the other four roots move to the vertical line
$\mathrm{Re}(u)=-\eta/2$ and become
\[
\left\{\pm\tilde z_3-\frac{\eta}{2},
\ \pm\tilde z_4-\frac{\eta}{2}\right\}.
\]
In the XXX spin chain, these two types of rearranged roots appear simultaneously
and together characterize a helical spin-singlet excitation.
In Fig.~\ref{fig:string}, the case \(k=3\) is shown.
This excitation corresponds to the sixth excited state.

\section{Bethe roots}
\label{sec:5}
\setcounter{equation}{0}
After obtaining $\Lambda(u)$, the Bethe roots can be determined by solving equations given in Eq.~\eqref{eq:T-Q relation}. The overall procedure follows a strategy similar to that described in Ref.~\cite{Belliard2018}.
However, since the $\Lambda(u)$ constructed from zero roots deviates slightly from the DMRG result, we use the DMRG-computed $\Lambda(u)$ to extract the Bethe roots.

We summarize the BA-DMRG procedure for obtaining the Bethe roots of $\Lambda(u)$ as follows:
\begin{enumerate}
  \item Obtain the ground state MPS of the Hamiltonian.
  \item Represent the transfer matrix $t(u)$ as an MPO.
  \item Select a set of interpolation points $\{u_j\}$ and compute the corresponding eigenvalues $\bar{\Lambda}(u_j)$ of the transfer matrix in the ground state.
  \item Express $\bar{Q}(u_j \pm \eta)$ as linear combinations of $\bar{Q}(u_j)$ using the Lagrange interpolation formula.
  \item Construct and solve the linear system for $\bar{Q}(u_j)$ using the $T-Q$ relation \eqref{eq:T-Q relation}.
  \item Reconstruct the polynomial $\bar{Q}(u)$ using the obtained values of $\bar{Q}(u_j)$.
  \item Determine the Bethe roots of $\bar{Q}(u)$ through numerical root-finding algorithms.
  \item Verify the correctness of the obtained roots (the specific procedures and principles will be presented in the following sections); if verification fails, return to step 3 and use the current results as new interpolation points.
  \item Using the Bethe roots obtained from the above procedure as initial guesses, we then solve the Bethe Ansatz equations (BAEs)~\eqref{eq:BAEs} directly to further improve the accuracy.
\end{enumerate}

In steps 3-7, we perform the normalization by dividing both sides of the $T-Q$ relation~\eqref{eq:inhomogeneous T-Q relation} by $u^{2N}$. The related functions are then replaced by $\bar{\Lambda}(u) = \frac{\Lambda(u)}{u^{2N}}$ and $\bar{Q}(u) = \frac{Q(u)}{u^{2N}}$. This normalization is necessary because, during the computation, some values of $\Lambda(u)$ may exceed the numerical range of the machine. Therefore, we normalize each variable throughout the process  to ensure numerical stability.

In step 4, $Q(u)$ is a degree-$2N$ polynomial with a unit leading coefficient~\eqref{eq:Q-function}, and hence it can be uniquely specified by its values at $2N$ distinct points.
We select $2N$ values $Q(u_j)$ as unknowns and use Lagrange interpolation to express $Q(u_j \pm \eta)$ as linear combinations of $Q(u_j)$.
Specifically, the interpolation takes the form
\begin{equation}
Q(u) = \prod_{j=1}^{2N} (u-u_j) + \sum_{j=1}^{2N} Q(u_j) \prod_{\substack{1 \leq k \leq 2N \\ k \ne j}} \frac{u - u_k}{u_j - u_k}.
\label{eq:lagrangeQ}
\end{equation}

In step 8, if verification fails, iteration is needed: return to step 3 and use the current results as new interpolation points, repeating this process until the desired accuracy is achieved.

The Bethe roots obtained in the preceding steps already satisfy the BAEs reasonably well. In step 9, we therefore use them as initial guesses and solve the BAEs~\eqref{eq:BAEs} directly to further improve the numerical precision.
It should be emphasized that, because the BAEs are highly nonlinear, the numerical solver can converge robustly only when the initial guesses have already reached a sufficiently high level of accuracy.

It is worth emphasizing that, unlike Ref.~\cite{Belliard2018}, where $\Lambda(u)$ is computed using high-precision arithmetic (e.g., VPA) and thus achieves significantly more digits than the sixteen available in standard double precision, the $\Lambda(u)$ obtained from the DMRG does not reach such accuracy.
High-precision DMRG implementations in ITensor are still immature; consequently, all DMRG calculations in this study are performed in double precision and are further limited by the prescribed convergence tolerance. In the homogeneous case, numerical verification shows that, under the parameter settings adopted in this work, the relative error of $\Lambda(u)$ at $N=100$ computed by DMRG is no worse than $10^{-10}$.
Consequently, the $2N$ interpolation nodes must be judiciously placed to guarantee numerical stability; otherwise, the iterative solver struggles to converge.
We should select the interpolation points for large lattices by following the known distribution pattern of Bethe roots on small lattices.

\subsection{Ground state solution with U(1) symmetry}
For the homogeneous case with $\xi = 0$, it has already been mentioned in Sec.~\ref{sec:homogeneous} that by using Eq.~\eqref{eq:logrithm}, the Bethe roots can be accurately determined even for large $N$.
The Bethe roots obtained from the BA-DMRG method can then be compared with these results to verify the validity of our approach.

\begin{figure}[htbp]
  \centering
  \subfloat[Global configuration of Bethe roots]{
    \includegraphics[width=0.47\textwidth]{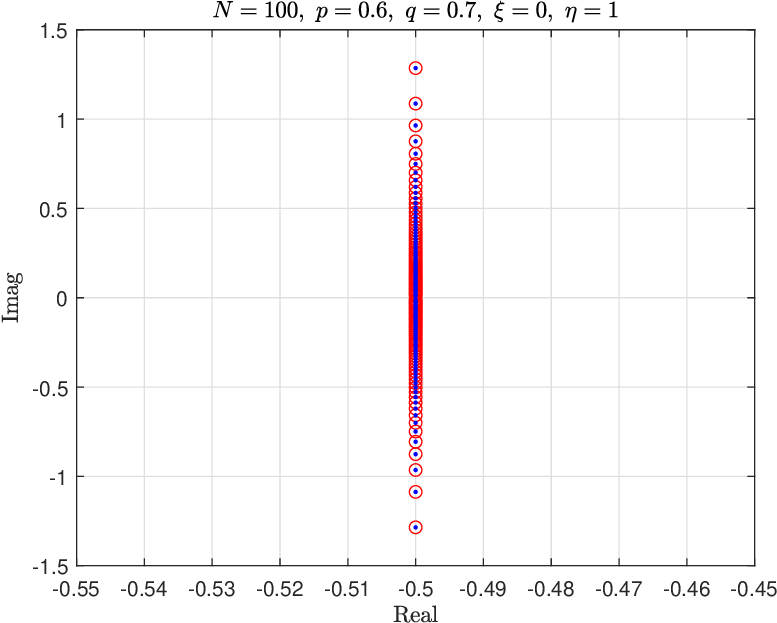}
  }
  \subfloat[Enlarged view of the central region]{
    \includegraphics[width=0.47\textwidth]{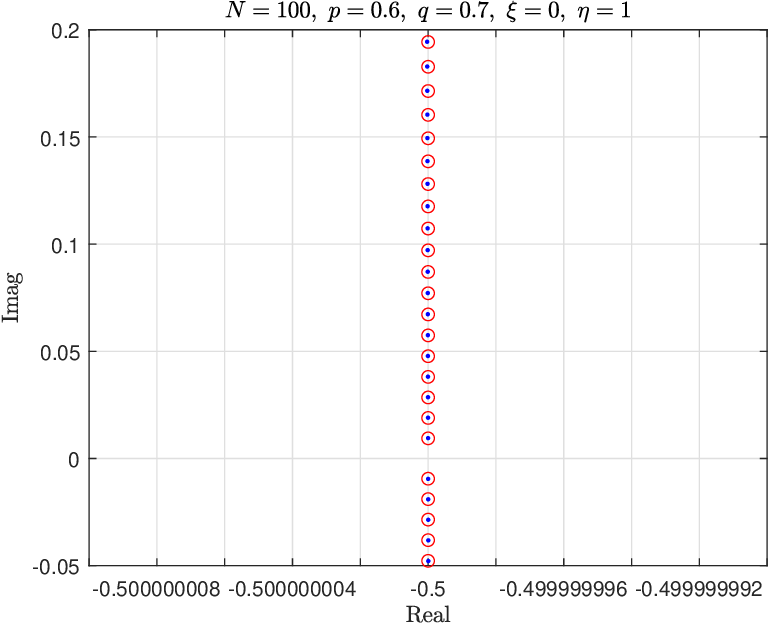}
  }
\caption{
Distribution of Bethe roots in the complex plane with parameters $N = 100$, $p = 0.7$, $q = 0.6$, $\xi = 0$, and $\eta = 1$.
Red circles denote Bethe roots obtained by solving the logarithmic form of the BAEs, while blue dots represent those computed using the BA-DMRG method.
}
\label{fig:xi0compare}
\end{figure}

As mentioned in Sec.~\ref{sec:homogeneous}, in the homogeneous case, the $Q(u)$ function for the ground state reduces from a $2N$-th degree polynomial to an $N$-th degree polynomial.
Starting from the initial $N$ interpolation nodes $\{u_j\}_{j=1}^{N}$, defined as $\{x_j, -x_j - \eta\mid j = 1, 2, \dots, \tfrac{N}{2}\}$ with $x_j = -\tfrac{\eta}{2} + \tfrac{i}{N}j^{k}$ and $k \in [1, 1.1]$, and performing iterative refinements, we obtain the comparison illustrated in Fig.~\ref{fig:xi0compare}.
Before being used as the initial guesses for solving the BAEs, the roots obtained from the BA-DMRG method differ from those obtained from the logarithmic BAEs by no more than $4.9\times 10^{-11}$ in absolute value.

These results show that, even for large systems at $\xi$ = 0, the Bethe roots obtained from the logarithmic BAEs coincide almost perfectly with those delivered by BA-DMRG.
Such consistency confirms the feasibility of the BA-DMRG approach as a reliable numerical tool for extracting Bethe roots.
Together, these findings provide strong evidence that our approach is both robust and well suited for extension to more general inhomogeneous systems.

\subsection{Ground state solution without U(1) symmetry}
In the inhomogeneous case, the ground-state $Q(u)$ becomes a degree-$2N$ polynomial, twice the degree of the homogeneous case. The corresponding Bethe-root pattern exhibits a more intricate structure.
Together these two effects push the solution difficulty sharply upward, so the attainable lattice size $N$ is therefore limited compared with the homogeneous case
\footnote{If the precision of BA-DMRG data could be raised beyond 16 significant digits, high-accuracy Bethe roots for larger systems would become accessible.}.

Following the same strategy as in the homogeneous case, but noting that $Q(u)$ is now a $2N$-th degree polynomial, the interpolation scheme needs to be modified accordingly.
Specifically, the $2N$ interpolation nodes $\{u_j\}_{j=1}^{2N}$ are constructed as $\{x_j, -x_j - \eta \mid j = 1, 2, \dots, N\}$.
The first $\tfrac{N}{2}$ $\{x_j\}$ nodes are defined as $x_j = -\tfrac{\eta}{2} + \tfrac{i}{N} j^{k}$, where $k \in [1, 1.1]$, while the remaining $\tfrac{N}{2}$ $\{x_j\}$ nodes are specified as $x_j = j + t$, with $j = \tfrac{N}{2} + 1, \tfrac{N}{2} + 2, \dots, N$ and $t \in (0, \tfrac{N}{4})$.
In addition, another effective choice for the remaining $\tfrac{N}{2}$ nodes is $x_j=t+(j-\tfrac{N}{2})i-\tfrac{N}{4}i$, where $t$ is chosen around $\tfrac{N}{4}$.
With the interpolation nodes placed as above, convergence is achieved in fewer than 10 iterations.

\begin{figure}[htbp]
  \centering
  \subfloat[Global configuration of Bethe roots]{
    \includegraphics[width=0.47\textwidth]{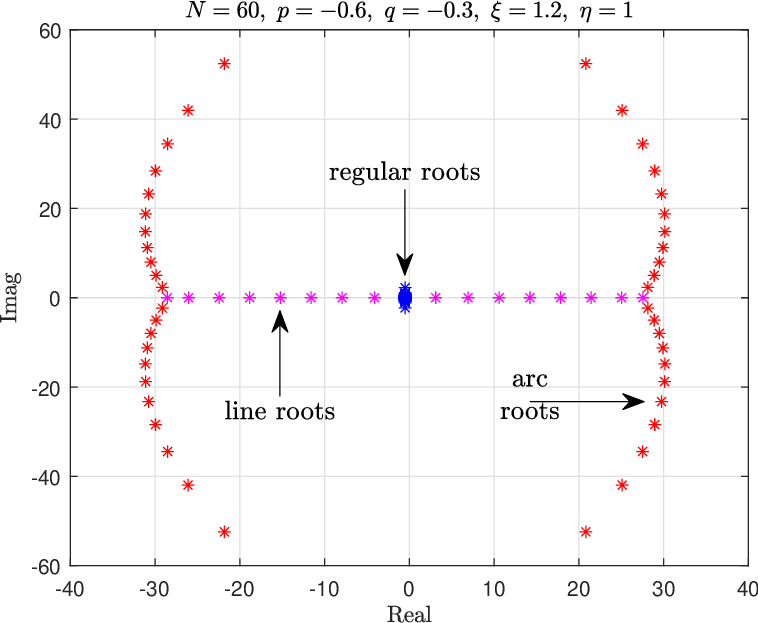}
  }
  \subfloat[Enlarged view of the central region]{
    \includegraphics[width=0.47\textwidth]{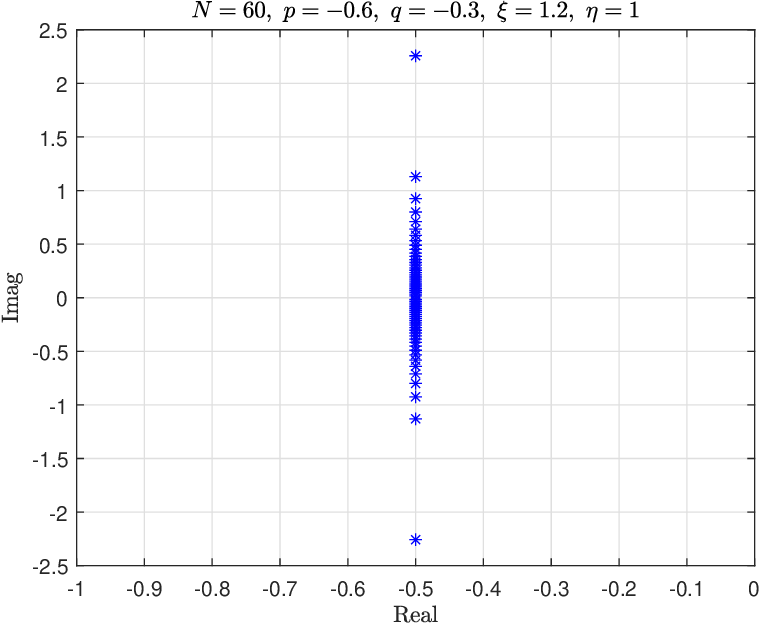}
  }
\caption{
Distribution of Bethe roots in the complex plane with parameters $N = 60$, $p = -0.6$, $q = -0.3$, $\xi = 1.2$, and $\eta = 1$.
The blue, magenta, and red points denote the regular, line, and arc roots, respectively.
}
\label{fig:III}
\end{figure}
With the interpolation nodes detailed above, we compute the Bethe roots for the non-U(1)-symmetric case; the resulting configuration is shown in Fig.~\ref{fig:III}.

We find that the ground-state Bethe roots $\{\lambda_j, -\lambda_j-\eta\mid j=1,\ldots,N\}$ fall into three classes:
\begin{enumerate}
\item \textbf{Regular roots $\{\lambda_j^ {(\mathrm r)}, -\lambda_j^ {(\mathrm r)}-\eta\mid j=1,\ldots,n^{(\mathrm r)}\}$:}
  \begin{itemize}
    \item \emph{Central roots:} $\pm i\eta\mu_j-\tfrac{\eta}{2}$, with all $\mu_{j}$ purely real;
    \item \emph{Boundary strings:} $\pm\bigl(p-\tfrac{\eta}{2}\bigr)-\tfrac{\eta}{2}$ and $\pm\bigl(\bar q-\tfrac{\eta}{2}\bigr)-\tfrac{\eta}{2}$, whose positions are fixed solely by the boundary parameters.
  \end{itemize}
  $n^{(\mathrm r)}$ is precisely $\tfrac{N}{2}$ or $\tfrac{N-2}{2}$; their distribution mirrors that of the homogeneous chain.

\item \textbf{Line roots $\{\lambda_j^ {(\mathrm l)}, -\lambda_j^ {(\mathrm l)}-\eta\mid j=1,\ldots,n^{(\mathrm l)}\}$:}
  The line roots take the form $\pm (\lambda_0^ {(\mathrm l)}+js)-\tfrac{\eta}{2}$.
  Here $\lambda_0^ {(\mathrm l)}$ and $s$ are positive real numbers, and $s$ denotes the spacing between neighboring line roots.
  The spacing $s\in (2\eta, \infty)$ depends mainly on $\xi$, only weakly on $p$ and $q$, and is independent of the system size $N$.
  Moreover, $n^ {(\mathrm l)}$ increases with the system size $N$.

\item \textbf{Arc roots $\{\lambda_j^ {(\mathrm a)}, -\lambda_j^ {(\mathrm a)}-\eta\mid j=1,\ldots,n^{(\mathrm a)}\}$:}
  The arc roots take the form $\{\pm\hat{\lambda}_j-\tfrac{\eta}{2}, \pm\hat{\lambda}_j^*-\tfrac{\eta}{2}\}$.
  Here $\hat{\lambda}_j$ is a complex number with nonzero real and imaginary parts, and $*$ denotes complex conjugation.
  $\{\lambda_j^ {(\mathrm a)}\}$ branch off from the outer ends of the line roots and form symmetric arcs in the upper and lower half-planes.
  Accordingly, $\{\lambda_j^ {(\mathrm a)}\}$ also move to infinity in the thermodynamic limit.
  Moreover, $n^{(\mathrm a)}$ increases with the system size $N$.
\end{enumerate}
For the present parameter set the pattern coincides with the periodic-boundary results of Ref.~\cite{Belliard2018} obtained from the inhomogeneous $T-Q$ relation.
Other parameter regions reveal richer structures, most notably the \textbf{Paired-line roots} $\{\lambda_j^ {(\mathrm p)}, -\lambda_j^ {(\mathrm p)}-\eta\mid j=1,\ldots,n^{(\mathrm p)}\}$ introduced in the next subsection.

Now, an important task is to verify the correctness of the obtained Bethe roots.
According to the BAEs~\eqref{eq:BAEs}, we verify the correctness of the obtained Bethe roots.
The BAEs can be rewritten in the form
\begin{equation}
  A(\lambda_j) = -B(\lambda_j) - C(\lambda_j), \qquad j=1,\ldots,N,
\end{equation}
where
\begin{align}
A(u)
&=
\left(\frac{u+\eta}{u}\right)^{2N+1}
\frac{(u+p)\bigl[(1+\xi^{2})^{\frac{1}{2}}u+q\bigr]}
     {(u-p+\eta)\bigl[(1+\xi^{2})^{\frac{1}{2}}(u+\eta)-q\bigr]},
\\[4pt]
B(u)
&=
\frac{Q(u+\eta)}{Q(u-\eta)},
\\[4pt]
C(u)
&=
\frac{\bigl[1-(1+\xi^{2})^{\frac{1}{2}}\bigr](2u+\eta)(u+\eta)^{2N+1}}
     {(u-p+\eta)\bigl[(1+\xi^{2})^{\frac{1}{2}}(u+\eta)-q\bigr]\,Q(u-\eta)}.
\end{align}
Define
\begin{equation}
  G(u) = \frac{A(u)+C(u)}{-B(u)}.
\end{equation}
The BAEs can be written in the equivalent form $G(\lambda_j)=1$, which in numerical computation is tested through the residual condition $\varepsilon_j=\left|G(\lambda_j)-1\right|=0$.
We evaluated $\varepsilon_j$ for all roots in step 8, and the maximal residual was $2.59 \times 10^{-5}$.
We then further refined the roots in step 9 by directly solving the BAEs with a numerical solver, taking the previously obtained roots as initial guesses.
In practice, the BAEs are solved in Matlab using fsolve with the Levenberg--Marquardt algorithm.
After refinement, the maximal residual is reduced to $1.19 \times 10^{-13}$.
The detailed numerical data are provided in Appendix~\ref{app:D}.

\subsection{Structure of ground state}

Unlike the zero-root case, the number of Bethe roots in the ground state already differs between the $U(1)$-symmetric and $U(1)$-symmetry-broken cases, leading to markedly distinct root distributions in the two situations.
Since the Bethe roots in the homogeneous case exhibit simple behavior, this section focuses on analyzing the structure of the inhomogeneous Bethe roots.
From the BAEs~\eqref{eq:BAEs}, it is evident that $p$ and $\bar{q}$ enter the equations symmetrically. Therefore, it suffices to analyze the variation with respect to only one of these parameters.
Given the considerable complexity of their configurations, we restrict our discussion here to the characteristic changes of the Bethe roots as $p$ vary and $\bar{q}$ fixed.

As $p$ varies, we systematically investigate the resurgence of $U(1)$ symmetry while $|p|$ is continuously increased from finite values to infinity.
In the asymptotic limit \(|p| \to \infty\), the system recovers \(U(1)\) symmetry: exactly \(N\) Bethe roots diverge to infinity, while the remaining \(N\) roots condense onto the vertical contour \(\mathrm{Re}(u) = -0.5\) (Fig.~\ref{fig:xi0compare}).
We now systematically track the reorganization of the Bethe-root configuration from the generic inhomogeneous distribution shown in Fig.~\ref{fig:III} to the regularized pattern exhibited in Fig.~\ref{fig:xi0compare} under the continuous amplification of \(|p|\).
The evolution exhibits two distinct paradigms, contingent upon the sign of the boundary parameter $p$.

\subsubsection{$p>0$ case}
\begin{figure}[htbp]
  \centering
  \subfloat[Global configuration of Bethe roots]{
    \includegraphics[width=0.47\textwidth]{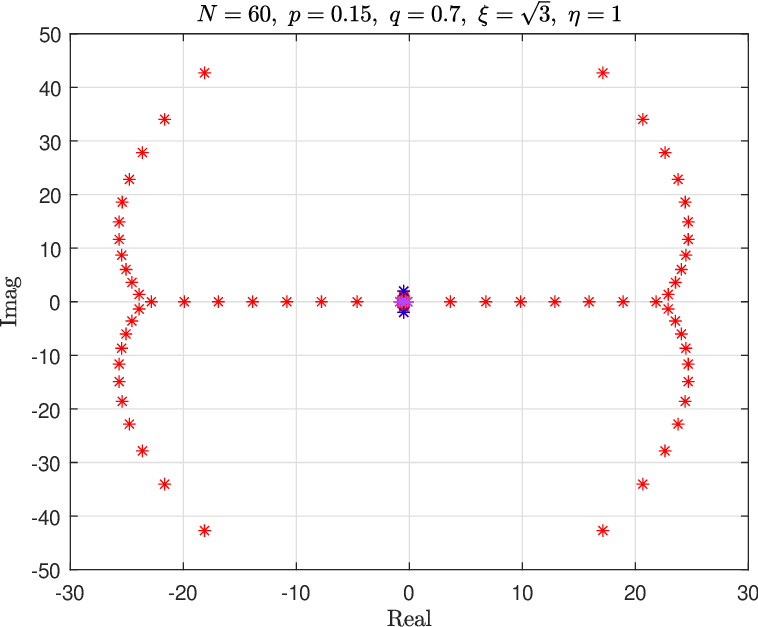}
  }
  \subfloat[Enlarged view of the central region]{
    \includegraphics[width=0.47\textwidth]{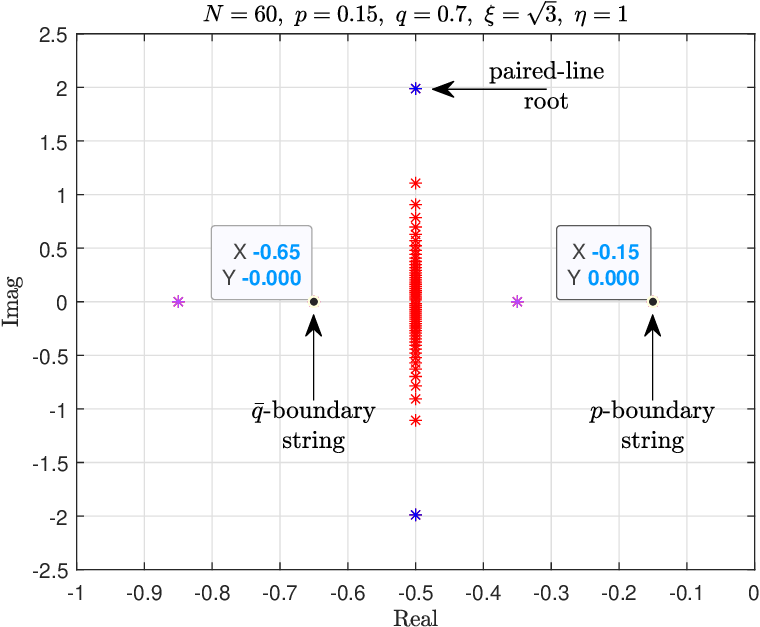}
  }
\caption{
Distribution of Bethe roots in the complex plane with parameters
$N = 60$, $p = 0.15$, $q = 0.7$, $\xi = \sqrt{3}$, and $\eta = 1$.
The magenta points represent the boundary strings,
and the blue points correspond to the paired-line roots, which here appear as a single conjugate pair.
}
\label{fig:+1}
\end{figure}

In this subsection we focus on the case where the boundary parameter satisfies $p>0$. For clarity, we first illustrate the case with \(\bar q=0.35\); the negative-\(\bar q\) scenario is qualitatively similar and its minor differences will be summarized at the end of this subsection.
To establish a baseline for the subsequent analysis, we first present the initial structure of the Bethe roots for small $p$.
For $0<p<\tfrac{\eta}{2}$ or $0<\bar{q}<\tfrac{\eta}{2}$, the Bethe roots exhibit a clear {\bf boundary-string} configuration that is identical to the one appearing in the homogeneous case.
The endpoints of these boundary strings are located at $\pm(p-\tfrac{\eta}{2})-\tfrac{\eta}{2}$ or $\pm(\bar{q}-\tfrac{\eta}{2})-\tfrac{\eta}{2}$, respectively, as illustrated in Fig.~\ref{fig:+1}.
We classify boundary strings as regular roots defined in this work.

As $p$ increases, both the line roots and the arc roots gradually move toward the central roots.
During this process, certain arc roots approach the real axis and transform into line roots.
Meanwhile, conjugate pairs aligned along directions parallel to the central roots begin to appear.
They initially occupy the positions of the line roots closest to the central roots, and this replacement then progresses outward to the more distant line roots.
We refer to these conjugate pairs as the \textbf{paired-line roots}, and such roots occur only when either $p>0$ or $\bar{q}>0$.

\paragraph{Quantisation rule.}
The number of paired-line pairs increases in unit steps. When $|\bar{q}|$ is small, and as $p$ increases, each time the combination $p+\bar{q}$ reaches an integer multiple of $\eta$, that is, $p+\bar{q}=k\eta$ with $k\in\mathbb{Z}^{+}$, one additional pair of paired-line roots appears.
For $\bar{q}>0$, the total number of paired-line root pairs is therefore given by $\max\bigl\{1,\,\lfloor p+\bar{q}\rfloor\bigr\}$,
where $\lfloor\,\rfloor$ denotes the floor function.
The number of regular roots remains unchanged and is always equal to $N$.

\begin{figure}[htbp]
  \centering
  \subfloat[Global configuration of Bethe roots]{
    \includegraphics[width=0.47\textwidth]{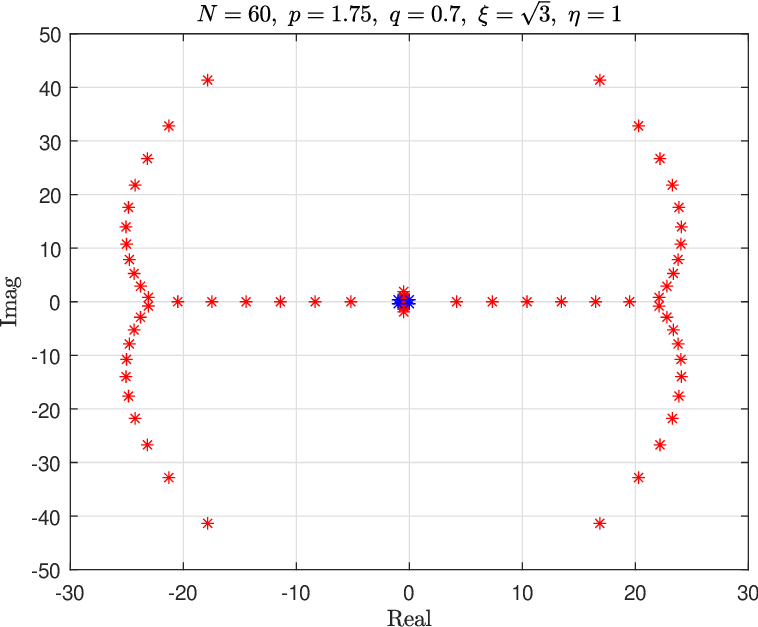}
  }
  \subfloat[Enlarged view of the central region]{
    \includegraphics[width=0.47\textwidth]{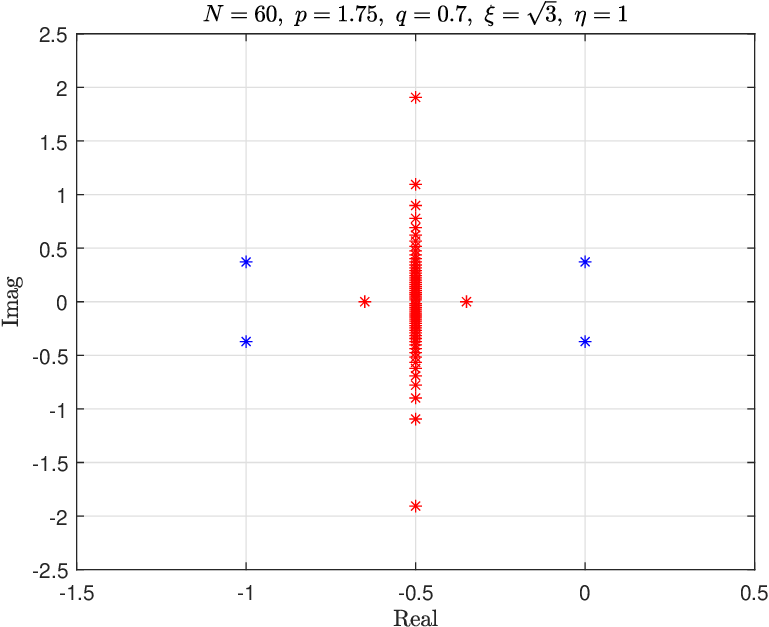}
  }
\caption{
Distribution of Bethe roots in the complex plane with parameters $N = 60$, $p = 1.75$, $q = 0.7$, $\xi = \sqrt3$, and $\eta = 1$.
The blue points correspond to the paired-line roots, which appear here as two paired-line root pairs.
}
\label{fig:+2}
\end{figure}

\begin{figure}[htbp]
  \centering
  \subfloat[Global configuration of Bethe roots]{
    \includegraphics[width=0.47\textwidth]{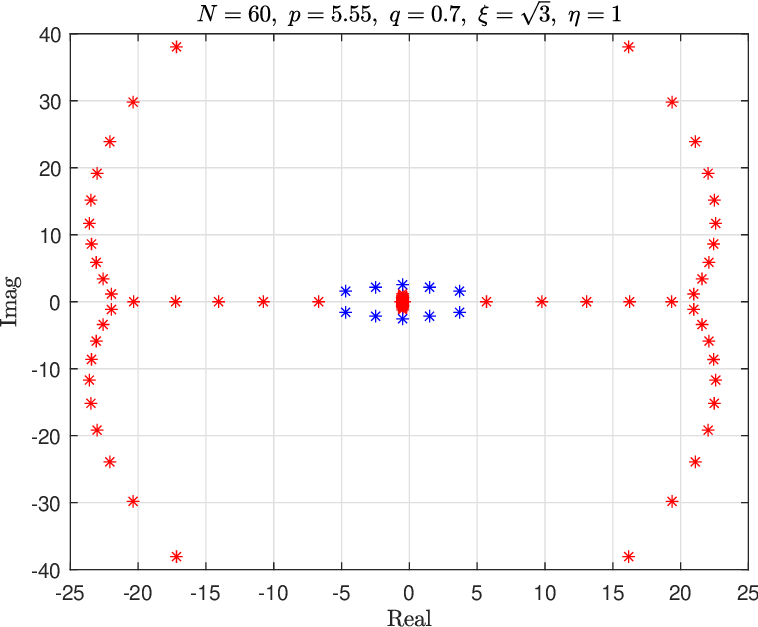}
  }
  \subfloat[Enlarged view of the central region]{
    \includegraphics[width=0.47\textwidth]{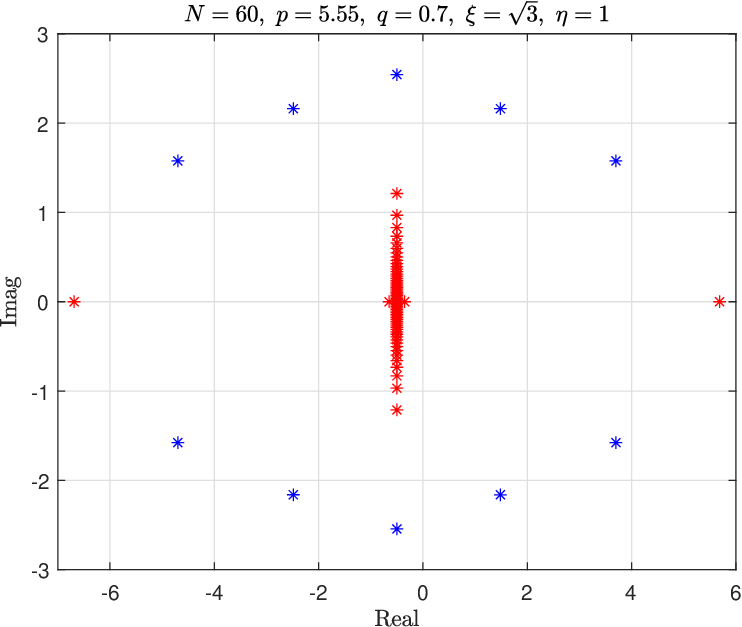}
  }
\caption{
Distribution of Bethe roots in the complex plane with parameters $N = 60$, $p = 5.55$, $q = 0.7$, $\xi = \sqrt3$, and $\eta = 1$.
The blue points correspond to the paired-line roots, which appear here as five paired-line root pairs.
}
\label{fig:+3}
\end{figure}

This minimal configuration is shown in Fig.~\ref{fig:+1}, where the single paired-line root pair lies on the vertical contour \(\mathrm{Re}(u) = -0.5\).
As concrete illustrations of this pattern, we present the cases $p=1.75$ and $p=5.55$, where the numbers of paired-line root pairs are two and five, respectively, as shown in Figs.~\ref{fig:+2} and \ref{fig:+3}.


The increase in the number of paired-line root pairs occurs within an extremely narrow interval, and within this crossover region the changes in the Bethe-root configuration are highly intricate.
As the system size $N$ grows, the width of this interval becomes even smaller, indicating that the crossover behavior is dominated by finite-size effects. We provide a detailed illustration of the crossover behavior in Appendix~\ref{app:E}.

\begin{figure}[htbp]
  \centering
  \subfloat[Global configuration of Bethe roots]{
    \includegraphics[width=0.47\textwidth]{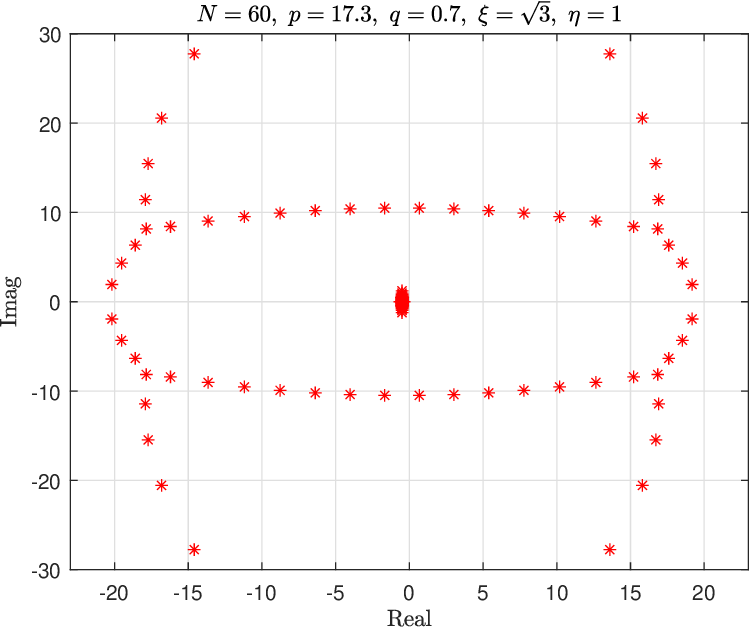}
  }
  \subfloat[Enlarged view of the central region]{
    \includegraphics[width=0.47\textwidth]{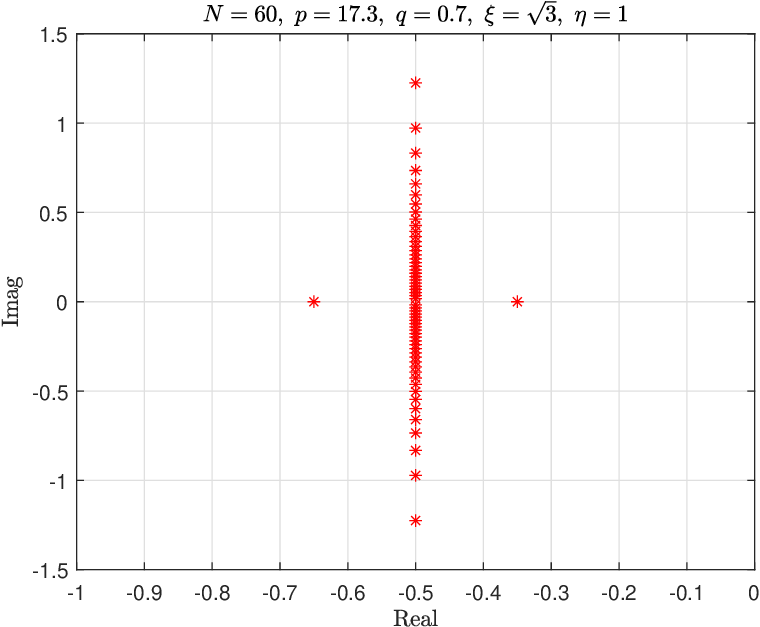}
  }
\caption{
Distribution of Bethe roots in the complex plane with parameters $N = 60$, $p = 17.3$, $q = 0.7$, $\xi = \sqrt3$, and $\eta = 1$.
}
\label{fig:+4}
\end{figure}

After all line roots have converted into paired-line root pairs, further increasing \(p\) drives the two members of each paired-line root to split apart: one drifts upward and the other downward, moving away from the real axis, as illustrated in Fig.~\ref{fig:+4}.

\begin{figure}[htbp]
  \centering
  \subfloat[Global configuration of Bethe roots]{
    \includegraphics[width=0.47\textwidth]{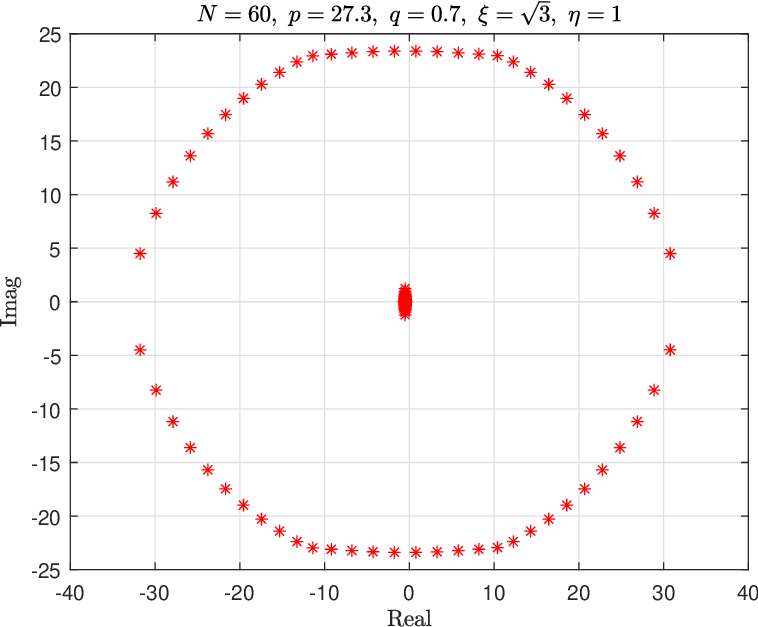}
  }
  \subfloat[Enlarged view of the central region]{
    \includegraphics[width=0.47\textwidth]{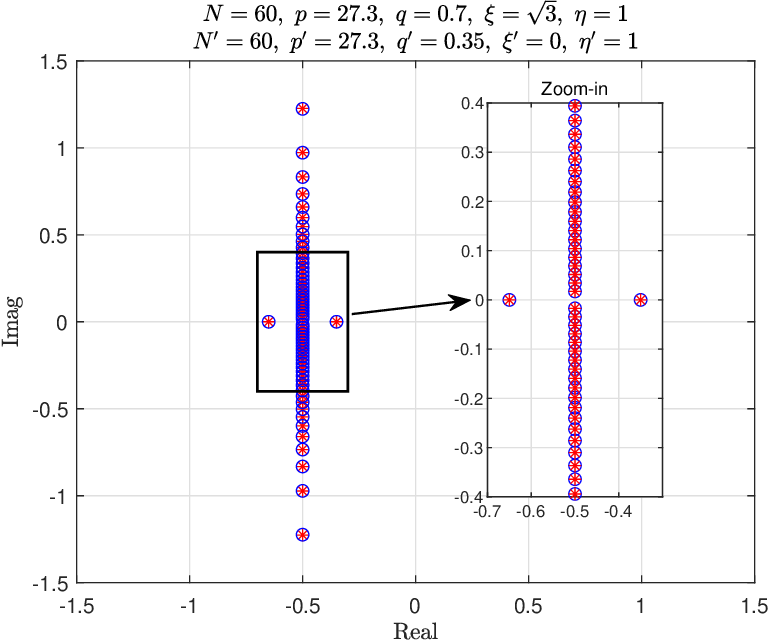}
  }
\caption{
Distribution of Bethe roots in the complex plane with parameters $N = 60$, $p = 27.3$, $q = 0.7$, $\xi = \sqrt{3}$, and $\eta = 1$.
Blue circles in panel~(b) indicate the Bethe roots obtained in the homogeneous case with parameters $N' = 60$, $p' = 27.3$, $q' = 0.35$, $\xi' = 0$, and $\eta' = 1$.
}
\label{fig:+5}
\end{figure}

As \(p\) is further increased, the arc roots and the paired-line root pairs formed by the $N$ Bethe roots move away from the center toward infinity, while the $N$ regular roots remain at the center.
This behavior arises because, for the Hamiltonian~\eqref{eq:Hamiltonian}, when either $p$ or $q$ tends to infinity, the magnetic field at one boundary approaches zero and the previously broken $U(1)$ symmetry is restored.
At this point, the Bethe Ansatz equations are expected to reduce to those of the homogeneous case.
By setting the boundary parameters to be identical-so that $q'$ in the homogeneous case equals $\bar{q}$ in the inhomogeneous case-the central Bethe roots obtained in the inhomogeneous case can be directly compared with those of the homogeneous case (see Fig.~\ref{fig:+5}).
The two sets of roots show only negligible discrepancies, which further demonstrates the validity of the algorithm.

For negative \(\bar q\), the evolution of the Bethe-root structure with increasing \(p\) closely parallels the positive-\(\bar q\) case, with only two quantitative differences:
(i) the number of paired-line root pairs is $\max\bigl\{1,\,\lfloor p+\bar{q}+1\rfloor\bigr\}$;
(ii) the number of regular roots is $N-2$ at small $p$, but once all line roots have transformed into paired-line root pairs, the number of regular roots becomes $N$ and remains fixed at this value throughout the subsequent evolution.
Apart from these differences, the rest of the analysis remains unchanged.

\subsubsection{$p<0$ case}
When $p<0$ and increase $|p|$ , the structural change is comparatively simpler, and the overall evolution is identical regardless of the sign of $\bar{q}$.
For consistency with the previous subsection, we likewise set $\bar{q}=0.35$ in the following analysis.
\begin{figure}[htbp]
  \centering
  \subfloat[Global configuration of Bethe roots]{
    \includegraphics[width=0.47\textwidth]{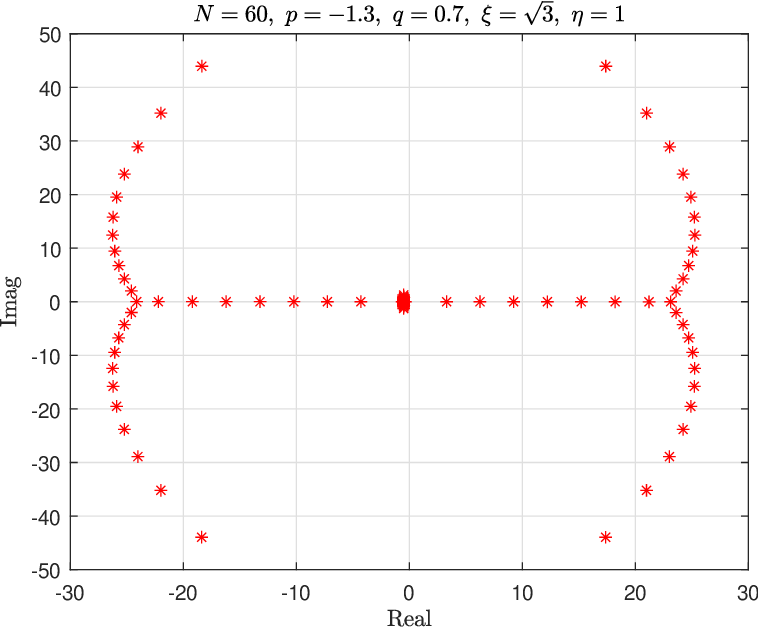}
  }
  \subfloat[Enlarged view of the central region]{
    \includegraphics[width=0.47\textwidth]{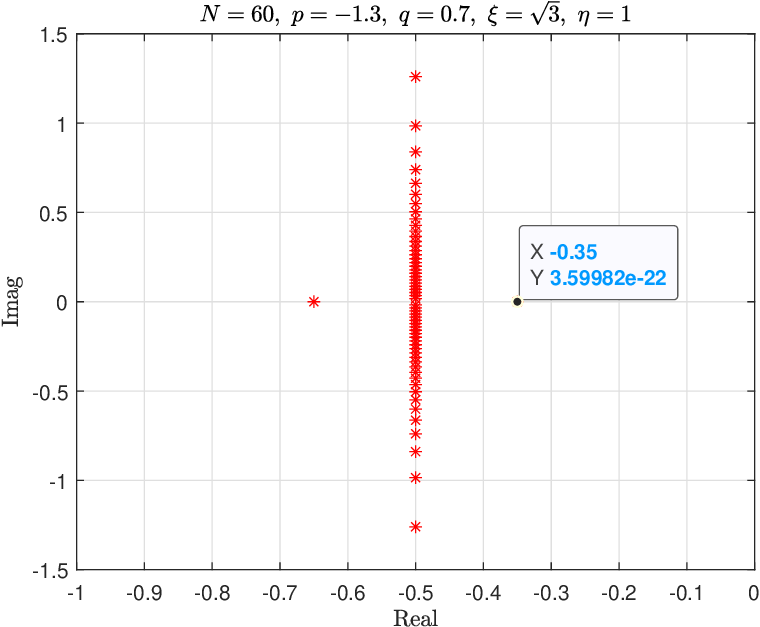}
  }
\caption{
Distribution of Bethe roots in the complex plane with parameters $N = 60$, $p = -1.3$, $q = 0.7$, $\xi = \sqrt3$, and $\eta = 1$.
}
\label{fig:II0}
\end{figure}

\begin{figure}[htbp]
  \centering
  \subfloat[Global configuration of Bethe roots]{
    \includegraphics[width=0.47\textwidth]{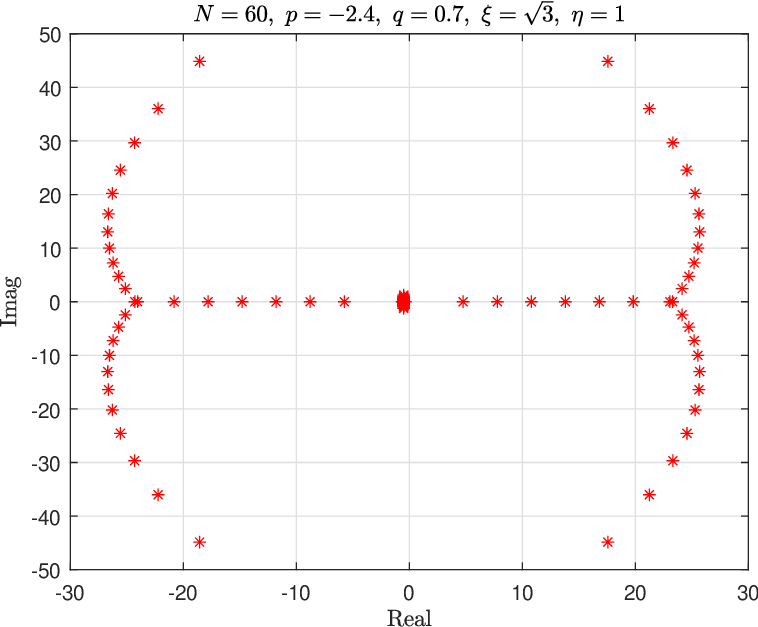}
  }
  \subfloat[Enlarged view of the region where the structural change occurs]{
    \includegraphics[width=0.47\textwidth]{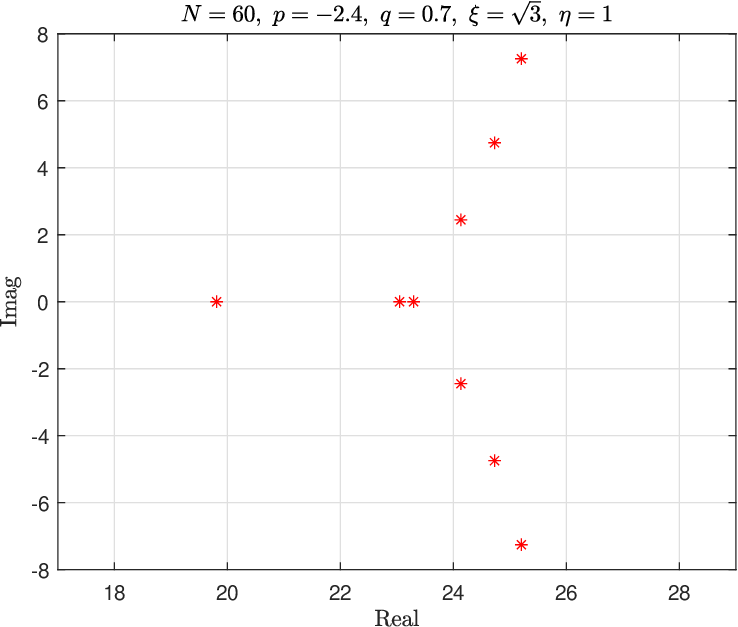}
  }
\caption{
Distribution of Bethe roots in the complex plane with parameters $N = 60$, $p = -2.4$, $q = 0.7$, $\xi = \sqrt3$, and $\eta = 1$.
}
\label{fig:II1}
\end{figure}

In contrast to the case of positive $p$, as the absolute value of the parameter increases, the inner line roots move outward away from the center, while the outermost line roots remain almost unchanged, as shown from Fig.~\ref{fig:II0} to Fig.~\ref{fig:II1}.

\begin{figure}[htbp]
  \centering
  \subfloat[Global configuration of Bethe roots]{
    \includegraphics[width=0.47\textwidth]{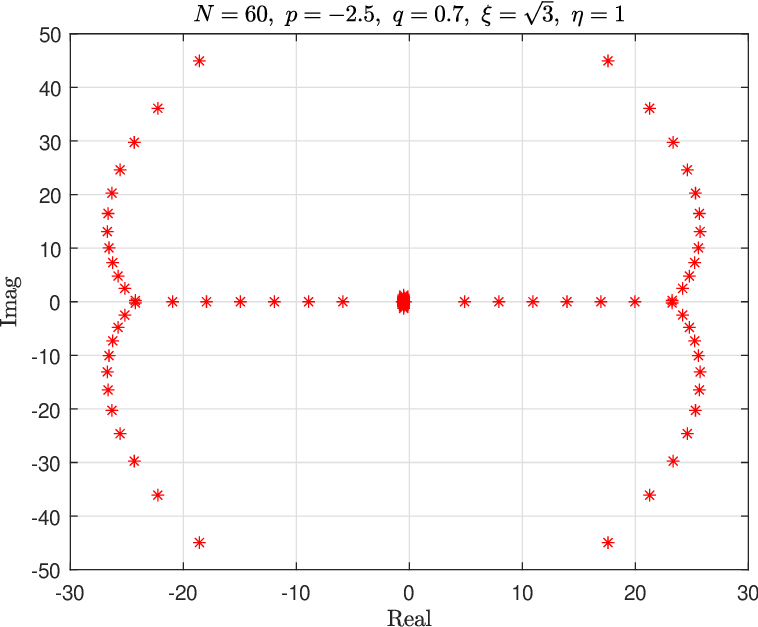}
  }
  \subfloat[Enlarged view of the region where the structural change occurs]{
    \includegraphics[width=0.47\textwidth]{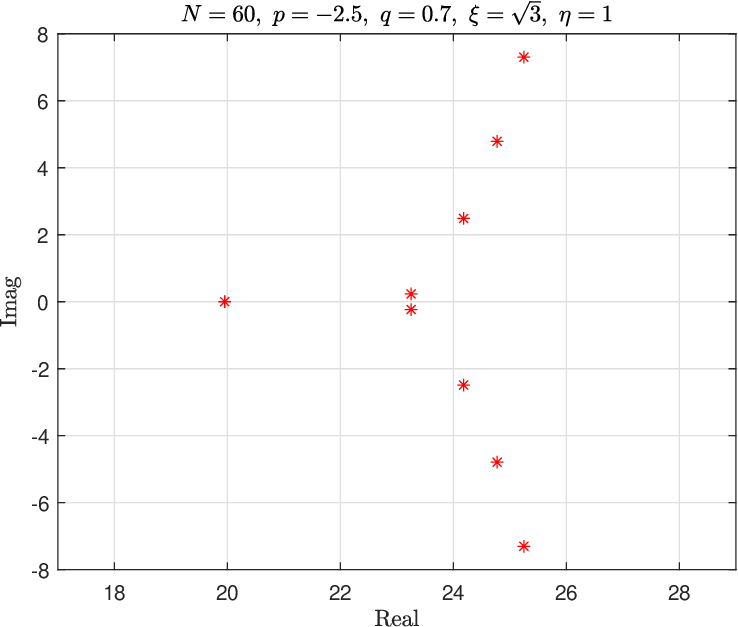}
  }
\caption{
Distribution of Bethe roots in the complex plane with parameters $N = 60$, $p = -2.5$, $q = 0.7$, $\xi = \sqrt3$, and $\eta = 1$.
}
\label{fig:II2}
\end{figure}

When the two outermost line roots on the real axis approach their limiting separation, they separate from that position and become part of the arc roots, as shown in Fig.~\ref{fig:II2}.

\begin{figure}[htbp]
  \centering
  \subfloat[Global configuration of Bethe roots]{
    \includegraphics[width=0.47\textwidth]{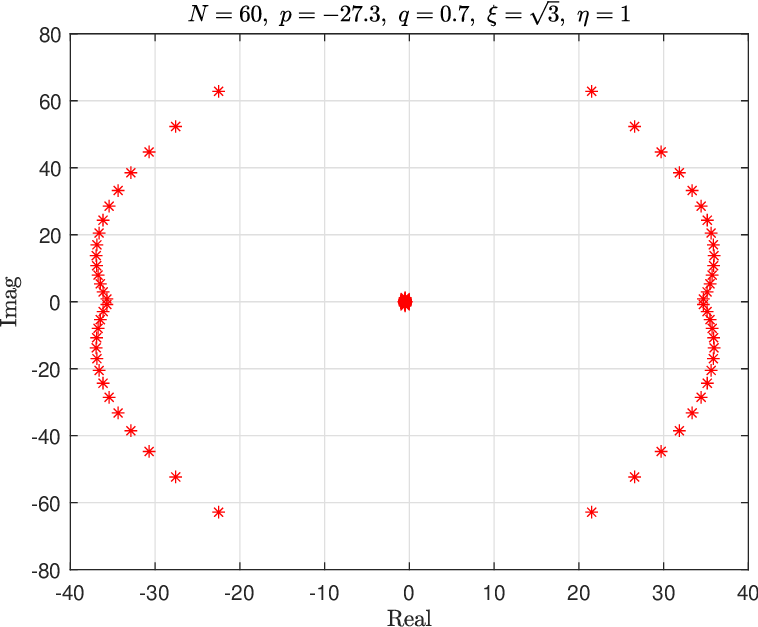}
  }
  \subfloat[Enlarged view of the central region]{
    \includegraphics[width=0.47\textwidth]{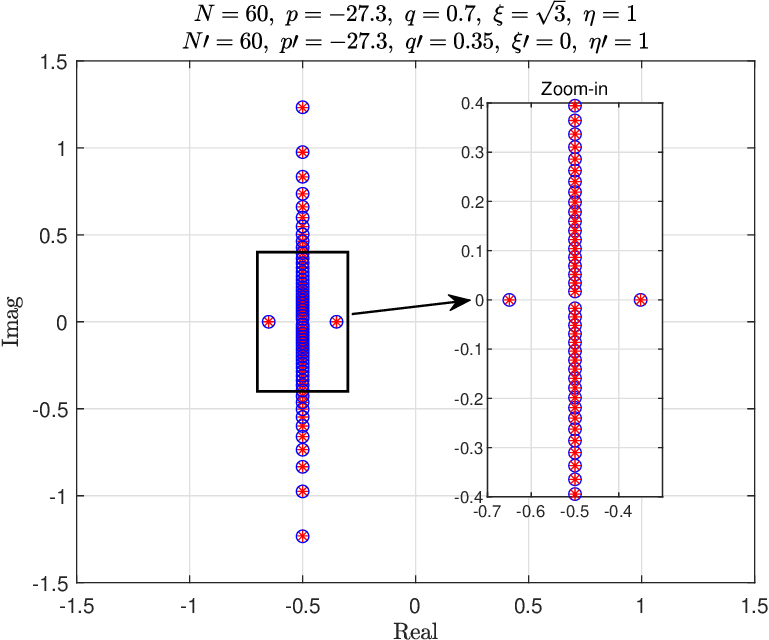}
  }
\caption{
Distribution of Bethe roots in the complex plane with parameters $N = 60$, $p = -27.3$, $q = 0.7$, $\xi = \sqrt3$, and $\eta = 1$.
Blue circles in panel~(b) indicate the Bethe roots obtained in the homogeneous case with parameters $N' = 60$, $p' = -27.3$, $q' = 0.35$, $\xi' = 0$, and $\eta' = 1$.
}
\label{fig:II3}
\end{figure}

Eventually, all line roots transform into the arc roots and continue to move toward infinity, totaling $N$, while the $N$ regular roots remain at the center.
As in the previous case, the Hamiltonian regains $U(1)$ symmetry, and the Bethe Ansatz equations reduce to the homogeneous form.
A comparison of these two situations, shown in Fig.~\ref{fig:II3}, exhibits excellent agreement.

\subsection{Surface energy}
\label{sec:surface energy}
After clarifying the structure of the Bethe roots, we calculate the surface energy through their distribution.
The surface energy $E_b$ is defined, in the thermodynamic limit, as the difference between the ground-state energy $E_g$ of the present system and the ground-state energy $E_p$ of the corresponding periodic chain~\cite{Gaudin1971,Wen2017}:
\begin{equation}
E_b = E_g - E_p.
\label{eq:surface-energy-definition}
\end{equation}

According to the four classes of Bethe roots introduced above, and since the energy expression in Eq.~\eqref{Eigenvalue1} shows that the contributions from all roots are additive, the calculation of the ground-state energy can likewise be decomposed into four parts.
The energy expression can be rewritten as
\begin{equation}
E_g = E^{(\mathrm r)} + E^{(\mathrm l)} + E^{(\mathrm p)} + E^{(\mathrm a)} + \frac{\eta}{p} + \frac{\eta}{\bar q} + N - 1,
\label{eq:E_g}
\end{equation}
where
\begin{equation}
  E^{(\mathrm \alpha)} = \sum_{j=1}^{n^{(\mathrm \alpha)}} \frac{2\eta^2}{\lambda_j^{(\mathrm \alpha)}(\lambda_j^{(\mathrm \alpha)}+\eta)}, \quad \mathrm \alpha={\mathrm{r,l,p,a}}.
\label{eq:Eigenvalue_alpha}
\end{equation}
Correspondingly, the $Q$-function can be factorized into four sectors as
\begin{equation}
Q(u)=Q^{(\mathrm r)}(u)Q^{(\mathrm l)}(u)Q^{(\mathrm p)}(u)Q^{(\mathrm a)}(u),
\end{equation}
where
\begin{align}
Q^{(\mathrm \alpha)}(u)&=\prod_{j=1}^{n^{(\mathrm \alpha)}}(u-\lambda_j^{(\mathrm \alpha)})(u+\lambda_j^{(\mathrm \alpha)}+\eta), \quad \mathrm \alpha={\mathrm{r,l,p,a}}.
\end{align}
In the thermodynamic limit, because the absolute values of $\{\lambda_j^{(\mathrm a)}\}$ all tend to infinity, the corresponding energy contribution vanishes:
\begin{equation}
E^{(\mathrm a)} = 0.
\end{equation}

We now evaluate $E^{(\mathrm r)}$.
Because their structure is similar to that of the periodic chain, the thermodynamic Bethe ansatz method can be applied.
Substituting $\{\lambda_j^{(\mathrm r)}\}$ into Eq.~\eqref{eq:BAEs}, we obtain
\begin{align}
&\left(\frac{\lambda_j^{(\mathrm r)}+\eta}{\lambda_j^{(\mathrm r)}}\right)^{2N+1}
\frac{(\lambda_j^{(\mathrm r)}+p)\big[(1+\xi^{2})^{\frac{1}{2}}\lambda_j^{(\mathrm r)}+q\big]}
{(\lambda_j^{(\mathrm r)}-p+\eta)\big[(1+\xi^{2})^{\frac{1}{2}}(\lambda_j^{(\mathrm r)}+\eta)-q\big]}
=-\prod_{\alpha\in\{\mathrm r,\mathrm l,\mathrm p,\mathrm a\}}
\frac{Q^{(\alpha)}(\lambda_j^{(\mathrm r)}+\eta)}
     {Q^{(\alpha)}(\lambda_j^{(\mathrm r)}-\eta)}
\nonumber\\
&-\frac{[1-(1+\xi^{2})^{\frac{1}{2}}](2\lambda_j^{(\mathrm r)}+\eta)(\lambda_j^{(\mathrm r)}+\eta)^{2N+1}}
{(\lambda_j^{(\mathrm r)}-p+\eta)\big[(1+\xi^{2})^{\frac{1}{2}}(\lambda_j^{(\mathrm r)}+\eta)-q\big]
\displaystyle\prod_{\alpha\in\{\mathrm r,\mathrm l,\mathrm p,\mathrm a\}}
Q^{(\alpha)}(\lambda_j^{(\mathrm r)}-\eta)},
\quad j=1,\ldots,n^{(\mathrm r)}.
\label{eq:reBAEs}
\end{align}
Since the absolute values of all $\{\lambda_j^{(\mathrm a)}\}$ tend to infinity, it follows that
\begin{equation}
\frac{Q^{(\mathrm a)}(\lambda_j^{(\mathrm r)}+\eta)}{Q^{(\mathrm a)}(\lambda_j^{(\mathrm r)}-\eta)}\to1.
\end{equation}
Moreover, all of $\{\lambda_j^{(\mathrm a)}\}$ have absolute values tending to infinity, the contribution of the inhomogeneous term can be neglected, which gives
\begin{equation}
\frac{[1-(1+\xi^{2})^{\frac{1}{2}}](2\lambda_j^{(\mathrm r)}+\eta)(\lambda_j^{(\mathrm r)}+\eta)^{2N+1}}
{(\lambda_j^{(\mathrm r)}-p+\eta)\big[(1+\xi^{2})^{\frac{1}{2}}(\lambda_j^{(\mathrm r)}+\eta)-q\big]
\displaystyle\prod_{\alpha\in\{\mathrm r,\mathrm l,\mathrm p,\mathrm a\}}
Q^{(\alpha)}(\lambda_j^{(\mathrm r)}-\eta)}\to0.
\end{equation}

At this stage, only $Q^{(\mathrm l)}(u)$ and $Q^{(\mathrm p)}(u)$ can affect the distribution of $\{\lambda_j^{(\mathrm r)}\}$.
For convenience, we write $\lambda^{(\mathrm l)}_{\beta}=\eta l_{\beta}-\frac{\eta}{2}$ with $\beta=1,\ldots,n^{(\mathrm l)}$, and $\lambda^{(\mathrm p)}_{2\gamma-1}=\eta a_{\gamma}+i\eta b_{\gamma}-\frac{\eta}{2}$, $\lambda^{(\mathrm p)}_{2\gamma}=\eta a_{\gamma}-i\eta b_{\gamma}-\frac{\eta}{2}$ with $\gamma=1,\ldots,\frac{n^{(\mathrm p)}}{2}$.
Here, $l_{\beta}>1$, $a_{\gamma}>1$, and $b_{\gamma}>0$; the remaining parameter regimes are discussed in Appendix~\ref{app:F}.
The corresponding $Q$-function is
\begin{align}
Q^{(\mathrm l)}(u)
&=
\prod_{\beta=1}^{n^{(\mathrm l)}}
\left(u-\eta l_{\beta}+\tfrac{\eta}{2}\right)
\left(u+\eta l_{\beta}+\tfrac{\eta}{2}\right),
\nonumber\\
Q^{(\mathrm p)}(u)
&=
\prod_{\gamma=1}^{\tfrac{n^{(\mathrm p)}}{2}}
\left[
\left(u-\eta a_{\gamma}-i\eta b_{\gamma}+\tfrac{\eta}{2}\right)
\left(u+\eta a_{\gamma}+i\eta b_{\gamma}+\tfrac{\eta}{2}\right)
\right.
\nonumber\\
&\quad\times
\left.
\left(u-\eta a_{\gamma}+i\eta b_{\gamma}+\tfrac{\eta}{2}\right)
\left(u+\eta a_{\gamma}-i\eta b_{\gamma}+\tfrac{\eta}{2}\right)
\right].
\label{eq:lab}
\end{align}
Then Eq.~\eqref{eq:reBAEs} can be rewritten as
\begin{align}
&\left(\frac{\lambda_j^{(\mathrm r)}+\eta}{\lambda_j^{(\mathrm r)}}\right)^{2N+1}
\frac{(\lambda_j^{(\mathrm r)}+p)(\lambda_j^{(\mathrm r)}+\bar{q})}
     {(\lambda_j^{(\mathrm r)}-p+\eta)(\lambda_j^{(\mathrm r)}-\bar{q}+\eta)}
\nonumber\\
&=
-\prod_{k=1}^{n^{(\mathrm r)}}
\frac{\big(\lambda_j^{(\mathrm r)}-\lambda_k^{(\mathrm r)}+\eta\big)
      \big(\lambda_j^{(\mathrm r)}+\lambda_k^{(\mathrm r)}+2\eta\big)}
     {\big(\lambda_j^{(\mathrm r)}-\lambda_k^{(\mathrm r)}-\eta\big)
      \big(\lambda_j^{(\mathrm r)}+\lambda_k^{(\mathrm r)}\big)}
\prod_{\beta=1}^{n^{(\mathrm l)}}
\frac{\big(\lambda_j^{(\mathrm r)}-\eta l_{\beta}+\tfrac{3\eta}{2}\big)
      \big(\lambda_j^{(\mathrm r)}+\eta l_{\beta}+\tfrac{3\eta}{2}\big)}
     {\big(\lambda_j^{(\mathrm r)}-\eta l_{\beta}-\tfrac{\eta}{2}\big)
      \big(\lambda_j^{(\mathrm r)}+\eta l_{\beta}-\tfrac{\eta}{2}\big)}
\nonumber\\
&\quad\times
\prod_{\gamma=1}^{\frac{n^{(\mathrm p)}}{2}}
\left[
\frac{\big(\lambda_j^{(\mathrm r)}-\eta a_{\gamma}-i\eta b_{\gamma}+\tfrac{3\eta}{2}\big)
      \big(\lambda_j^{(\mathrm r)}+\eta a_{\gamma}+i\eta b_{\gamma}+\tfrac{3\eta}{2}\big)}
     {\big(\lambda_j^{(\mathrm r)}-\eta a_{\gamma}-i\eta b_{\gamma}-\tfrac{\eta}{2}\big)
      \big(\lambda_j^{(\mathrm r)}+\eta a_{\gamma}+i\eta b_{\gamma}-\tfrac{\eta}{2}\big)}
\right.
\nonumber\\
&\quad\left.\times
\frac{\big(\lambda_j^{(\mathrm r)}-\eta a_{\gamma}+i\eta b_{\gamma}+\tfrac{3\eta}{2}\big)
      \big(\lambda_j^{(\mathrm r)}+\eta a_{\gamma}-i\eta b_{\gamma}+\tfrac{3\eta}{2}\big)}
     {\big(\lambda_j^{(\mathrm r)}-\eta a_{\gamma}+i\eta b_{\gamma}-\tfrac{\eta}{2}\big)
      \big(\lambda_j^{(\mathrm r)}+\eta a_{\gamma}-i\eta b_{\gamma}-\tfrac{\eta}{2}\big)}
\right],
\quad j=1,\ldots,n^{(\mathrm r)}.
\label{eq:rreBAEs}
\end{align}
For convenience, we put $\lambda_j^{(\mathrm r)}=i\eta\mu_j-\tfrac{\eta}{2}$, $\hat{p}=\tfrac{p}{\eta}-\tfrac{1}{2}$, $\hat{q}=\tfrac{\bar{q}}{\eta}-\tfrac{1}{2}$, $\eta=1$.
Eq.~\eqref{eq:rreBAEs} can be rewritten as
\begin{align}
&-\prod_{k=1}^{n^{(\mathrm r)}}
\frac{(\mu_j-\mu_k-i)(\mu_j+\mu_k-i)}
     {(\mu_j-\mu_k+i)(\mu_j+\mu_k+i)}
\nonumber\\
&=
\left(\frac{\mu_j-\frac{i}{2}}{\mu_j+\frac{i}{2}}\right)^{2N+1}
\frac{(\mu_j-i\hat q)(\mu_j-i\hat p)}
     {(\mu_j+i\hat q)(\mu_j+i\hat p)}
\prod_{\beta=1}^{n^{(\mathrm l)}}
\frac{\left[\mu_j-i(l_{\beta}-1)\right]\left[\mu_j+i(l_{\beta}+1)\right]}
     {\left[\mu_j+i(l_{\beta}-1)\right]\left[\mu_j-i(l_{\beta}+1)\right]}
\nonumber\\
&\quad\times
\prod_{\gamma=1}^{\frac{n^{(\mathrm p)}}{2}}
\left\{
\frac{\left[\mu_j-b_{\gamma}-i(a_{\gamma}-1)\right]\left[\mu_j-b_{\gamma}+i(a_{\gamma}+1)\right]}
     {\left[\mu_j-b_{\gamma}+i(a_{\gamma}-1)\right]\left[\mu_j-b_{\gamma}-i(a_{\gamma}+1)\right]}
\right.
\nonumber\\
&\quad\left.\times
\frac{\left[\mu_j+b_{\gamma}-i(a_{\gamma}-1)\right]\left[\mu_j+b_{\gamma}+i(a_{\gamma}+1)\right]}
     {\left[\mu_j+b_{\gamma}+i(a_{\gamma}-1)\right]\left[\mu_j+b_{\gamma}-i(a_{\gamma}+1)\right]}
\right\},
\quad j=1,\ldots,n^{(\mathrm r)}.
\label{eq:rrreBAEs}
\end{align}
The contribution of $\{\mu_j\}$ to the ground-state energy is thus given by
\begin{equation}
E^{(\mathrm r)}=-\sum_{j=1}^{n^{(\mathrm r)}}\frac{2}{\mu_j^2+\frac{1}{4}}.
\end{equation}

Let us consider the case where both $p$ and $\bar q$ are greater than $\tfrac{\eta}{2}$.
In this regime, one has $n^{(\mathrm r)}=\tfrac{N}{2}$ in the ground state, and no boundary string is present.
Moreover, since the roots appear in pairs $\pm\mu_j$ on the real axis, we take
$\mu_j>0$ without loss of generality.
By taking the logarithm of Eq.~\eqref{eq:rrreBAEs}, we obtain
\begin{align}
&2\pi I_j
+\sum_{k=1}^{n^{(\mathrm r)}}\left[\theta_{2}(\mu_j-\mu_k)+\theta_{2}(\mu_j+\mu_k)\right]
-\theta_{1}(\mu_j)
\nonumber\\
&=
2N\theta_{1}(\mu_j)
+\theta_{2\hat{p}}(\mu_j)
+\theta_{2\hat{q}}(\mu_j)
+\sum_{\beta=1}^{n^{(\mathrm l)}}
\left[\theta_{2(l_{\beta}-1)}(\mu_j)-\theta_{2(l_{\beta}+1)}(\mu_j)\right]
\nonumber\\
&\quad
+\sum_{\gamma=1}^{\frac{n^{(\mathrm p)}}{2}}
\Big[
\theta_{2(a_{\gamma}-1)}(\mu_j-b_{\gamma})
-\theta_{2(a_{\gamma}+1)}(\mu_j-b_{\gamma})
\nonumber\\
&\quad
+\theta_{2(a_{\gamma}-1)}(\mu_j+b_{\gamma})
-\theta_{2(a_{\gamma}+1)}(\mu_j+b_{\gamma})
\Big],
\label{eq:BAEslog}
\end{align}
where $\theta_n(x) = 2 \arctan(2x / n)$, and $\{I_j\}$ are integers.
Then, we define the counting function
\begin{align}
Z(u)&=\frac{1}{2\pi}\Bigg\{
\theta_1(u)
+\frac{1}{2N}\Bigg[
\theta_1(u)
+\theta_{2\hat p}(u)
+\theta_{2\hat q}(u)
+\sum_{\beta=1}^{n^{(\mathrm l)}}
\bigl(\theta_{2(l_{\beta}-1)}(u)-\theta_{2(l_{\beta}+1)}(u)\bigr)
\nonumber\\
&\quad
+\sum_{\gamma=1}^{\frac{n^{(\mathrm p)}}{2}}
\Big(
\theta_{2(a_{\gamma}-1)}(u-b_{\gamma})
-\theta_{2(a_{\gamma}+1)}(u-b_{\gamma})
+\theta_{2(a_{\gamma}-1)}(u+b_{\gamma})
-\theta_{2(a_{\gamma}+1)}(u+b_{\gamma})
\Big)
\nonumber\\
&\quad
-\sum_{k=1}^{n^{(\mathrm r)}}\bigl(\theta_2(u-\mu_k)+\theta_2(u+\mu_k)\bigr)
\Bigg]
\Bigg\}.
\end{align}
It is obvious that $Z(\mu_j) = I_j /(2N)$. In the thermodynamic limit, the density distributions are determined by
\begin{equation}
  \rho(u)+\rho^{h}(u)=\frac{dZ(u)}{du}.
\end{equation}
Taking the derivative of $Z(u)$, we obtain
\begin{align}
\rho(u)+\rho^{h}(u)
&=a_1(u)
+\frac{1}{2N}\Bigg[
a_1(u)
+a_{2\hat p}(u)
+a_{2\hat q}(u)
+\sum_{\beta=1}^{n^{(\mathrm l)}}
\bigl(a_{2(l_{\beta}-1)}(u)-a_{2(l_{\beta}+1)}(u)\bigr)
\nonumber\\
&\quad
+\sum_{\gamma=1}^{\frac{n^{(\mathrm p)}}{2}}
\Big(
a_{2(a_{\gamma}-1)}(u-b_{\gamma})
-a_{2(a_{\gamma}+1)}(u-b_{\gamma})
\nonumber\\&\quad
+a_{2(a_{\gamma}-1)}(u+b_{\gamma})
-a_{2(a_{\gamma}+1)}(u+b_{\gamma})
\Big)
\Bigg]
\nonumber\\
&\quad
-\int_{-\infty}^{\infty}a_2(u-v)\rho(v)\,dv.
\end{align}
Numerical analysis shows that, as in the case with $U(1)$ symmetry, $I_j$ are consecutive in the ground state.
The hole density takes the form $\rho^{h}(u)=\frac{1}{2N}\delta(u)$.
The appearance of $\delta(u)$ is due to the hole at $I_j=0$, which is a solution of the BAEs but yields a vanishing wave function~\cite{book}.
This means that the density of roots in the ground state $\rho_g(u)$ satisfies
\begin{align}
\rho_g(u)
&=a_1(u)
+\frac{1}{2N}\Bigg[
a_1(u)
+a_{2\hat p}(u)
+a_{2\hat q}(u)
-\delta(u)
+\sum_{\beta=1}^{n^{(\mathrm l)}}
\bigl(a_{2(l_{\beta}-1)}(u)-a_{2(l_{\beta}+1)}(u)\bigr)
\nonumber\\
&\quad
+\sum_{\gamma=1}^{\frac{n^{(\mathrm p)}}{2}}
\Big(
a_{2(a_{\gamma}-1)}(u-b_{\gamma})
-a_{2(a_{\gamma}+1)}(u-b_{\gamma})
+a_{2(a_{\gamma}-1)}(u+b_{\gamma})
-a_{2(a_{\gamma}+1)}(u+b_{\gamma})
\Big)
\Bigg]
\nonumber\\
&\quad
-\int_{-\infty}^{\infty}a_2(u-v)\rho_g(v)\,dv,
\label{eq:rhog}
\end{align}
where $a_n(u)=\frac{1}{2\pi}\frac{n}{u^2+n^2/4}$.

For comparison, the root density of the corresponding periodic chain satisfies~\cite{tak71}
\begin{equation}
\rho_p(u)=a_1(u)
-\int_{-\infty}^{\infty}a_2(u-v)\rho_p(v)\,dv.
\label{eq:periodic-density-equation}
\end{equation}
Subtracting Eq.~\eqref{eq:periodic-density-equation} from Eq.~\eqref{eq:rhog}, we obtain
\begin{align}
\delta\rho(u)&=\rho_g(u)-\rho_p(u) \nonumber \\
&=
\frac{1}{2N}\Bigg[
a_1(u)
+a_{2\hat p}(u)
+a_{2\hat q}(u)
-\delta(u)
+\sum_{\beta=1}^{n^{(\mathrm l)}}
\bigl(a_{2(l_{\beta}-1)}(u)-a_{2(l_{\beta}+1)}(u)\bigr)
\nonumber\\
&\quad
+\sum_{\gamma=1}^{\frac{n^{(\mathrm p)}}{2}}
\Big(
a_{2(a_{\gamma}-1)}(u-b_{\gamma})
-a_{2(a_{\gamma}+1)}(u-b_{\gamma})
+a_{2(a_{\gamma}-1)}(u+b_{\gamma})
-a_{2(a_{\gamma}+1)}(u+b_{\gamma})
\Big)
\Bigg]
\nonumber\\
&\quad
-\int_{-\infty}^{\infty}a_2(u-v)\delta\rho(v)\,dv,
\end{align}
Taking the Fourier transform of $\delta\rho(u)$, we have
\begin{align}
\delta \tilde{\rho}(\omega)
&=
\frac{1}{2N}\,
\frac{
e^{-\hat{p}|\omega|}
+e^{-\hat{q}|\omega|}
+e^{-|\omega|/2}
-1
+\displaystyle\sum_{\beta=1}^{n^{(\mathrm l)}}
\left[
e^{-(l_{\beta}-1)|\omega|}
-e^{-(l_{\beta}+1)|\omega|}
\right]
}{
1+e^{-|\omega|}}
\nonumber\\
&\quad
+\frac{1}{N}\,
\frac{
\displaystyle\sum_{\gamma=1}^{\frac{n^{(\mathrm p)}}{2}}
\cos(\omega b_{\gamma})
\left[
e^{-(a_{\gamma}-1)|\omega|}
-e^{-(a_{\gamma}+1)|\omega|}
\right]
}{
1+e^{-|\omega|}}.
\end{align}
Then, the contribution of the regular roots to the surface energy is given by
\begin{align}
E^{(\mathrm r)}-E_p
&=-N
-4\pi N\int_{-\infty}^{\infty} a_1(u)\delta\rho(u)\,du
\nonumber\\
&=-N
-2N\int_{-\infty}^{\infty}\tilde a_1(\omega)\delta\tilde\rho(\omega)\,d\omega
\nonumber\\
&=-N
+\pi-2\ln2
-2\int_{0}^{\infty}
\frac{
e^{-\frac p \eta\omega}
+e^{-\frac {\bar q}{ \eta}\omega}
}{
1+e^{-\omega}
}\,d\omega
-\sum_{\beta=1}^{n^{(\mathrm l)}}
\frac{2\eta^2}{\left(\eta l_{\beta}-\tfrac\eta 2\right)\left(\eta l_{\beta}+\tfrac\eta 2\right)}
\nonumber\\
&\quad
-\sum_{\gamma=1}^{\frac{n^{(\mathrm p)}}{2}}
\left[
\frac{2\eta^2}{(\eta a_{\gamma}-i\eta b_{\gamma}-\tfrac\eta 2)(\eta a_{\gamma}-i\eta b_{\gamma}+\tfrac\eta 2)}
+
\frac{2\eta^2}{(\eta a_{\gamma}+i\eta b_{\gamma}-\tfrac\eta 2)(\eta a_{\gamma}+i\eta b_{\gamma}+\tfrac\eta 2)}
\right]
\nonumber\\
&=-N
+\pi-2\ln2
-2\int_{0}^{\infty}
\frac{
e^{-\frac p \eta\omega}
+e^{-\frac {\bar q}{ \eta}\omega}
}{
1+e^{-\omega}
}\,d\omega
\nonumber\\
&\quad
-\sum_{\beta=1}^{n^{(\mathrm l)}}
\frac{2\eta^2}{\lambda^{(\mathrm l)}_{\beta}\bigl(\lambda^{(\mathrm l)}_{\beta}+\eta\bigr)}
-\sum_{\gamma=1}^{n^{(\mathrm p)}}
\frac{2\eta^2}{\lambda^{(\mathrm p)}_{\gamma}\bigl(\lambda^{(\mathrm p)}_{\gamma}+\eta\bigr)}.
\label{eq:eb0}
\end{align}
According to Eq.~\eqref{eq:Eigenvalue_alpha}, the last two terms on the
right-hand side of Eq.~\eqref{eq:eb0} can be identified with
$E^{(\mathrm l)}$ and $E^{(\mathrm p)}$. These contributions originate from
the back-flow effects induced by the line roots and paired-line roots,
respectively.
Thus, Eq.~\eqref{eq:eb0} can be rewritten as
\begin{align}
E^{(\mathrm r)}-E_p=-N
+\pi-2\ln2
-2\int_{0}^{\infty}
\frac{
e^{-\frac p \eta\omega}
+e^{-\frac {\bar q}{ \eta}\omega}
}{
1+e^{-\omega}
}\,d\omega-E^{(\mathrm l)}-E^{(\mathrm p)}.
\end{align}
Using Eq.~\eqref{eq:E_g}, the surface energy expression in Eq.~\eqref{eq:surface-energy-definition} can be written as
\begin{align}
E_b&=\frac{\eta}{p} + \frac{\eta}{\bar q} - 1
+\pi-2\ln2
-2\int_{0}^{\infty}
\frac{
e^{-\frac {p}{\eta}\omega}
+e^{-\frac {\bar q}{ \eta}\omega}
}{
1+e^{-\omega}
}d\omega.
\end{align}

For $p\bar{q}<0$, the relation $n^{(\mathrm r)}=\tfrac{N-1}{2}$ implies that the missing regular root may be viewed as a hole at infinity, which gives no contribution to the surface energy.
The possible boundary string also leaves the energy contribution determined by the root-density equation unchanged.
With all cases summarized, the surface energy is finally obtained in the following unified form:
\begin{align}
E_b&=\frac{\eta}{|p|} + \frac{\eta}{|\bar q|} - 1
+\pi-2\ln2
-2\int_{0}^{\infty}
\frac{
e^{-\frac {|p|}{\eta}\omega}
+e^{-\frac {|\bar q|}{ \eta}\omega}
}{
1+e^{-\omega}
}d\omega.
\end{align}
The present result also coincides exactly with that obtained by the zero-root method in Ref.~\cite{Dong2023}.

\subsection{Structure of excited states}
For the \(U(1)\)-symmetric case, the excitation structure of the Bethe roots has been thoroughly discussed in Ref.~\cite{Takahashi1999}.
In this subsection, we focus on the excitation structure of the Bethe roots once the \(U(1)\) symmetry is broken.

As shown in Sec.~\ref{sec:zero-excited}, there is a clear correspondence between the two types of elementary excitations and the distributions of zero roots.
This correspondence allows us to investigate the Bethe-root structure of the corresponding elementary excitations through the structure of the zero roots.

\begin{figure}[htbp]
  \centering
  \subfloat[Global configuration of Bethe roots]{
    \includegraphics[width=0.47\textwidth]{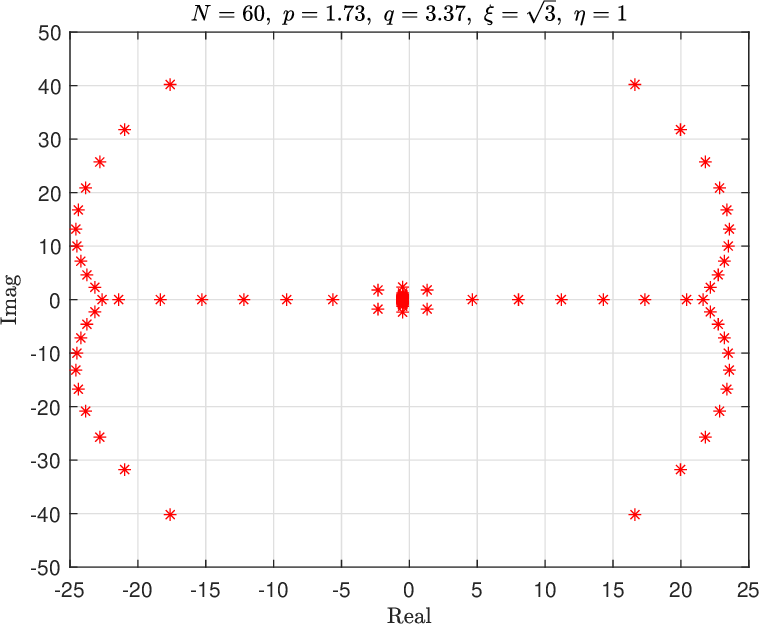}
  }
  \subfloat[Enlarged view of the central region]{
    \includegraphics[width=0.47\textwidth]{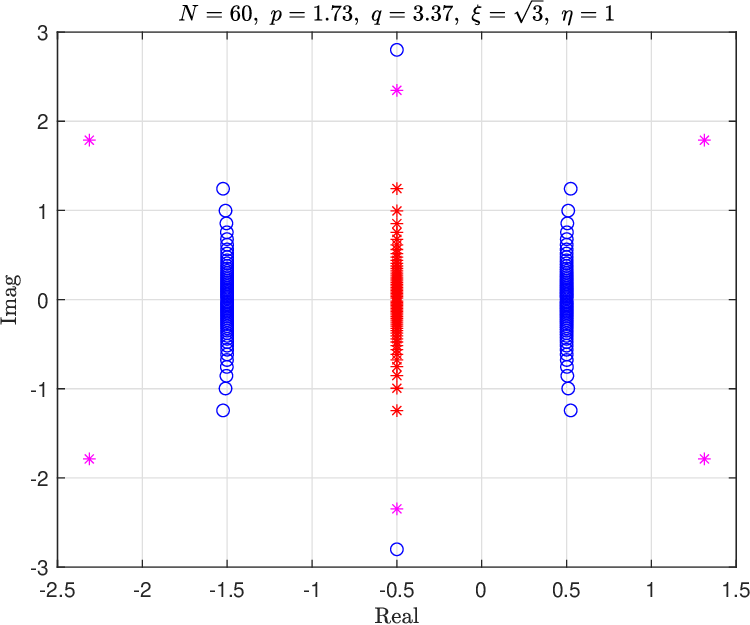}
  }
\caption{
Ground-state distribution of Bethe roots in the complex plane with parameters $N = 60$, $p = 1.73$, $q = 3.37$, $\xi = \sqrt{3}$, and $\eta = 1$.
In panel~(b), the zero roots are marked by blue circles, the regular roots by red asterisks, and the paired-line roots by magenta asterisks.
}
\label{fig:bground}
\end{figure}

Numerically, the regular roots \(\{\lambda_j^{(\mathrm r)}\}\) in the ground state are in close correspondence with the zero roots \(\{z_j\}\): the roots in the central region correspond to the bulk strings, while those near the boundaries match the boundary strings one by one.
As an example, for the case \(p\bar q>0\), we use the same parameters as in
Sec.~\ref{sec:zero-excited}. The corresponding ground-state configuration is
shown in Fig.~\ref{fig:bground}. To facilitate the comparison between the two
sets of roots, we overlay the zero roots from Fig.~\ref{fig:ground} onto
Fig.~\ref{fig:bground} (b), together with the Bethe roots.

\subsubsection{Helical spinon excitation}
\begin{figure}[htbp]
  \centering
  \subfloat[Global configuration of Bethe roots]{
    \includegraphics[width=0.47\textwidth]{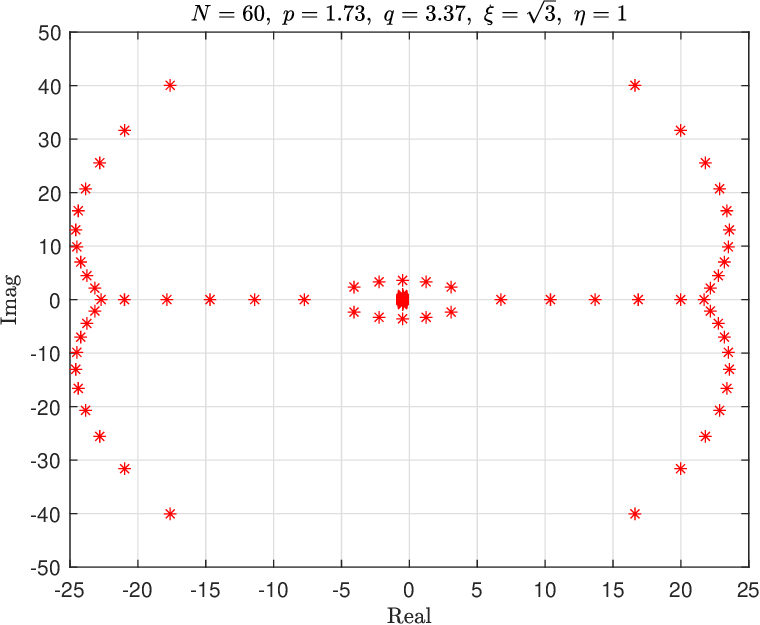}
  }
  \subfloat[Enlarged view of the central region]{
    \includegraphics[width=0.47\textwidth]{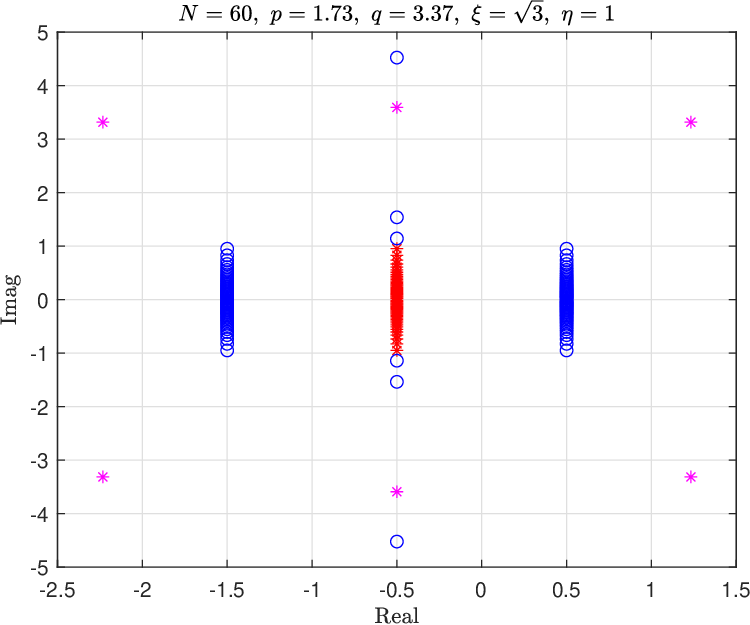}
  }
\caption{
Third excited-state distribution of Bethe roots in the complex plane with parameters $N = 60$, $p = 1.73$, $q = 3.37$, $\xi = \sqrt{3}$, and $\eta = 1$.
In panel~(b), the zero roots are marked by blue circles, the regular roots by red asterisks, and the paired-line roots by magenta asterisks.
}
\label{fig:bspinon}
\end{figure}

In the Bethe-root configuration, the helical spinon excitation is characterized
by the disappearance of four central regular roots, which are replaced by either
line roots or paired-line roots. Figure~\ref{fig:bspinon} shows the Bethe-root
configuration corresponding to the third excited state. We overlay the zero-root
structure obtained for the same set of parameters, shown in Fig.~\ref{fig:spinon},
onto Fig.~\ref{fig:bspinon} (b). Relative to the ground-state configuration, the newly generated zero roots
\[
\left\{\pm\tilde z_1-\frac{\eta}{2},
\ \pm\tilde z_2-\frac{\eta}{2}\right\}
\]
correspond to four holes in the regular-root distribution.

Using the $\mu$-parametrization introduced in Sec.~\ref{sec:surface energy},
$\lambda_j^{(\mathrm r)}=-\frac{\eta}{2}+i\eta\mu_j$,
the positions of these holes can be written as
\[
\{\pm\mu_1^h,\ \pm\mu_2^h\},
\]
where $\tilde z_j=i\eta\mu_j^h$ and $\mu_j^h>0$. The remaining regular roots
occupy the other available quantum-number positions.

According to Sec.~\ref{sec:surface energy}, the energy change can be characterized by the change in the density of the central roots.
Owing to the reflection symmetry of the roots, we therefore parameterize the
independent holes by the two positive quantities $\mu_1^h$ and $\mu_2^h$.
In this case, the density of holes is
\begin{equation}
\rho_{sp}^{h}(u)=\frac{1}{2N}\delta\!\left(u-\mu_{1}^{h}\right)+\frac{1}{2N}\delta\!\left(u-\mu_{2}^{h}\right).
\end{equation}
The corresponding change in the root density is given by
\begin{equation}
  \delta\rho(u)+\rho_{sp}^{h}(u)=-\int_{-\infty}^{\infty} a_{2}(u-v)\,\delta\rho(v)\,dv.
\end{equation}
With the help of Fourier transformation, we obtain
\begin{equation}
  \delta\tilde{\rho}(\omega)
=
-\frac{1}{2N}\,
\frac{e^{i\omega\mu_1^h}+e^{i\omega\mu_2^h}}
{1+e^{-|\omega|}}.
\end{equation}
The energy of the helical spinon excitation is
\begin{align}
\delta E_{sp}
&=-2N\int_{-\infty}^{\infty}\tilde a_1(\omega)\delta\tilde\rho(\omega)d\omega
\nonumber\\
&=\frac{\pi}{\cosh(\pi\mu_1^h)}+\frac{\pi}{\cosh(\pi\mu_2^h)}.
\end{align}

\subsubsection{String excitation}
\begin{figure}[htbp]
  \centering
  \subfloat[Global configuration of Bethe roots]{
    \includegraphics[width=0.47\textwidth]{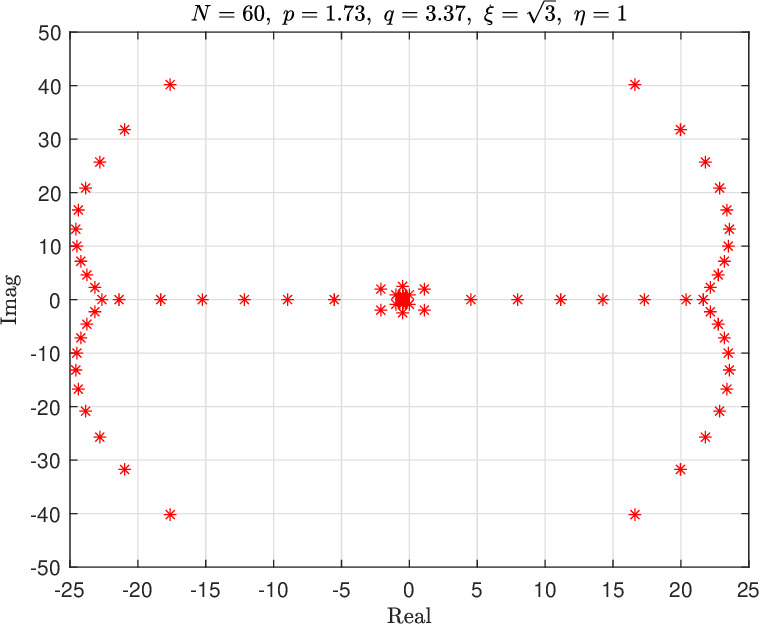}
  }
  \subfloat[Enlarged view of regular roots]{
    \includegraphics[width=0.47\textwidth]{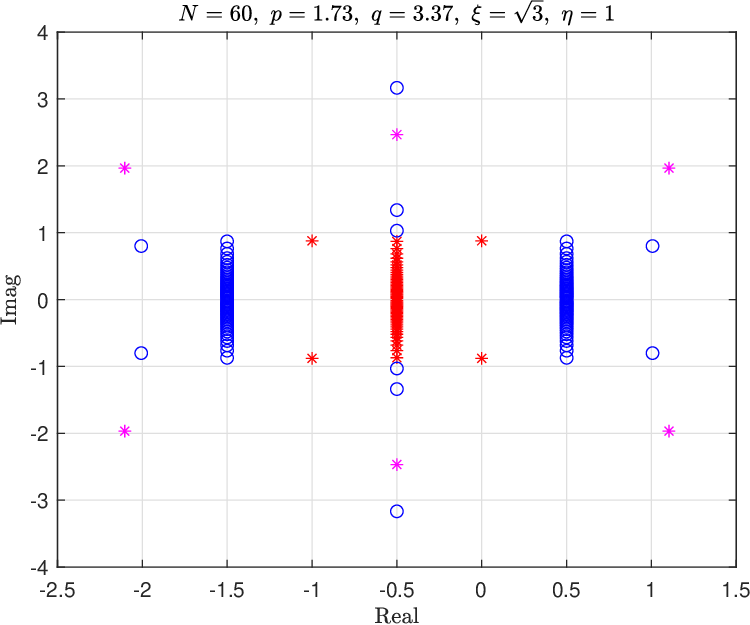}
  }
\caption{
Sixth excited-state distribution of Bethe roots in the complex plane with parameters $N = 60$, $p = 1.73$, $q = 3.37$, $\xi = \sqrt{3}$, and $\eta = 1$.
In panel~(b), the zero roots are marked by blue circles, the regular roots by red asterisks, and the paired-line roots by magenta asterisks.
}
\label{fig:bstring}
\end{figure}

Under the string excitation, \(2n\) central roots
\(\{\pm \mu_j^\prime\}_{j=1}^{n}\) disappear and are replaced by a pair of
symmetric \(n\)-strings \(\{\pm \mu_{j,s}^{(n)}\}_{j=1}^{n}\), where
\[
\mu_{j,s}^{(n)}=\mu_s^{(n)}-\frac{i}{2}(n+1-2j).
\]
Here, \(\mu_s^{(n)}\) denotes the center of the positive \(n\)-string on the
real axis. Figure~\ref{fig:bstring} shows the case \(n=2\), corresponding to
the sixth excited state.
Similarly, we overlay the zero-root structure obtained for the same set of
parameters, shown in Fig.~\ref{fig:string}, onto Fig.~\ref{fig:bstring} (b).
One can see that the newly generated zero roots
\[
\left\{\pm\tilde z_3-\frac{\eta}{2},
\ \pm\tilde z_4-\frac{\eta}{2}\right\}
\]
mark the holes in the regular Bethe-root distribution.

We now calculate the excitation energy of the simplest excited state, consisting of a 2-string at $\mu^{(2)}_s$ and two holes on the real axis.
As discussed above, the corresponding density functions are then given by
\begin{align}
\rho_{1,st}^{h}(u)&=\frac{1}{2N}\left[\delta\!\left(u-\mu_{1}^{h}\right)+\delta\!\left(u-\mu_{2}^{h}\right)\right],\\
\rho_2(u)&=\frac{1}{2N}\delta\!\left(u-\mu^{(2)}_s\right),
\end{align}
where $\rho_{1,\mathrm{st}}^{h}(u)$ and $\rho_2(u)$ are the densities of 2-string holes and 2-strings, respectively.
The deviation of the density from that of the ground state is given by
\begin{align}
\delta\rho_{1}(u)
&=
-\rho_{1,st}^{h}(u)
-\int_{-\infty}^{\infty} a_{2}(u-v)\,\delta\rho_{1}(v)\,dv\nonumber\\
&\quad
-\int_{-\infty}^{\infty} \bigl[a_{1}(u-v)+a_{3}(u-v)\bigr]\rho_{2}(v)\,dv.
\end{align}
By taking the Fourier transform, we obtain
\begin{equation}
  \delta \tilde{\rho}_1(\omega) = -\frac{1}{2N} \left[ \frac{e^{i\omega \mu_1^h} + e^{i\omega \mu_2^h}}{1 + e^{-|\omega|}} + e^{-|\omega|/2} e^{i\omega \mu^{(2)}_s} \right].
\end{equation}
This allows us to derive the string excitation energy as
\begin{align}
\delta E_{st}
&=-2N\int_{-\infty}^{\infty}\tilde a_1(\omega)\delta\tilde\rho_{1}(\omega)d\omega - 2\pi\left[a_1\left(\mu^{(2)}_s+\tfrac{i}{2}\right)+a_1\left(\mu^{(2)}_s-\tfrac{i}{2}\right)\right]
\nonumber\\
&=\frac{\pi}{\cosh(\pi\mu_1^h)}+\frac{\pi}{\cosh(\pi\mu_2^h)}.
\end{align}

For the case \(p\bar q<0\), the elementary excitations still follow the same
patterns discussed above. It should be noted that, regardless of the sign of
\(p\bar q\), paired-line-root structures may appear in both types of elementary
excitations. Such structures are often difficult to distinguish numerically from
string structures. Therefore, in the analysis of excited states, zero roots
provide a clearer structural characterization than Bethe roots.

\section{Conclusion and outlook}
\label{sec:6}

By combining tensor-network techniques with the Bethe Ansatz, we have carried out a systematic study of the open XXX spin-\(\tfrac12\) chain on large lattices, and high-precision results for the zero roots and Bethe roots have been obtained for the ground state as well as for several representative excited states.

We find that, regardless of whether the \(U(1)\) symmetry is preserved or broken, the ground-state zero roots can be classified into three types: bulk strings, boundary strings, and additional roots. To ensure the reliability of the results, we further examined the zero roots using rigorous complex-analysis methods. At large system sizes, we verified the zero-root phase diagram proposed in Ref.~\cite{Dong2023} and tracked in detail the evolution of the zero roots as the parameters cross the phase boundaries.

In the \(U(1)\)-symmetry-broken regime, the ground-state Bethe roots naturally split into four geometric classes, namely regular roots, line roots, paired-line roots, and arc roots. As the boundary parameters are increased from finite values to infinity, the \(U(1)\) symmetry is gradually restored, and we systematically tracked the evolutionary trajectories and transition rules of these four classes of Bethe roots throughout the entire crossover process. In addition, we analyzed the structures of zero roots and Bethe roots in several representative excited states.

Overall, in the symmetry-broken regime, the zero-root configuration is much simpler than the full Bethe-root configuration, thus providing a more efficient entry point for analytical studies. As an example of the analytical utility of the nondiagonal Bethe-root configuration, we further derived analytic expressions for the surface energy and excitation energies based on its geometric structure.

With further improvements in the precision of DMRG algorithms, together with techniques such as parallel computing, the BA-DMRG method proposed in this work is expected to be extended to even larger systems. The numerical framework established here may also be generalized to other quantum integrable models with higher spin or higher rank. The distribution structures of zero roots and Bethe roots obtained here in the nondiagonal case also provide a foundation for further analytical studies of the thermodynamic properties and correlation functions of \(U(1)\)-symmetry-broken systems.

\section*{Acknowledgments}

We would like to thank Prof. Y. Wang for his valuable discussions and continuous encouragement. Yi Qiao especially thanks Dr. K. Wang for helpful discussions and suggestions on tensor-network numerics. The financial supports from the National Key R$\&$D Program of China (Grant No.2021YFA1402104), the National Natural Science Foundation of China (Grant Nos. 12305005, 12547107, 12434006 and 12247103), Major Basic Research Program of Natural Science of Shaanxi Province (Grant Nos. 2021JCW-19 and 2017ZDJC-32), Young Talent Fund of Xi'an Association for Science and Technology (Grant No. 959202313086), and Shaanxi Fundamental Science Research Project for Mathematics and Physics (Grant No. 22JSZ005).

\appendix
\section{MPO Representation of the Transfer Matrix}
\label{app:A}
\setcounter{equation}{0}
\renewcommand{\theequation}{A.\arabic{equation}}
To facilitate the derivation of analytical expressions, the $R$-matrix in Eq.~\eqref{eq:R-matrix} can be rewritten as a $2\times 2$ operator-valued matrix acting on the space labeled by $0$:
\begin{equation}
R_{0,j}(u) :=
\begin{pmatrix}
a_j(u) & b_j(u) \\
c_j(u) & d_j(u)
\end{pmatrix}_0 .
\label{eq:R'-matrix}
\end{equation}
For the present $R$-matrix, $R_{j,0}(u) = R_{0,j}(u)$.
Using the projection notation introduced later, the matrix elements in the basis $\{|0\rangle_0,|1\rangle_0\}$ read
\begin{equation}
\begin{aligned}
a_j(u) &= R_{0,j}(u)\big|^{0}_{0}
= \begin{pmatrix} u+\eta & 0 \\ 0 & u \end{pmatrix}_j , \\[4pt]
b_j(u) &= R_{0,j}(u)\big|^{0}_{1}
= \begin{pmatrix} 0 & 0 \\ \eta & 0 \end{pmatrix}_j , \\[4pt]
c_j(u) &= R_{0,j}(u)\big|^{1}_{0}
= \begin{pmatrix} 0 & \eta \\ 0 & 0 \end{pmatrix}_j , \\[4pt]
d_j(u) &= R_{0,j}(u)\big|^{1}_{1}
= \begin{pmatrix} u & 0 \\ 0 & u+\eta \end{pmatrix}_j .
\end{aligned}
\label{eq:R-elements}
\end{equation}

The analytical derivation is presented below.
For clarity we set $N=3$, with the general case obtained analogously. For simplicity, we will omit the parameter $(u)$ in the subsequent derivation:
\begin{equation}
\begin{aligned}
&\operatorname{tr}_0\!\left\{
K_0^+ R_{0,3} R_{0,2} R_{0,1}
K_0^- R_{1,0} R_{2,0} R_{3,0}
\right\}
\\[4pt]
&=\operatorname{tr}_0\!\left\{
K_0^+ R_{0,3} R_{0,2} R_{0,1}
K_0^- R_{0,1} R_{0,2} R_{0,3}
\right\}
\\[4pt]
&= K_0^+ R_{0,3} R_{0,2} R_{0,1}
   K_0^- R_{0,1} R_{0,2} R_{0,3}\big|^{y}_{y}
\\
&= K_0^+\big|^{y}_{a}
   R_{0,3}\big|^{a}_{b}
   R_{0,2}\big|^{b}_{c}
   R_{0,1}\big|^{c}_{d}
   K_0^-\big|^{d}_{e}
   R_{0,1}\big|^{e}_{f}
   R_{0,2}\big|^{f}_{g}
   R_{0,3}\big|^{g}_{y}
\\
&=
  R_{0,1}\big|^{c}_{d} K_0^-\big|^{d}_{e} R_{0,1}\big|^{e}_{f}
  \cdot
  R_{0,2}\big|^{b}_{c} R_{0,2}\big|^{f}_{g}
  \cdot
  R_{0,3}\big|^{a}_{b} K_0^+\big|^{y}_{a} R_{0,3}\big|^{g}_{y}
\\
&=
  R_{0,1}\big|^{c}_{d} K_0^-\big|^{d}_{e} R_{0,1}\big|^{e}_{f}
  \cdot
  R_{0,2}^{t_0}\big|^{c}_{b} R_{0,2}\big|^{f}_{g}
  \cdot
  R_{0,3}^{t_0}\big|^{b}_{a} K_0^{+t_0}\big|^{a}_{y}
  R_{0,3}^{t_0}\big|^{y}_{g}
\\
&=
 (R_{0,1}K_0^-R_{0,1})\big|^{c}_{f}
 \cdot
 R_{0,2}^{t_0}\big|^{c}_{b} R_{0',2}\big|^{f}_{g}
 \cdot
 (R_{0,3}^{t_0} K_0^{+t_0} R_{0,3}^{t_0})\big|^{b}_{g}
\\
&=
 (R_{0,1}K_0^-R_{0,1})\big|^{cf}
 \cdot
 (R_{0,2}^{t_0}R_{0',2})\big|^{bg}_{cf}
 \cdot
 (R_{0,3}^{t_0}K_0^{+t_0}R_{0,3}^{t_0})\big|_{bg}
\\
&= W_1 \cdot W_2 \cdot W_3 .
\end{aligned}
\label{eq:transform}
\end{equation}
In this way, each tensor in Fig.~\ref{fig:transform} can be explicitly written out:
\begin{equation}
W_1 =
\begin{pmatrix}
(p+u)a_1 a_1 + (p-u)b_1 c_1 \\ (p+u)a_1 b_1 + (p-u)b_1 d_1 \\
(p+u)c_1 a_1 + (p-u)d_1 c_1 \\ (p+u)c_1 b_1 + (p-u)d_1 d_1
\end{pmatrix}^t,
\label{eq:W1}
\end{equation}
\begin{equation}
W_j =
\begin{pmatrix}
a_j a_j & a_j b_j & c_j a_j & c_j b_j \\
a_j c_j & a_j d_j & c_j c_j & c_j d_j \\
b_j a_j & b_j b_j & d_j a_j & d_j b_j \\
b_j c_j & b_j d_j & d_j c_j & d_j d_j
\end{pmatrix},
\label{eq:Wj}
\end{equation}
\begin{equation}
W_N =
\begin{pmatrix}
(q+u+\eta)a_N a_N + \xi(u+\eta)(a_N b_N + c_N a_N) + (q-u-\eta)c_N b_N \\
(q+u+\eta)a_N c_N + \xi(u+\eta)(a_N d_N + c_N c_N) + (q-u-\eta)c_N d_N \\
(q+u+\eta)b_N a_N + \xi(u+\eta)(b_N b_N + d_N a_N) + (q-u-\eta)d_N b_N \\
(q+u+\eta)b_N c_N + \xi(u+\eta)(b_N d_N + d_N c_N) + (q-u-\eta)d_N d_N
\end{pmatrix}.
\label{eq:WN}
\end{equation}

\bigskip
\noindent\textbf{Notation.}
Let $X_{0,j}$ be a linear operator on $\mathcal{H}_0 \otimes \mathcal{H}_j$, and $\{|\alpha\rangle_0\}$ a fixed basis of $\mathcal{H}_0$.
Its matrix element with respect to the $0$-th space is defined as
\begin{equation}
X_{0,j}\big|^{\alpha}_{\beta}
:= (\langle\alpha|_0 \otimes \mathbb{I}_j)\,
   X_{0,j}\,
   (|\beta\rangle_0 \otimes \mathbb{I}_j),
\label{eq:single-index}
\end{equation}
which is an operator on $\mathcal{H}_j$.

For two operators $X_{0,j}$ and $Y_{0,k}$, the component of their tensor product is defined by
\begin{equation}
(X_{0,j} Y_{0,k})\big|^{\alpha\gamma}_{\beta\delta}
:= X_{0,j}\big|^{\alpha}_{\beta}\;
   Y_{0,k}\big|^{\gamma}_{\delta},
\label{eq:fused-index}
\end{equation}
where $\alpha\gamma$ and $\beta\delta$ denote ordered pairs of basis indices.

In the derivation above, repeated upper-lower index pairs are implicitly summed.
When composite labels such as $\alpha\gamma$ appear, the summation convention is applied to each component index separately.

The transpose in the $0$-th space is defined directly at the level of matrix elements by
\begin{equation}
X_{0,j}^{\,t_0}\big|^{\alpha}_{\beta}
:= (\langle\beta|_0 \otimes \mathbb{I}_j)\,
   X_{0,j}\,
   (|\alpha\rangle_0 \otimes \mathbb{I}_j)
 = X_{0,j}\big|^{\beta}_{\alpha}.
\label{eq:t0}
\end{equation}

\section{Derivation from the Maximum Modulus Principle}
\label{app:B}
\setcounter{equation}{0}
\renewcommand{\theequation}{B.\arabic{equation}}
Let $f(z)$ be analytic on the closed disk $|z| \leq R$. Suppose there exists a constant $a > 0$ such that
\begin{equation}
|f(z)| \geq a \quad \text{for all } |z| = R,
\end{equation}
and
\begin{equation}
|f(0)| < a.
\end{equation}
Then $f(z)$ must have at least one zero inside the open disk $|z| < R$.

Assume to the contrary that $f(z)$ has no zeros in the interior of the disk. Since $f(z)$ is analytic and nonzero on $|z| \leq R$, the reciprocal function
\begin{equation}
\varphi(z) = \frac{1}{f(z)}
\end{equation}
is also analytic on the closed disk.

By the assumption, we have
\begin{equation}
|\varphi(0)| = \left| \frac{1}{f(0)} \right| > \frac{1}{a},
\end{equation}
and on the boundary,
\begin{equation}
|\varphi(z)| = \left| \frac{1}{f(z)} \right| \leq \frac{1}{a} \quad \text{for all } |z| = R.
\end{equation}

This implies that $|\varphi(z)|$ attains a strict maximum at an interior point $z = 0$, namely
\begin{equation}
|\varphi(z)| \leq \frac{1}{a} < |\varphi(0)|, \quad \text{for all } |z| = R.
\end{equation}
This contradicts the maximum modulus principle, which states that the maximum of $|\varphi(z)|$ on a closed domain is achieved on the boundary unless $\varphi(z)$ is constant.

Therefore, the assumption must be false, and $f(z)$ has at least one zero in $|z| < R$.

\section{Argument Principle: Validation Data}
\label{app:C}
\setcounter{equation}{0}
\setcounter{table}{0}
\renewcommand{\theequation}{C.\arabic{equation}}
\renewcommand{\thetable}{C.\arabic{table}}
This appendix presents the numerical verification of all detected zero roots via the argument principle.
For each candidate, a closed loop composed of five evaluation points as shown in Fig. \ref{fig:u0} is used to compute the total argument change of the function $\Lambda(u)$.
The chosen offset $\delta$ ensures that only one zero is enclosed per loop, and the result $\Delta = 1$ confirms the presence of a zero.
\begin{longtable}{cccccccccc}
\caption
{
Validation of candidate zero roots for the ground state based on the argument principle. The model parameters are $N = 100$, $p = -0.6$, $q = -0.3$, $\xi = 1.2$, and $\eta = 1$.
Each row corresponds to one candidate surrounded by a closed path with offset $\delta$.
The columns $\arg_1$-$\arg_5$ list the phase-unwrapped argument values at the five evaluation points.
$\Delta = \frac{1}{2\pi} \Delta_C \arg f(z)$ represents the total argument change along the loop.
} \\
\toprule
Index & Real($z_j$) & Imag($z_j$) & $\delta$ & $\arg_1$ & $\arg_2$ & $\arg_3$ & $\arg_4$ & $\arg_5$ & $\Delta$ \\
\midrule
\endfirsthead
\multicolumn{8}{c}{{\tablename\ \thetable{} -- continued from previous page}} \\
\toprule
Index & Real($z_j$) &Imag($z_j$) & $\delta$ & $\arg_1$ & $\arg_2$ & $\arg_3$ & $\arg_4$ & $\arg_5$ & $\Delta$ \\
\midrule
\endhead
\multicolumn{8}{r}{{Continued on next page}} \\
\endfoot
\endlastfoot
1 & -1.537797484405 & 0.000000000000 & $10^{-12}$ &-1.57 &-0.00 &1.57 &3.14 &4.71 &1.0\\\midrule
2 & -1.499950624342 & 0.010265201348 & $10^{-12}$ &-0.32 &1.25 &2.82 &4.39 &5.96 &1.0\\\midrule
3 & -1.499950624342 & -0.010265201348 & $10^{-12}$ &-2.82 &-1.25 &0.32 &1.89 &3.46 &1.0\\\midrule
4 & -1.499826137427 & 0.020485525878 & $10^{-12}$ &2.47 &4.04 &5.61 &7.18 &8.75 &1.0\\\midrule
5 & -1.499826137427 & -0.020485525878 & $10^{-12}$ &0.67 &2.24 &3.81 &5.38 &6.95 &1.0\\\midrule
6 & -1.499668437249 & 0.030659439714 & $10^{-12}$ &-1.05 &0.52 &2.09 &3.66 &5.23 &1.0\\\midrule
7 & -1.499668437249 & -0.030659439714 & $10^{-12}$ &-2.09 &-0.52 &1.05 &2.62 &4.19 &1.0\\\midrule
8 & -1.499508982879 & 0.040805650649 & $10^{-12}$ &1.69 &3.26 &4.83 &6.40 &7.97 &1.0\\\midrule
9 & -1.499508982879 & -0.040805650649 & $10^{-12}$ &1.45 &3.03 &4.60 &6.17 &7.74 &1.0\\\midrule
10 & -1.499362827910 & 0.050950483670 & $10^{-12}$ &-1.88 &-0.31 &1.26 &2.83 &4.40 &1.0\\\midrule
11 & -1.499362827910 & -0.050950483670 & $10^{-12}$ &-1.26 &0.31 &1.88 &3.45 &5.03 &1.0\\\midrule
12 & -1.499234729326 & 0.061118328765 & $10^{-12}$ &2.33 &3.90 &5.47 &7.04 &8.61 &1.0\\\midrule
13 & -1.499234729326 & -0.061118328765 & $10^{-12}$ &0.81 &2.38 &3.95 &5.52 &7.09 &1.0\\\midrule
14 & -1.499124661078 & 0.071329849178 & $10^{-12}$ &-0.35 &1.22 &2.79 &4.36 &5.93 &1.0\\\midrule
15 & -1.499124661078 & -0.071329849178 & $10^{-12}$ &-2.79 &-1.22 &0.35 &1.92 &3.49 &1.0\\\midrule
16 & -1.499030724545 & 0.081602732169 & $10^{-12}$ &-3.02 &-1.45 &0.13 &1.70 &3.27 &1.0\\\midrule
17 & -1.499030724545 & -0.081602732169 & $10^{-12}$ &-0.13 &1.45 &3.02 &4.59 &6.16 &1.0\\\midrule
18 & -1.498950502407 & 0.091952693540 & $10^{-12}$ &0.61 &2.18 &3.75 &5.32 &6.89 &1.0\\\midrule
19 & -1.498950502407 & -0.091952693540 & $10^{-12}$ &2.53 &4.10 &5.67 &7.24 &8.81 &1.0\\\midrule
20 & -1.498881624162 & 0.102394318961 & $10^{-12}$ &-2.03 &-0.46 &1.11 &2.68 &4.25 &1.0\\\midrule
21 & -1.498881624162 & -0.102394318961 & $10^{-12}$ &-1.11 &0.46 &2.03 &3.61 &5.18 &1.0\\\midrule
22 & -1.498821967495 & 0.112941674848 & $10^{-12}$ &1.53 &3.10 &4.67 &6.24 &7.81 &1.0\\\midrule
23 & -1.498821967495 & -0.112941674848 & $10^{-12}$ &1.61 &3.18 &4.75 &6.32 &7.90 &1.0\\\midrule
24 & -1.498769723486 & 0.123608744783 & $10^{-12}$ &-2.13 &-0.56 &1.01 &2.59 &4.16 &1.0\\\midrule
25 & -1.498769723486 & -0.123608744783 & $10^{-12}$ &-1.01 &0.56 &2.13 &3.70 &5.27 &1.0\\\midrule
26 & -1.498723364861 & 0.134409764676 & $10^{-12}$ &0.49 &2.06 &3.63 &5.20 &6.77 &1.0\\\midrule
27 & -1.498723364861 & -0.134409764676 & $10^{-12}$ &2.65 &4.22 &5.79 &7.36 &8.94 &1.0\\\midrule
28 & -1.498681620170 & -0.145359467945 & $10^{-12}$ &0.04 &1.62 &3.19 &4.76 &6.33 &1.0\\\midrule
29 & -1.498681620170 & 0.145359467945 & $10^{-12}$ &3.10 &4.67 &6.24 &7.81 &9.38 &1.0\\\midrule
30 & -1.498643427402 & -0.156473301820 & $10^{-12}$ &-2.55 &-0.98 &0.59 &2.16 &3.73 &1.0\\\midrule
31 & -1.498643427402 & 0.156473301820 & $10^{-12}$ &-0.59 &0.98 &2.55 &4.12 &5.69 &1.0\\\midrule
32 & -1.498607890702 & -0.167767617858 & $10^{-12}$ &1.14 &2.71 &4.28 &5.86 &7.43 &1.0\\\midrule
33 & -1.498607890702 & 0.167767617858 & $10^{-12}$ &2.00 &3.57 &5.14 &6.71 &8.28 &1.0\\\midrule
34 & -1.498574252042 & -0.179259858798 & $10^{-12}$ &-1.71 &-0.14 &1.43 &3.00 &4.58 &1.0\\\midrule
35 & -1.498574252042 & 0.179259858798 & $10^{-12}$ &-1.43 &0.14 &1.71 &3.28 &4.85 &1.0\\\midrule
36 & -1.498541855015 & -0.190968750517 & $10^{-12}$ &2.28 &3.86 &5.43 &7.00 &8.57 &1.0\\\midrule
37 & -1.498541855015 & 0.190968750517 & $10^{-12}$ &0.86 &2.43 &4.00 &5.57 &7.14 &1.0\\\midrule
38 & -1.498510123234 & -0.202914506180 & $10^{-12}$ &-2.87 &-1.30 &0.27 &1.84 &3.41 &1.0\\\midrule
39 & -1.498510123234 & 0.202914506180 & $10^{-12}$ &-0.27 &1.30 &2.87 &4.44 &6.01 &1.0\\\midrule
40 & -1.498478538264 & -0.215119055622 & $10^{-12}$ &-2.81 &-1.24 &0.33 &1.90 &3.48 &1.0\\\midrule
41 & -1.498478538264 & 0.215119055622 & $10^{-12}$ &-0.33 &1.24 &2.81 &4.38 &5.95 &1.0\\\midrule
42 & -1.498446620333 & -0.227606305494 & $10^{-12}$ &2.19 &3.76 &5.33 &6.90 &8.48 &1.0\\\midrule
43 & -1.498446620333 & 0.227606305494 & $10^{-12}$ &0.95 &2.52 &4.09 &5.66 &7.23 &1.0\\\midrule
44 & -1.498413915387 & -0.240402437913 & $10^{-12}$ &-1.56 &0.01 &1.58 &3.15 &4.72 &1.0\\\midrule
45 & -1.498413915387 & 0.240402437913 & $10^{-12}$ &-1.58 &-0.01 &1.56 &3.13 &4.70 &1.0\\\midrule
46 & -1.498379979920 & 0.253536262645 & $10^{-12}$ &2.23 &3.80 &5.37 &6.94 &8.51 &1.0\\\midrule
47 & -1.498379979920 & -0.253536262645 & $10^{-12}$ &0.92 &2.49 &4.06 &5.63 &7.20 &1.0\\\midrule
48 & -1.498344367511 & 0.267039629271 & $10^{-12}$ &-0.25 &1.32 &2.89 &4.46 &6.03 &1.0\\\midrule
49 & -1.498344367511 & -0.267039629271 & $10^{-12}$ &-2.89 &-1.32 &0.25 &1.82 &3.39 &1.0\\\midrule
50 & -1.498306617468 & 0.280947921906 & $10^{-12}$ &-2.71 &-1.14 &0.43 &2.00 &3.57 &1.0\\\midrule
51 & -1.498306617468 & -0.280947921906 & $10^{-12}$ &-0.43 &1.14 &2.71 &4.28 &5.85 &1.0\\\midrule
52 & -1.498266239594 & 0.295300650461 & $10^{-12}$ &2.01 &3.58 &5.15 &6.72 &8.29 &1.0\\\midrule
53 & -1.498266239594 & -0.295300650461 & $10^{-12}$ &1.14 &2.71 &4.28 &5.85 &7.42 &1.0\\\midrule
54 & -1.498222698566 & -0.310142165508 & $10^{-12}$ &-1.86 &-0.29 &1.28 &2.85 &4.42 &1.0\\\midrule
55 & -1.498222698566 & 0.310142165508 & $10^{-12}$ &-1.28 &0.29 &1.86 &3.43 &5.00 &1.0\\\midrule
56 & -1.498175399285 & -0.325522534071 & $10^{-12}$ &2.61 &4.19 &5.76 &7.33 &8.90 &1.0\\\midrule
57 & -1.498175399285 & 0.325522534071 & $10^{-12}$ &0.53 &2.10 &3.67 &5.24 &6.81 &1.0\\\midrule
58 & -1.498123665480 & -0.341498614506 & $10^{-12}$ &0.25 &1.82 &3.39 &4.96 &6.53 &1.0\\\midrule
59 & -1.498123665480 & 0.341498614506 & $10^{-12}$ &2.89 &4.46 &6.03 &7.60 &9.17 &1.0\\\midrule
60 & -1.498066714188 & 0.358135397291 & $10^{-12}$ &-1.06 &0.51 &2.08 &3.65 &5.22 &1.0\\\midrule
61 & -1.498066714188 & -0.358135397291 & $10^{-12}$ &-2.08 &-0.51 &1.06 &2.63 &4.20 &1.0\\\midrule
62 & -1.498003623959 & -0.375507684154 & $10^{-12}$ &1.24 &2.81 &4.38 &5.95 &7.52 &1.0\\\midrule
63 & -1.498003623959 & 0.375507684154 & $10^{-12}$ &1.90 &3.47 &5.05 &6.62 &8.19 &1.0\\\midrule
64 & -1.497933293140 & -0.393702223808 & $10^{-12}$ &-0.36 &1.21 &2.78 &4.35 &5.93 &1.0\\\midrule
65 & -1.497933293140 & 0.393702223808 & $10^{-12}$ &-2.78 &-1.21 &0.36 &1.93 &3.50 &1.0\\\midrule
66 & -1.497854383995 & -0.412820455349 & $10^{-12}$ &-0.56 &1.01 &2.58 &4.15 &5.72 &1.0\\\midrule
67 & -1.497854383926 & 0.412820455322 & $10^{-12}$ &-2.58 &-1.01 &0.56 &2.14 &3.71 &1.0\\\midrule
68 & -1.497765248390 & 0.432982083595 & $10^{-12}$ &1.61 &3.18 &4.75 &6.32 &7.89 &1.0\\\midrule
69 & -1.497765245778 & -0.432982082050 & $10^{-12}$ &1.53 &3.10 &4.68 &6.25 &7.82 &1.0\\\midrule
70 & -1.497663849835 & -0.454329835627 & $10^{-12}$ &-2.56 &-0.99 &0.59 &2.16 &3.73 &1.0\\\midrule
71 & -1.497663817867 & 0.454329812915 & $10^{-12}$ &-0.59 &0.99 &2.56 &4.13 &5.70 &1.0\\\midrule
72 & -1.497547476813 & 0.477035682510 & $10^{-12}$ &-0.50 &1.07 &2.64 &4.21 &5.79 &1.0\\\midrule
73 & -1.497547289313 & -0.477035535699 & $10^{-12}$ &-2.64 &-1.07 &0.50 &2.07 &3.64 &1.0\\\midrule
74 & -1.497413408143 & -0.501310359159 & $10^{-12}$ &1.65 &3.22 &4.79 &6.36 &7.93 &1.0\\\midrule
75 & -1.497412793089 & 0.501309849137 & $10^{-12}$ &1.49 &3.06 &4.63 &6.20 &7.77 &1.0\\\midrule
76 & -1.497255294626 & 0.527412667506 & $10^{-12}$ &-2.88 &-1.31 &0.26 &1.83 &3.40 &1.0\\\midrule
77 & -1.497254083110 & -0.527411625020 & $10^{-12}$ &-0.26 &1.31 &2.88 &4.46 &6.03 &1.0\\\midrule
78 & -1.497070337484 & -0.555673660500 & $10^{-12}$ &-2.07 &-0.50 &1.07 &2.64 &4.21 &1.0\\\midrule
79 & -1.497068864609 & 0.555672362108 & $10^{-12}$ &-1.07 &0.50 &2.07 &3.64 &5.21 &1.0\\\midrule
80 & -1.496845058431 & 0.586510751527 & $10^{-12}$ &2.51 &4.09 &5.66 &7.23 &8.80 &1.0\\\midrule
81 & -1.496843960123 & -0.586509770845 & $10^{-12}$ &0.63 &2.20 &3.77 &5.34 &6.91 &1.0\\\midrule
82 & -1.496572258922 & -0.620483259367 & $10^{-12}$ &2.19 &3.76 &5.33 &6.90 &8.47 &1.0\\\midrule
83 & -1.496571777498 & 0.620482830425 & $10^{-12}$ &0.95 &2.52 &4.09 &5.66 &7.23 &1.0\\\midrule
84 & -1.496231071195 & 0.658339515148 & $10^{-12}$ &-0.44 &1.13 &2.70 &4.27 &5.84 &1.0\\\midrule
85 & -1.496230958626 & -0.658339417530 & $10^{-12}$ &-2.70 &-1.13 &0.44 &2.02 &3.59 &1.0\\\midrule
86 & -1.495795097477 & -0.701132402982 & $10^{-12}$ &-1.63 &-0.06 &1.51 &3.08 &4.65 &1.0\\\midrule
87 & -1.495795086062 & 0.701132393799 & $10^{-12}$ &-1.51 &0.06 &1.63 &3.20 &4.77 &1.0\\\midrule
88 & -1.495218118382 & 0.750399337671 & $10^{-12}$ &-0.60 &0.97 &2.54 &4.11 &5.68 &1.0\\\midrule
89 & -1.495218118079 & -0.750399337475 & $10^{-12}$ &-2.54 &-0.97 &0.60 &2.17 &3.74 &1.0\\\midrule
90 & -1.494419555125 & -0.808519758323 & $10^{-12}$ &-3.08 &-1.51 &0.06 &1.63 &3.21 &1.0\\\midrule
91 & -1.494419554811 & 0.808519758535 & $10^{-12}$ &-0.06 &1.51 &3.08 &4.65 &6.22 &1.0\\\midrule
92 & -1.493242561897 & 0.879467798895 & $10^{-12}$ &-3.09 &-1.52 &0.06 &1.63 &3.20 &1.0\\\midrule
93 & -1.493242557769 & -0.879467802695 & $10^{-12}$ &-0.06 &1.52 &3.09 &4.66 &6.23 &1.0\\\midrule
94 & -1.491336013025 & -0.970656104707 & $10^{-12}$ &-2.27 &-0.70 &0.87 &2.44 &4.01 &1.0\\\midrule
95 & -1.491335998112 & 0.970656121093 & $10^{-12}$ &-0.87 &0.70 &2.27 &3.84 &5.41 &1.0\\\midrule
96 & -1.487719917574 & 1.098563225209 & $10^{-12}$ &0.01 &1.59 &3.16 &4.73 &6.30 &1.0\\\midrule
97 & -1.487719897134 & -1.098563251194 & $10^{-12}$ &3.13 &4.70 &6.27 &7.84 &9.41 &1.0\\\midrule
98 & -1.478277809193 & -1.314410092648 & $10^{-12}$ &-2.59 &-1.02 &0.55 &2.12 &3.69 &1.0\\\midrule
99 & -1.478277799023 & 1.314410108052 & $10^{-12}$ &-0.55 &1.02 &2.59 &4.16 &5.73 &1.0\\\midrule
100 & -1.192055320267 & 0.000000000000 & $10^{-12}$ &1.57 &3.14 &4.71 &6.28 &7.85 &1.0\\\midrule
101 & -0.500000001273 & 2.182890275359 & $10^{-12}$ &-3.14 &-1.57 &-0.00 &1.57 &3.15 &1.0\\\midrule
102 & -0.499999998727 & -2.182890275359 & $10^{-12}$ &-0.00 &1.57 &3.14 &4.71 &6.28 &1.0\\\midrule
103 & 0.192055320267 & -0.000000000000 & $10^{-12}$ &-1.57 &-0.00 &1.57 &3.14 &4.71 &1.0\\\midrule
104 & 0.478277799023 & -1.314410108052 & $10^{-12}$ &2.59 &4.16 &5.73 &7.30 &8.87 &1.0\\\midrule
105 & 0.478277809193 & 1.314410092648 & $10^{-12}$ &0.55 &2.12 &3.69 &5.26 &6.83 &1.0\\\midrule
106 & 0.487719897134 & 1.098563251194 & $10^{-12}$ &-0.01 &1.56 &3.13 &4.70 &6.27 &1.0\\\midrule
107 & 0.487719917574 & -1.098563225209 & $10^{-12}$ &-3.13 &-1.56 &0.01 &1.59 &3.16 &1.0\\\midrule
108 & 0.491335998112 & -0.970656121093 & $10^{-12}$ &2.27 &3.84 &5.41 &6.98 &8.55 &1.0\\\midrule
109 & 0.491336013025 & 0.970656104707 & $10^{-12}$ &0.87 &2.44 &4.01 &5.58 &7.15 &1.0\\\midrule
110 & 0.493242557769 & 0.879467802695 & $10^{-12}$ &3.09 &4.66 &6.23 &7.80 &9.37 &1.0\\\midrule
111 & 0.493242561897 & -0.879467798895 & $10^{-12}$ &0.06 &1.63 &3.20 &4.77 &6.34 &1.0\\\midrule
112 & 0.494419554811 & -0.808519758535 & $10^{-12}$ &3.08 &4.65 &6.22 &7.79 &9.36 &1.0\\\midrule
113 & 0.494419555125 & 0.808519758323 & $10^{-12}$ &0.06 &1.63 &3.21 &4.78 &6.35 &1.0\\\midrule
114 & 0.495218118079 & 0.750399337475 & $10^{-12}$ &0.60 &2.17 &3.74 &5.31 &6.89 &1.0\\\midrule
115 & 0.495218118382 & -0.750399337671 & $10^{-12}$ &2.54 &4.11 &5.68 &7.25 &8.82 &1.0\\\midrule
116 & 0.495795086062 & -0.701132393799 & $10^{-12}$ &1.63 &3.20 &4.77 &6.34 &7.91 &1.0\\\midrule
117 & 0.495795097477 & 0.701132402982 & $10^{-12}$ &1.51 &3.08 &4.65 &6.23 &7.80 &1.0\\\midrule
118 & 0.496230958626 & 0.658339417530 & $10^{-12}$ &0.44 &2.02 &3.59 &5.16 &6.73 &1.0\\\midrule
119 & 0.496231071195 & -0.658339515148 & $10^{-12}$ &2.70 &4.27 &5.84 &7.41 &8.98 &1.0\\\midrule
120 & 0.496571777498 & -0.620482830425 & $10^{-12}$ &-2.19 &-0.62 &0.95 &2.52 &4.09 &1.0\\\midrule
121 & 0.496572258922 & 0.620483259367 & $10^{-12}$ &-0.95 &0.62 &2.19 &3.76 &5.33 &1.0\\\midrule
122 & 0.496843960123 & 0.586509770845 & $10^{-12}$ &-2.51 &-0.94 &0.63 &2.20 &3.77 &1.0\\\midrule
123 & 0.496845058431 & -0.586510751527 & $10^{-12}$ &-0.63 &0.94 &2.51 &4.09 &5.66 &1.0\\\midrule
124 & 0.497068864609 & -0.555672362108 & $10^{-12}$ &2.07 &3.64 &5.21 &6.78 &8.35 &1.0\\\midrule
125 & 0.497070337484 & 0.555673660500 & $10^{-12}$ &1.07 &2.64 &4.21 &5.78 &7.36 &1.0\\\midrule
126 & 0.497254083110 & 0.527411625020 & $10^{-12}$ &2.88 &4.46 &6.03 &7.60 &9.17 &1.0\\\midrule
127 & 0.497255294626 & -0.527412667506 & $10^{-12}$ &0.26 &1.83 &3.40 &4.97 &6.54 &1.0\\\midrule
128 & 0.497412793089 & -0.501309849137 & $10^{-12}$ &-1.65 &-0.08 &1.49 &3.06 &4.63 &1.0\\\midrule
129 & 0.497413408143 & 0.501310359159 & $10^{-12}$ &-1.49 &0.08 &1.65 &3.22 &4.79 &1.0\\\midrule
130 & 0.497547289313 & 0.477035535699 & $10^{-12}$ &0.50 &2.07 &3.64 &5.21 &6.78 &1.0\\\midrule
131 & 0.497547476813 & -0.477035682510 & $10^{-12}$ &2.64 &4.21 &5.79 &7.36 &8.93 &1.0\\\midrule
132 & 0.497663817867 & -0.454329812915 & $10^{-12}$ &2.56 &4.13 &5.70 &7.27 &8.84 &1.0\\\midrule
133 & 0.497663849835 & 0.454329835627 & $10^{-12}$ &0.59 &2.16 &3.73 &5.30 &6.87 &1.0\\\midrule
134 & 0.497765245778 & 0.432982082050 & $10^{-12}$ &-1.61 &-0.04 &1.53 &3.10 &4.68 &1.0\\\midrule
135 & 0.497765248390 & -0.432982083595 & $10^{-12}$ &-1.53 &0.04 &1.61 &3.18 &4.75 &1.0\\\midrule
136 & 0.497854383926 & -0.412820455322 & $10^{-12}$ &0.56 &2.14 &3.71 &5.28 &6.85 &1.0\\\midrule
137 & 0.497854383995 & 0.412820455349 & $10^{-12}$ &2.58 &4.15 &5.72 &7.29 &8.86 &1.0\\\midrule
138 & 0.497933293140 & -0.393702223808 & $10^{-12}$ &0.36 &1.93 &3.50 &5.07 &6.64 &1.0\\\midrule
139 & 0.497933293140 & 0.393702223808 & $10^{-12}$ &2.78 &4.35 &5.93 &7.50 &9.07 &1.0\\\midrule
140 & 0.498003623959 & -0.375507684154 & $10^{-12}$ &-1.24 &0.33 &1.90 &3.47 &5.05 &1.0\\\midrule
141 & 0.498003623959 & 0.375507684154 & $10^{-12}$ &-1.90 &-0.33 &1.24 &2.81 &4.38 &1.0\\\midrule
142 & 0.498066714188 & 0.358135397291 & $10^{-12}$ &1.06 &2.63 &4.20 &5.77 &7.34 &1.0\\\midrule
143 & 0.498066714188 & -0.358135397291 & $10^{-12}$ &2.08 &3.65 &5.22 &6.79 &8.36 &1.0\\\midrule
144 & 0.498123665480 & -0.341498614506 & $10^{-12}$ &-0.25 &1.32 &2.89 &4.46 &6.03 &1.0\\\midrule
145 & 0.498123665480 & 0.341498614506 & $10^{-12}$ &-2.89 &-1.32 &0.25 &1.82 &3.39 &1.0\\\midrule
146 & 0.498175399285 & -0.325522534071 & $10^{-12}$ &-2.61 &-1.04 &0.53 &2.10 &3.67 &1.0\\\midrule
147 & 0.498175399285 & 0.325522534071 & $10^{-12}$ &-0.53 &1.04 &2.61 &4.19 &5.76 &1.0\\\midrule
148 & 0.498222698566 & -0.310142165508 & $10^{-12}$ &1.86 &3.43 &5.00 &6.58 &8.15 &1.0\\\midrule
149 & 0.498222698566 & 0.310142165508 & $10^{-12}$ &1.28 &2.85 &4.42 &5.99 &7.56 &1.0\\\midrule
150 & 0.498266239594 & 0.295300650461 & $10^{-12}$ &-2.01 &-0.43 &1.14 &2.71 &4.28 &1.0\\\midrule
151 & 0.498266239594 & -0.295300650461 & $10^{-12}$ &-1.14 &0.43 &2.01 &3.58 &5.15 &1.0\\\midrule
152 & 0.498306617468 & 0.280947921906 & $10^{-12}$ &2.71 &4.28 &5.85 &7.42 &8.99 &1.0\\\midrule
153 & 0.498306617468 & -0.280947921906 & $10^{-12}$ &0.43 &2.00 &3.57 &5.14 &6.72 &1.0\\\midrule
154 & 0.498344367511 & 0.267039629271 & $10^{-12}$ &0.25 &1.82 &3.39 &4.96 &6.53 &1.0\\\midrule
155 & 0.498344367511 & -0.267039629271 & $10^{-12}$ &2.89 &4.46 &6.03 &7.60 &9.17 &1.0\\\midrule
156 & 0.498379979920 & 0.253536262645 & $10^{-12}$ &-2.23 &-0.66 &0.92 &2.49 &4.06 &1.0\\\midrule
157 & 0.498379979920 & -0.253536262645 & $10^{-12}$ &-0.92 &0.66 &2.23 &3.80 &5.37 &1.0\\\midrule
158 & 0.498413915387 & -0.240402437913 & $10^{-12}$ &1.56 &3.13 &4.70 &6.27 &7.85 &1.0\\\midrule
159 & 0.498413915387 & 0.240402437913 & $10^{-12}$ &1.58 &3.15 &4.72 &6.29 &7.86 &1.0\\\midrule
160 & 0.498446620333 & -0.227606305494 & $10^{-12}$ &-2.19 &-0.62 &0.95 &2.52 &4.09 &1.0\\\midrule
161 & 0.498446620333 & 0.227606305494 & $10^{-12}$ &-0.95 &0.62 &2.19 &3.76 &5.33 &1.0\\\midrule
162 & 0.498478538264 & -0.215119055622 & $10^{-12}$ &2.81 &4.38 &5.95 &7.52 &9.09 &1.0\\\midrule
163 & 0.498478538264 & 0.215119055622 & $10^{-12}$ &0.33 &1.90 &3.48 &5.05 &6.62 &1.0\\\midrule
164 & 0.498510123234 & -0.202914506180 & $10^{-12}$ &2.87 &4.44 &6.01 &7.59 &9.16 &1.0\\\midrule
165 & 0.498510123234 & 0.202914506180 & $10^{-12}$ &0.27 &1.84 &3.41 &4.98 &6.55 &1.0\\\midrule
166 & 0.498541855015 & -0.190968750517 & $10^{-12}$ &-2.28 &-0.71 &0.86 &2.43 &4.00 &1.0\\\midrule
167 & 0.498541855015 & 0.190968750517 & $10^{-12}$ &-0.86 &0.71 &2.28 &3.86 &5.43 &1.0\\\midrule
168 & 0.498574252042 & -0.179259858798 & $10^{-12}$ &1.71 &3.28 &4.85 &6.42 &7.99 &1.0\\\midrule
169 & 0.498574252042 & 0.179259858798 & $10^{-12}$ &1.43 &3.00 &4.57 &6.15 &7.72 &1.0\\\midrule
170 & 0.498607890702 & -0.167767617858 & $10^{-12}$ &-1.14 &0.43 &2.00 &3.57 &5.14 &1.0\\\midrule
171 & 0.498607890702 & 0.167767617858 & $10^{-12}$ &-2.00 &-0.43 &1.14 &2.71 &4.28 &1.0\\\midrule
172 & 0.498643427402 & -0.156473301820 & $10^{-12}$ &2.55 &4.12 &5.69 &7.27 &8.84 &1.0\\\midrule
173 & 0.498643427402 & 0.156473301820 & $10^{-12}$ &0.59 &2.16 &3.73 &5.30 &6.87 &1.0\\\midrule
174 & 0.498681620170 & -0.145359467945 & $10^{-12}$ &-0.04 &1.53 &3.10 &4.67 &6.24 &1.0\\\midrule
175 & 0.498681620170 & 0.145359467945 & $10^{-12}$ &-3.10 &-1.53 &0.04 &1.62 &3.19 &1.0\\\midrule
176 & 0.498723364861 & 0.134409764676 & $10^{-12}$ &-0.49 &1.08 &2.65 &4.22 &5.79 &1.0\\\midrule
177 & 0.498723364861 & -0.134409764676 & $10^{-12}$ &-2.65 &-1.08 &0.49 &2.06 &3.63 &1.0\\\midrule
178 & 0.498769723486 & 0.123608744783 & $10^{-12}$ &2.13 &3.70 &5.27 &6.84 &8.41 &1.0\\\midrule
179 & 0.498769723486 & -0.123608744783 & $10^{-12}$ &1.01 &2.59 &4.16 &5.73 &7.30 &1.0\\\midrule
180 & 0.498821967495 & 0.112941674848 & $10^{-12}$ &-1.53 &0.04 &1.61 &3.18 &4.75 &1.0\\\midrule
181 & 0.498821967495 & -0.112941674848 & $10^{-12}$ &-1.61 &-0.04 &1.53 &3.10 &4.67 &1.0\\\midrule
182 & 0.498881624162 & 0.102394318961 & $10^{-12}$ &2.03 &3.61 &5.18 &6.75 &8.32 &1.0\\\midrule
183 & 0.498881624162 & -0.102394318961 & $10^{-12}$ &1.11 &2.68 &4.25 &5.82 &7.39 &1.0\\\midrule
184 & 0.498950502407 & 0.091952693540 & $10^{-12}$ &-0.61 &0.96 &2.53 &4.10 &5.67 &1.0\\\midrule
185 & 0.498950502407 & -0.091952693540 & $10^{-12}$ &-2.53 &-0.96 &0.61 &2.18 &3.75 &1.0\\\midrule
186 & 0.499030724545 & 0.081602732169 & $10^{-12}$ &3.02 &4.59 &6.16 &7.73 &9.30 &1.0\\\midrule
187 & 0.499030724545 & -0.081602732169 & $10^{-12}$ &0.13 &1.70 &3.27 &4.84 &6.41 &1.0\\\midrule
188 & 0.499124661078 & 0.071329849178 & $10^{-12}$ &0.35 &1.92 &3.49 &5.06 &6.63 &1.0\\\midrule
189 & 0.499124661078 & -0.071329849178 & $10^{-12}$ &2.79 &4.36 &5.93 &7.51 &9.08 &1.0\\\midrule
190 & 0.499234729326 & 0.061118328765 & $10^{-12}$ &-2.33 &-0.76 &0.81 &2.38 &3.95 &1.0\\\midrule
191 & 0.499234729326 & -0.061118328765 & $10^{-12}$ &-0.81 &0.76 &2.33 &3.90 &5.47 &1.0\\\midrule
192 & 0.499362827910 & 0.050950483670 & $10^{-12}$ &1.88 &3.45 &5.03 &6.60 &8.17 &1.0\\\midrule
193 & 0.499362827910 & -0.050950483670 & $10^{-12}$ &1.26 &2.83 &4.40 &5.97 &7.54 &1.0\\\midrule
194 & 0.499508982879 & 0.040805650649 & $10^{-12}$ &-1.69 &-0.12 &1.45 &3.03 &4.60 &1.0\\\midrule
195 & 0.499508982879 & -0.040805650649 & $10^{-12}$ &-1.45 &0.12 &1.69 &3.26 &4.83 &1.0\\\midrule
196 & 0.499668437249 & 0.030659439714 & $10^{-12}$ &1.05 &2.62 &4.19 &5.76 &7.33 &1.0\\\midrule
197 & 0.499668437249 & -0.030659439714 & $10^{-12}$ &2.09 &3.66 &5.24 &6.81 &8.38 &1.0\\\midrule
198 & 0.499826137427 & 0.020485525878 & $10^{-12}$ &-2.47 &-0.90 &0.67 &2.24 &3.81 &1.0\\\midrule
199 & 0.499826137427 & -0.020485525878 & $10^{-12}$ &-0.67 &0.90 &2.47 &4.04 &5.61 &1.0\\\midrule
200 & 0.499950624342 & 0.010265201348 & $10^{-12}$ &0.32 &1.89 &3.46 &5.04 &6.61 &1.0\\\midrule
201 & 0.499950624342 & -0.010265201348 & $10^{-12}$ &2.82 &4.39 &5.96 &7.53 &9.10 &1.0\\\midrule
202 & 0.537797484405 & -0.000000000000 & $10^{-12}$ &1.57 &3.14 &4.71 &6.28 &7.85 &1.0\\\bottomrule
\label{tab:tablec}
\end{longtable}

\section{Complete Numerical Data of Bethe roots}
\label{app:D}
\setcounter{equation}{0}
\setcounter{table}{0}
\renewcommand{\theequation}{D.\arabic{equation}}
\renewcommand{\thetable}{D.\arabic{table}}
This appendix provides the numerical data of the Bethe roots shown in Fig.~\ref{fig:III}.
To verify the solutions, we evaluate
\begin{equation}
G(\lambda_j)=-\frac{A(\lambda_j)+C(\lambda_j)}{B(\lambda_j)},
\end{equation}
for all roots $\lambda_j$.
The Bethe Ansatz equations are satisfied when the residual condition
\begin{equation}
\varepsilon_j=\left|G(\lambda_j)-1\right|=0
\end{equation}
holds.
\begin{longtable}{cccc}
\caption{Complete numerical data of the Bethe roots shown in Fig.~\ref{fig:III}, including their real and imaginary parts and the residual $\varepsilon_j$.}\\
\toprule
Index & Real($\lambda_j$) & Imag($\lambda_j$) & $\varepsilon_j$  \\
\midrule
1 & -31.1446156423028775 & -14.7965725453649206 & $9.116 \times 10^{-14}$\\\midrule
2 & -31.1446156423026785 & 14.7965725453651107 & $7.890 \times 10^{-14}$\\\midrule
3 & -31.1121965369405586 & -18.7766800279121107 & $7.915 \times 10^{-14}$\\\midrule
4 & -31.1121965369402744 & 18.7766800279122421 & $9.223 \times 10^{-14}$\\\midrule
5 & -30.9198209099740140 & -11.2191369792878319 & $2.310 \times 10^{-14}$\\\midrule
6 & -30.9198209099738719 & 11.2191369792880487 & $3.635 \times 10^{-14}$\\\midrule
7 & -30.7502266206089132 & -23.2575792941639818 & $1.953 \times 10^{-14}$\\\midrule
8 & -30.7502266206086858 & 23.2575792941640493 & $2.579 \times 10^{-14}$\\\midrule
9 & -30.4881990800943505 & -7.9770706560502527 & $6.667 \times 10^{-14}$\\\midrule
10 & -30.4881990800942688 & 7.9770706560504454 & $1.187 \times 10^{-13}$\\\midrule
11 & -29.9481021517954424 & -28.3924428210070445 & $9.825 \times 10^{-15}$\\\midrule
12 & -29.9481021517952009 & 28.3924428210070765 & $3.981 \times 10^{-14}$\\\midrule
13 & -29.8864553286501220 & -5.0218487293887204 & $1.073 \times 10^{-13}$\\\midrule
14 & -29.8864553286500083 & 5.0218487293889007 & $6.955 \times 10^{-14}$\\\midrule
15 & -29.1398296141751274 & -2.3241303778903890 & $4.638 \times 10^{-14}$\\\midrule
16 & -29.1398296141750528 & 2.3241303778905946 & $6.465 \times 10^{-14}$\\\midrule
17 & -28.5545378345147896 & 0.0000000000001053 & $2.328 \times 10^{-14}$\\\midrule
18 & -28.5207150810533037 & -34.4455727791949187 & $1.048 \times 10^{-14}$\\\midrule
19 & -28.5207150810531012 & 34.4455727791949400 & $1.403 \times 10^{-14}$\\\midrule
20 & -26.1107497590830917 & -41.9531362701105692 & $3.181 \times 10^{-15}$\\\midrule
21 & -26.1107497590828821 & 41.9531362701105124 & $1.666 \times 10^{-14}$\\\midrule
22 & -26.0484347809020846 & 0.0000000000001638 & $3.393 \times 10^{-15}$\\\midrule
23 & -22.4470123572345699 & 0.0000000000001460 & $6.397 \times 10^{-15}$\\\midrule
24 & -21.8138177669619893 & 52.4083065115725049 & $3.707 \times 10^{-14}$\\\midrule
25 & -21.8138177669621207 & -52.4083065115724906 & $8.752 \times 10^{-15}$\\\midrule
26 & -18.8428782899300096 & 0.0000000000001282 & $5.174 \times 10^{-15}$\\\midrule
27 & -15.2300136488844622 & 0.0000000000001073 & $6.089 \times 10^{-15}$\\\midrule
28 & -11.6000778134918558 & 0.0000000000000842 & $4.699 \times 10^{-15}$\\\midrule
29 & -7.9291439513302979 & 0.0000000000000570 & $6.077 \times 10^{-15}$\\\midrule
30 & -4.1104471023195828 & 0.0000000000000258 & $1.154 \times 10^{-14}$\\\midrule
31 & -0.5000000000000002 & 0.4897941556864582 & $3.708 \times 10^{-14}$\\\midrule
32 & -0.5000000000000218 & -2.2575092853124419 & $3.133 \times 10^{-14}$\\\midrule
33 & -0.5000000000000002 & 0.5328668128698998 & $4.565 \times 10^{-14}$\\\midrule
34 & -0.5000000000000000 & 0.3852049749550474 & $1.287 \times 10^{-14}$\\\midrule
35 & -0.4999999999999999 & -0.4514616060666452 & $2.768 \times 10^{-14}$\\\midrule
36 & -0.5000000000000000 & 0.4168436592608959 & $8.428 \times 10^{-15}$\\\midrule
37 & -0.5000000000000001 & -1.1297422089002436 & $1.105 \times 10^{-14}$\\\midrule
38 & -0.4999999999999998 & -0.9240971928172551 & $1.586 \times 10^{-14}$\\\midrule
39 & -0.4999999999999999 & -0.7997464185400411 & $5.480 \times 10^{-15}$\\\midrule
40 & -0.4999999999999999 & -0.6400088998757980 & $3.118 \times 10^{-14}$\\\midrule
41 & -0.5000000000000001 & -0.5821725344845270 & $3.113 \times 10^{-14}$\\\midrule
42 & -0.5000000000000002 & -0.7101869900295184 & $3.507 \times 10^{-14}$\\\midrule
43 & -0.5000000000000000 & -0.3559998288879291 & $8.967 \times 10^{-15}$\\\midrule
44 & -0.5000000000000000 & -0.3288115991246061 & $2.101 \times 10^{-14}$\\\midrule
45 & -0.5000000000000000 & -0.3033145579798743 & $3.449 \times 10^{-15}$\\\midrule
46 & -0.5000000000000000 & -0.2792487723783211 & $1.107 \times 10^{-14}$\\\midrule
47 & -0.5000000000000000 & -0.2564030554296224 & $9.771 \times 10^{-15}$\\\midrule
48 & -0.4999999999999999 & -0.2346030502916976 & $1.529 \times 10^{-14}$\\\midrule
49 & -0.5000000000000000 & -0.2137026909316591 & $1.963 \times 10^{-15}$\\\midrule
50 & -0.4999999999999999 & -0.1935779461780248 & $1.557 \times 10^{-14}$\\\midrule
51 & -0.5000000000000000 & -0.1741221444611151 & $8.675 \times 10^{-15}$\\\midrule
52 & -0.5000000000000000 & -0.1552424154212793 & $1.391 \times 10^{-14}$\\\midrule
53 & -0.5000000000000000 & -0.1368569346675031 & $1.305 \times 10^{-15}$\\\midrule
54 & -0.5000000000000000 & -0.1188927547550130 & $1.259 \times 10^{-14}$\\\midrule
55 & -0.5000000000000000 & -0.1012840692651649 & $1.452 \times 10^{-15}$\\\midrule
56 & -0.5000000000000000 & -0.0839707997688026 & $5.486 \times 10^{-15}$\\\midrule
57 & -0.4999999999999999 & -0.0668974247660786 & $1.517 \times 10^{-14}$\\\midrule
58 & -0.4999999999999999 & -0.0500119899825411 & $1.798 \times 10^{-14}$\\\midrule
59 & -0.5000000000000000 & -0.0332652535569321 & $1.773 \times 10^{-14}$\\\midrule
60 & -0.5000000000000000 & -0.0166099295451636 & $1.627 \times 10^{-14}$\\\midrule
61 & -0.5000000000000000 & 0.0166099295451636 & $1.627 \times 10^{-14}$\\\midrule
62 & -0.5000000000000000 & 0.0332652535569321 & $1.773 \times 10^{-14}$\\\midrule
63 & -0.5000000000000000 & 0.0500119899825411 & $1.798 \times 10^{-14}$\\\midrule
64 & -0.5000000000000000 & 0.0668974247660786 & $1.517 \times 10^{-14}$\\\midrule
65 & -0.5000000000000000 & 0.0839707997688026 & $5.486 \times 10^{-15}$\\\midrule
66 & -0.5000000000000000 & 0.1012840692651649 & $1.452 \times 10^{-15}$\\\midrule
67 & -0.5000000000000000 & 0.1188927547550130 & $1.259 \times 10^{-14}$\\\midrule
68 & -0.5000000000000000 & 0.1368569346675031 & $1.305 \times 10^{-15}$\\\midrule
69 & -0.5000000000000000 & 0.1552424154212793 & $1.391 \times 10^{-14}$\\\midrule
70 & -0.5000000000000000 & 0.1741221444611151 & $8.675 \times 10^{-15}$\\\midrule
71 & -0.5000000000000000 & 0.1935779461780248 & $1.557 \times 10^{-14}$\\\midrule
72 & -0.5000000000000000 & 0.2137026909316591 & $1.963 \times 10^{-15}$\\\midrule
73 & -0.5000000000000000 & 0.2346030502916976 & $1.529 \times 10^{-14}$\\\midrule
74 & -0.5000000000000000 & 0.2564030554296224 & $9.771 \times 10^{-15}$\\\midrule
75 & -0.5000000000000000 & 0.2792487723783211 & $1.107 \times 10^{-14}$\\\midrule
76 & -0.5000000000000000 & 0.3033145579798743 & $3.449 \times 10^{-15}$\\\midrule
77 & -0.5000000000000000 & 0.3288115991246061 & $2.101 \times 10^{-14}$\\\midrule
78 & -0.5000000000000000 & 0.3559998288879291 & $8.967 \times 10^{-15}$\\\midrule
79 & -0.4999999999999998 & 0.7101869900295184 & $3.507 \times 10^{-14}$\\\midrule
80 & -0.4999999999999999 & 0.5821725344845270 & $3.113 \times 10^{-14}$\\\midrule
81 & -0.5000000000000000 & 0.6400088998757980 & $3.118 \times 10^{-14}$\\\midrule
82 & -0.5000000000000001 & 0.7997464185400411 & $5.480 \times 10^{-15}$\\\midrule
83 & -0.5000000000000002 & 0.9240971928172551 & $1.586 \times 10^{-14}$\\\midrule
84 & -0.4999999999999999 & 1.1297422089002436 & $1.105 \times 10^{-14}$\\\midrule
85 & -0.5000000000000000 & -0.4168436592608959 & $8.428 \times 10^{-15}$\\\midrule
86 & -0.5000000000000001 & 0.4514616060666452 & $2.768 \times 10^{-14}$\\\midrule
87 & -0.5000000000000000 & -0.3852049749550474 & $1.287 \times 10^{-14}$\\\midrule
88 & -0.4999999999999998 & -0.5328668128698998 & $4.565 \times 10^{-14}$\\\midrule
89 & -0.4999999999999782 & 2.2575092853124419 & $3.133 \times 10^{-14}$\\\midrule
90 & -0.4999999999999998 & -0.4897941556864582 & $3.708 \times 10^{-14}$\\\midrule
91 & 3.1104471023195828 & -0.0000000000000258 & $1.154 \times 10^{-14}$\\\midrule
92 & 6.9291439513302979 & -0.0000000000000570 & $6.077 \times 10^{-15}$\\\midrule
93 & 10.6000778134918558 & -0.0000000000000842 & $4.699 \times 10^{-15}$\\\midrule
94 & 14.2300136488844622 & -0.0000000000001073 & $6.089 \times 10^{-15}$\\\midrule
95 & 17.8428782899300096 & -0.0000000000001282 & $5.174 \times 10^{-15}$\\\midrule
96 & 20.8138177669621207 & 52.4083065115724906 & $8.752 \times 10^{-15}$\\\midrule
97 & 20.8138177669619893 & -52.4083065115725049 & $3.707 \times 10^{-14}$\\\midrule
98 & 21.4470123572345699 & -0.0000000000001460 & $6.397 \times 10^{-15}$\\\midrule
99 & 25.0484347809020846 & -0.0000000000001638 & $3.393 \times 10^{-15}$\\\midrule
100 & 25.1107497590828821 & -41.9531362701105124 & $1.666 \times 10^{-14}$\\\midrule
101 & 25.1107497590830917 & 41.9531362701105692 & $3.181 \times 10^{-15}$\\\midrule
102 & 27.5207150810531012 & -34.4455727791949400 & $1.403 \times 10^{-14}$\\\midrule
103 & 27.5207150810533037 & 34.4455727791949187 & $1.048 \times 10^{-14}$\\\midrule
104 & 27.5545378345147896 & -0.0000000000001053 & $2.328 \times 10^{-14}$\\\midrule
105 & 28.1398296141750528 & -2.3241303778905946 & $6.465 \times 10^{-14}$\\\midrule
106 & 28.1398296141751274 & 2.3241303778903890 & $4.638 \times 10^{-14}$\\\midrule
107 & 28.8864553286500083 & -5.0218487293889007 & $6.955 \times 10^{-14}$\\\midrule
108 & 28.8864553286501220 & 5.0218487293887204 & $1.073 \times 10^{-13}$\\\midrule
109 & 28.9481021517952009 & -28.3924428210070765 & $3.981 \times 10^{-14}$\\\midrule
110 & 28.9481021517954424 & 28.3924428210070445 & $9.825 \times 10^{-15}$\\\midrule
111 & 29.4881990800942688 & -7.9770706560504454 & $1.187 \times 10^{-13}$\\\midrule
112 & 29.4881990800943505 & 7.9770706560502527 & $6.667 \times 10^{-14}$\\\midrule
113 & 29.7502266206086858 & -23.2575792941640493 & $2.579 \times 10^{-14}$\\\midrule
114 & 29.7502266206089132 & 23.2575792941639818 & $1.953 \times 10^{-14}$\\\midrule
115 & 29.9198209099738719 & -11.2191369792880487 & $3.635 \times 10^{-14}$\\\midrule
116 & 29.9198209099740140 & 11.2191369792878319 & $2.310 \times 10^{-14}$\\\midrule
117 & 30.1121965369402744 & -18.7766800279122421 & $9.223 \times 10^{-14}$\\\midrule
118 & 30.1121965369405586 & 18.7766800279121107 & $7.915 \times 10^{-14}$\\\midrule
119 & 30.1446156423026785 & -14.7965725453651107 & $7.890 \times 10^{-14}$\\\midrule
120 & 30.1446156423028775 & 14.7965725453649206 & $9.116 \times 10^{-14}$\\\bottomrule
\label{tab:tabled}
\end{longtable}

\section{Crossover Behavior of the Paired-Line Roots}
\label{app:E}
\setcounter{figure}{0}
\renewcommand{\thefigure}{E.\arabic{figure}}
This appendix provides a detailed account of how the Bethe-root structure evolves as the number of paired-line root pairs increases for $p>0$. To illustrate the emergence and development of paired-line roots, this section presents their evolution from a single pair to three.
\begin{figure}[htbp]
  \centering
  \subfloat[Global configuration of Bethe roots]{
    \includegraphics[width=0.47\textwidth]{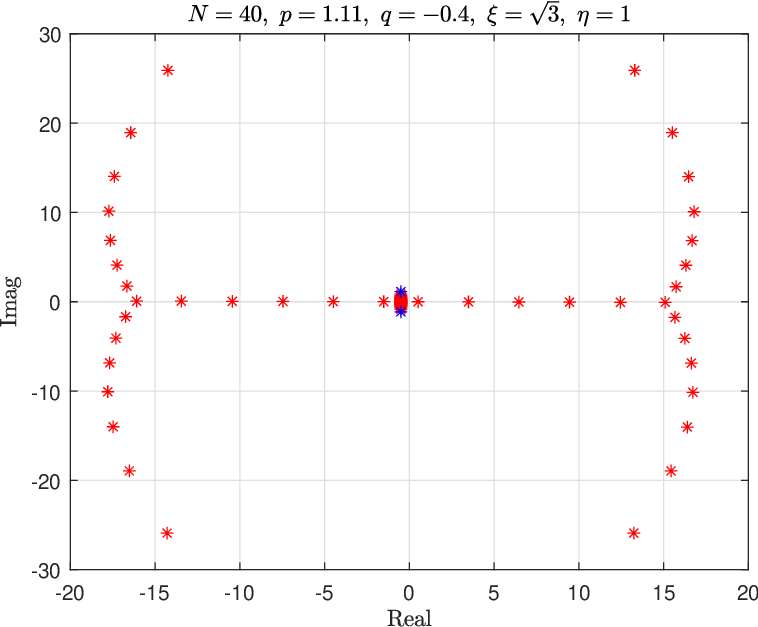}
  }
  \subfloat[Enlarged view of the central region]{
    \includegraphics[width=0.47\textwidth]{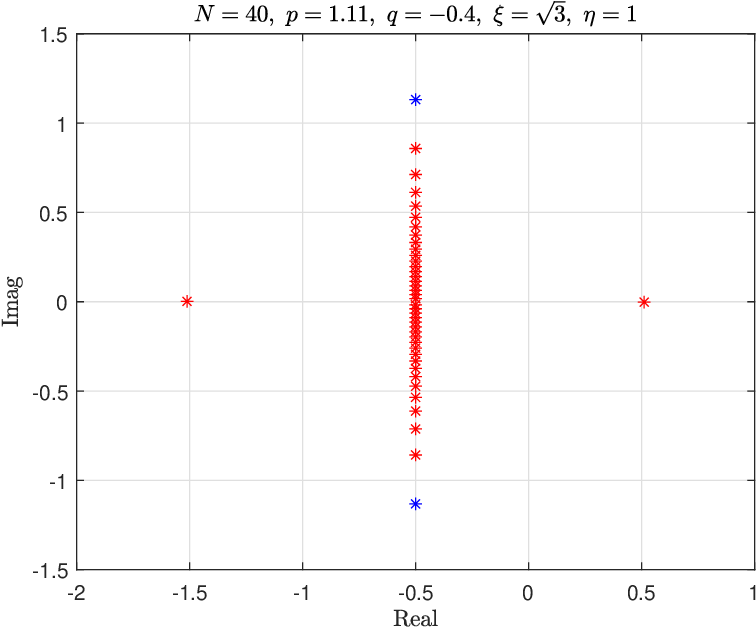}
  }
\caption{
Distribution of Bethe roots in the complex plane with parameters $N = 40$, $p = 1.11$, $q = -0.4$, $\xi = \sqrt3$, and $\eta = 1$. The blue points correspond to the paired-line roots.
}
\label{fig:cross1}
\end{figure}

\begin{figure}[htbp]
  \centering
  \subfloat[Global configuration of Bethe roots]{
    \includegraphics[width=0.47\textwidth]{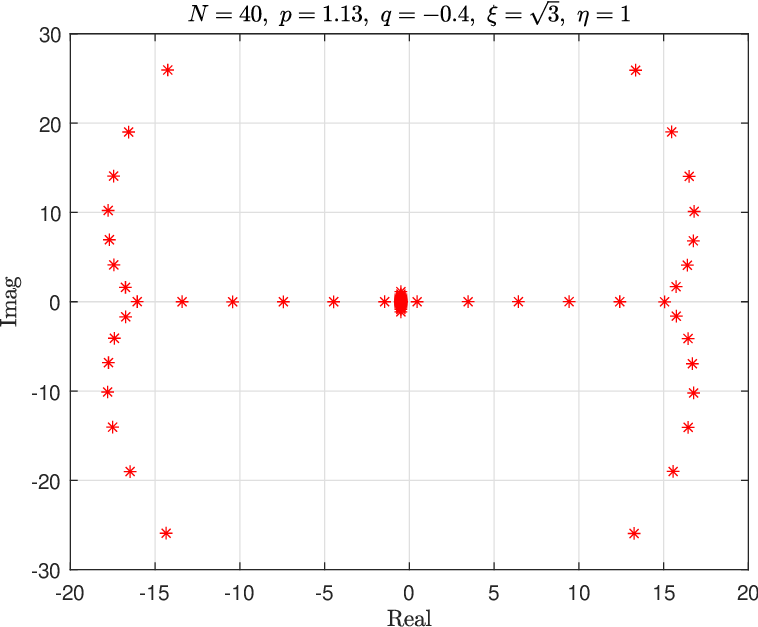}
  }
  \subfloat[Enlarged view of the central region]{
    \includegraphics[width=0.47\textwidth]{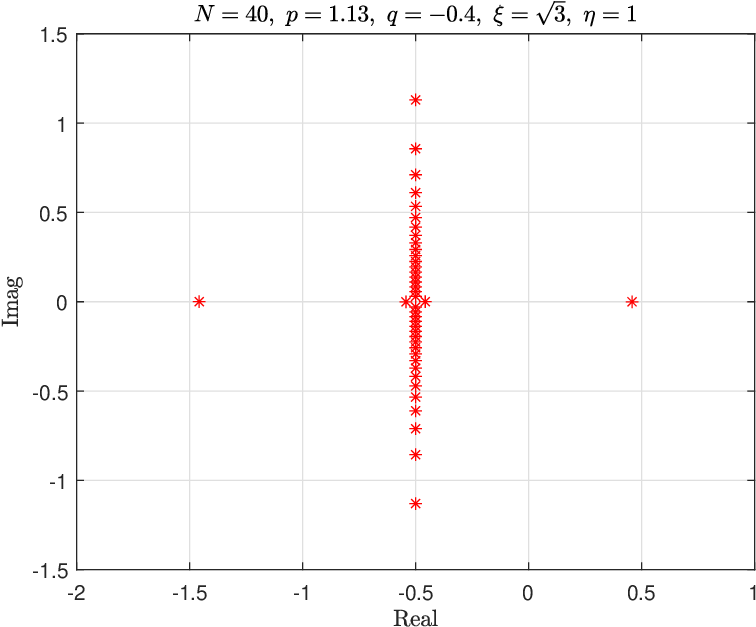}
  }
\caption{
Distribution of Bethe roots in the complex plane with parameters $N = 40$, $p = 1.13$, $q = -0.4$, $\xi = \sqrt3$, and $\eta = 1$.
}
\label{fig:cross2}
\end{figure}

First, as the innermost central roots approach the real axis, a pair of new line roots is generated (Figs.~\ref{fig:cross1}--\ref{fig:cross2}).

\begin{figure}[htbp]
  \centering
  \subfloat[Global configuration of Bethe roots]{
    \includegraphics[width=0.47\textwidth]{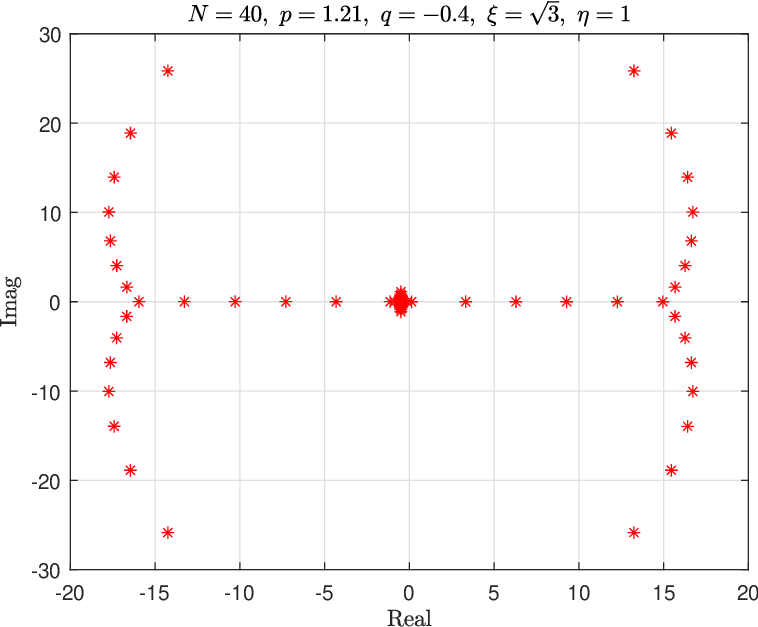}
  }
  \subfloat[Enlarged view of the central region]{
    \includegraphics[width=0.47\textwidth]{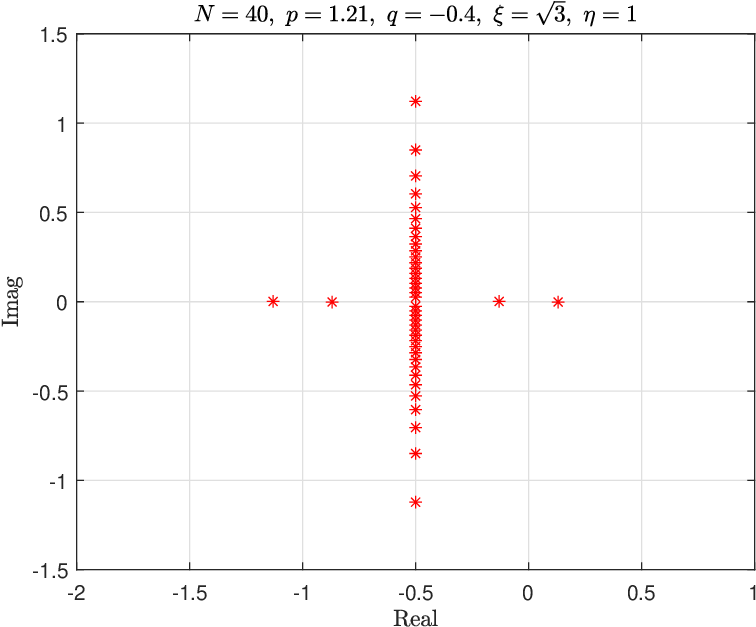}
  }
\caption{
Distribution of Bethe roots in the complex plane with parameters $N = 40$, $p = 1.21$, $q = -0.4$, $\xi = \sqrt3$, and $\eta = 1$.
}
\label{fig:cross3}
\end{figure}

Subsequently, the two new line roots move outward along the real axis, while the outer ones move inward (Fig.~\ref{fig:cross3}).

\begin{figure}[htbp]
  \centering
  \subfloat[Global configuration of Bethe roots]{
    \includegraphics[width=0.47\textwidth]{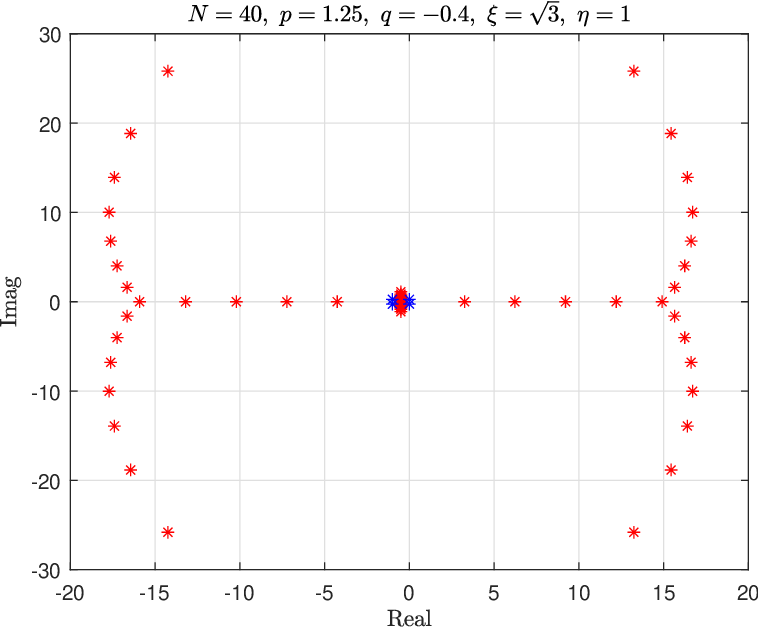}
  }
  \subfloat[Enlarged view of the central region]{
    \includegraphics[width=0.47\textwidth]{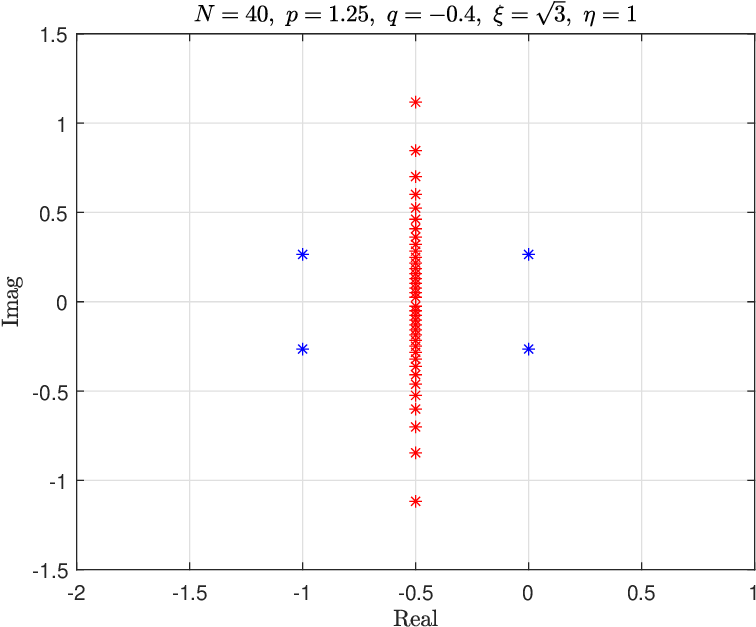}
  }
\caption{
Distribution of Bethe roots in the complex plane with parameters $N = 40$, $p = 1.23$, $q = -0.4$, $\xi = \sqrt3$, and $\eta = 1$. The blue points correspond to the paired-line roots.
}
\label{fig:cross4}
\end{figure}

When the two line roots meet, they coalesce into a paired-line root pair (Fig.~\ref{fig:cross4})

\begin{figure}[htbp]
  \centering
  \subfloat[Global configuration of Bethe roots]{
    \includegraphics[width=0.47\textwidth]{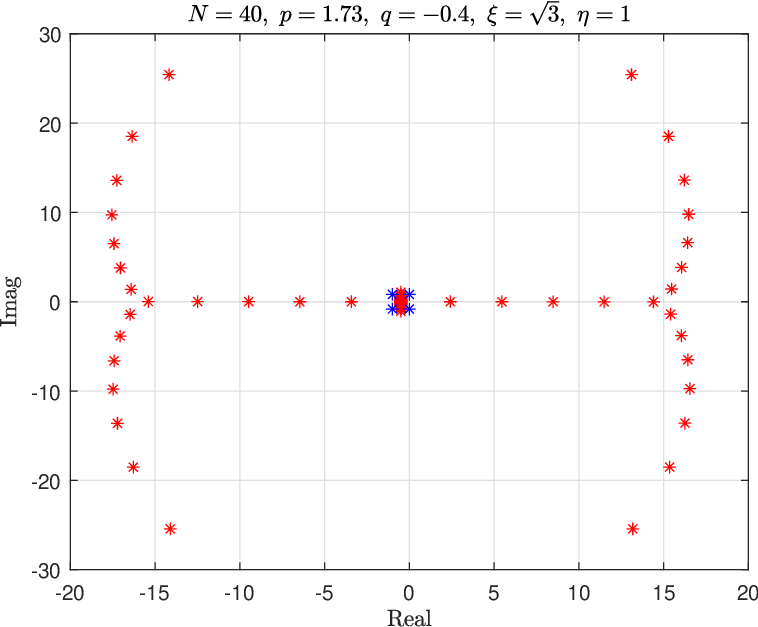}
  }
  \subfloat[Enlarged view of the central region]{
    \includegraphics[width=0.47\textwidth]{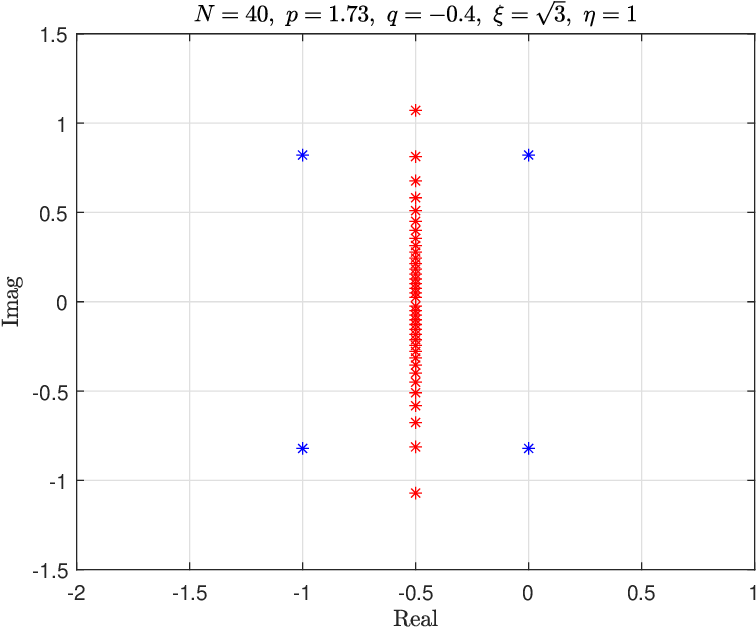}
  }
\caption{
Distribution of Bethe roots in the complex plane with parameters $N = 40$, $p = 1.73$, $q = -0.4$, $\xi = \sqrt3$, and $\eta = 1$. The blue points correspond to the paired-line roots.
}
\label{fig:cross5}
\end{figure}

As evolution proceeds, the two points of the pair continue to separate (Fig.~\ref{fig:cross5}).

\begin{figure}[htbp]
  \centering
  \subfloat[Global configuration of Bethe roots]{
    \includegraphics[width=0.47\textwidth]{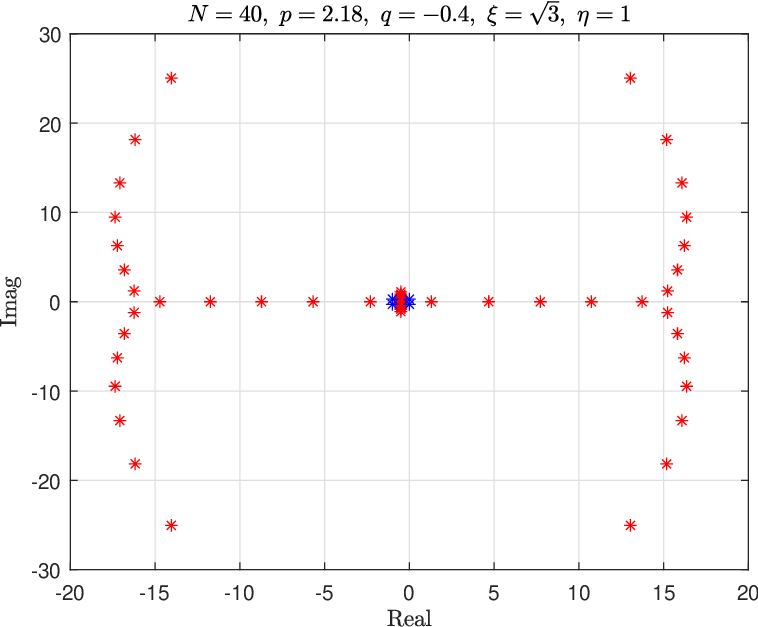}
  }
  \subfloat[Enlarged view of the central region]{
    \includegraphics[width=0.47\textwidth]{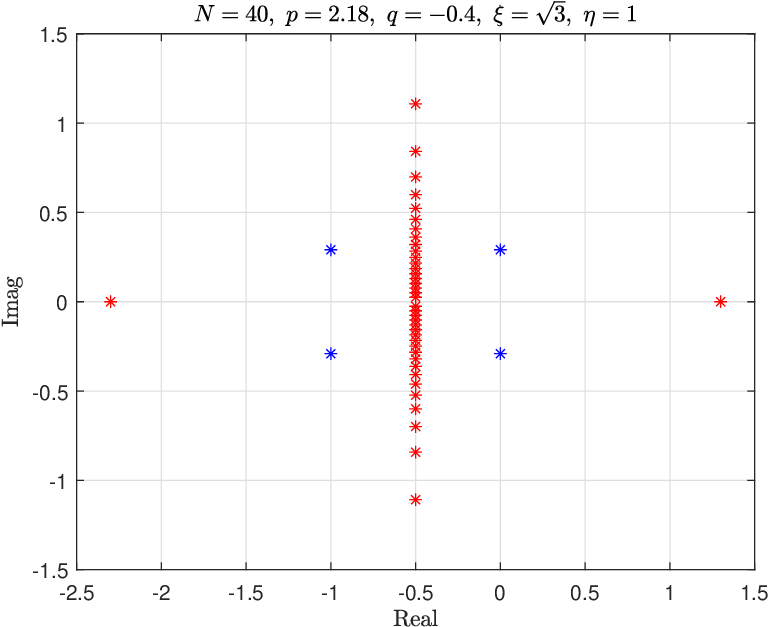}
  }
\caption{
Distribution of Bethe roots in the complex plane with parameters $N = 40$, $p = 2.18$, $q = -0.4$, $\xi = \sqrt3$, and $\eta = 1$. The blue points correspond to the paired-line roots.
}
\label{fig:cross6}
\end{figure}

However, as the system approaches the next crossover region, the separation between the two points of the paired-line roots contract again (Fig.~\ref{fig:cross6}).

\begin{figure}[htbp]
  \centering
  \subfloat[Global configuration of Bethe roots]{
    \includegraphics[width=0.47\textwidth]{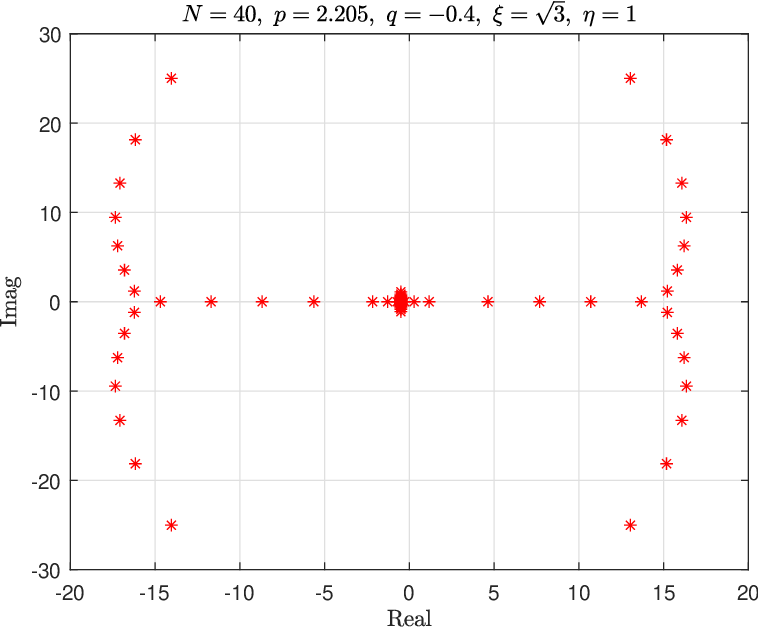}
  }
  \subfloat[Enlarged view of the central region]{
    \includegraphics[width=0.47\textwidth]{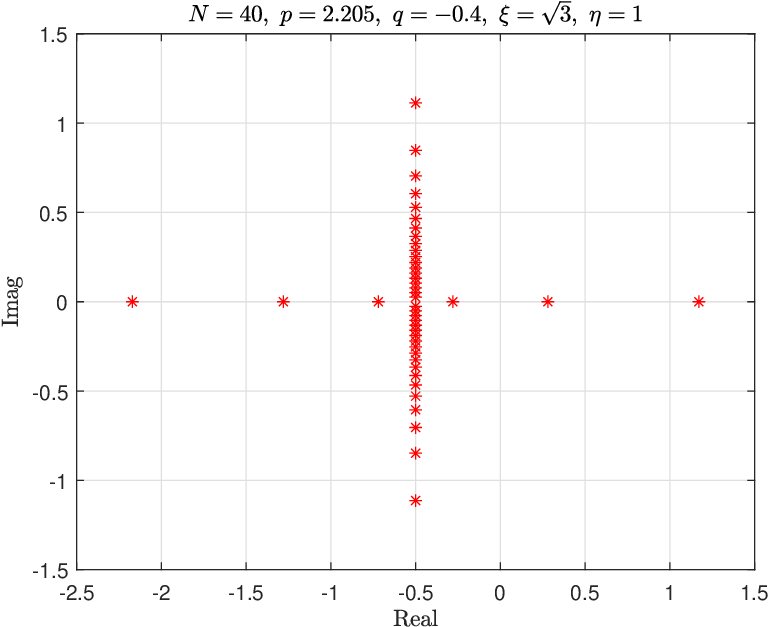}
  }
\caption{
Distribution of Bethe roots in the complex plane with parameters $N = 40$, $p = 2.2$, $q = -0.4$, $\xi = \sqrt3$, and $\eta = 1$.
}
\label{fig:cross7}
\end{figure}

After gradually approaching each other, the two points of the paired-line roots merge and transform back into line roots (Fig.~\ref{fig:cross7}).

\begin{figure}[htbp]
  \centering
  \subfloat[Global configuration of Bethe roots]{
    \includegraphics[width=0.47\textwidth]{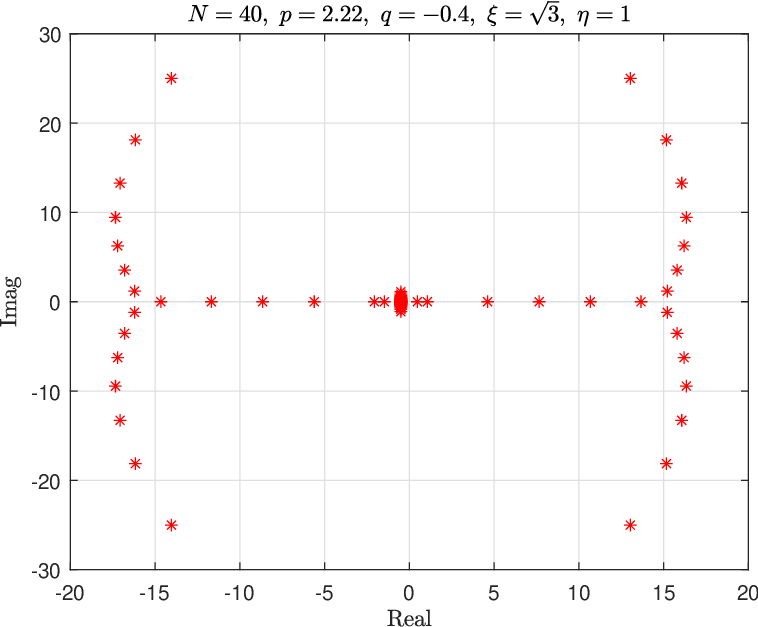}
  }
  \subfloat[Enlarged view of the central region]{
    \includegraphics[width=0.47\textwidth]{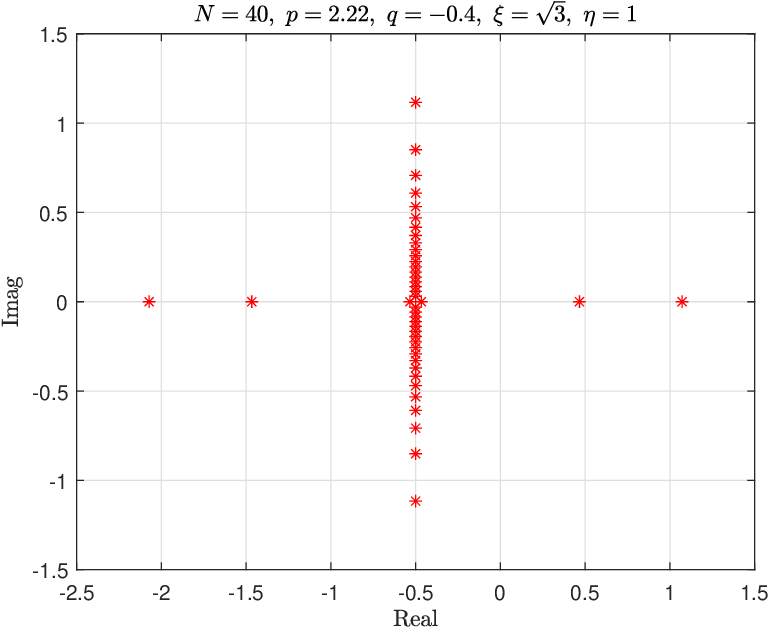}
  }
\caption{
Distribution of Bethe roots in the complex plane with parameters $N = 40$, $p = 2.22$, $q = -0.4$, $\xi = \sqrt3$, and $\eta = 1$.
}
\label{fig:cross8}
\end{figure}

In contrast to their initial formation, the two separated points of the line roots move toward each other and eventually meet the adjacent line roots (Fig.~\ref{fig:cross8}).

\begin{figure}[htbp]
  \centering
  \subfloat[Global configuration of Bethe roots]{
    \includegraphics[width=0.47\textwidth]{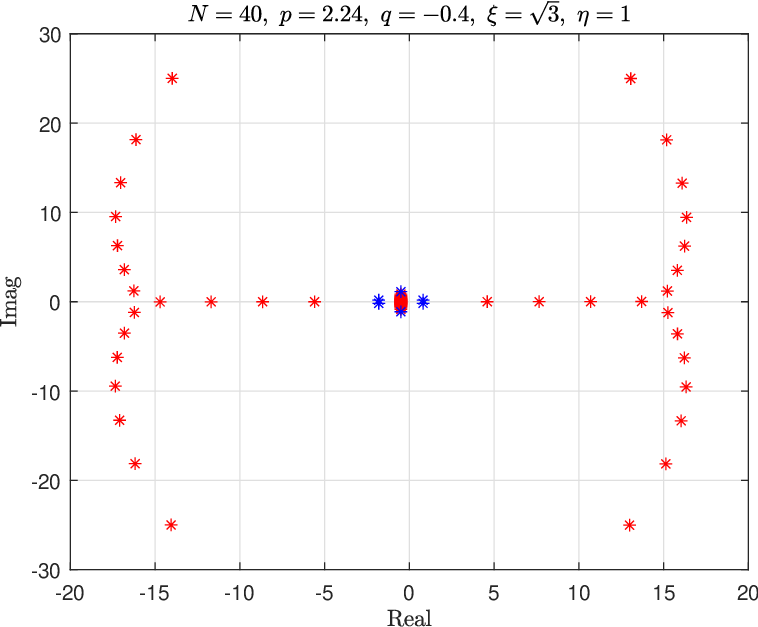}
  }
  \subfloat[Enlarged view of the central region]{
    \includegraphics[width=0.47\textwidth]{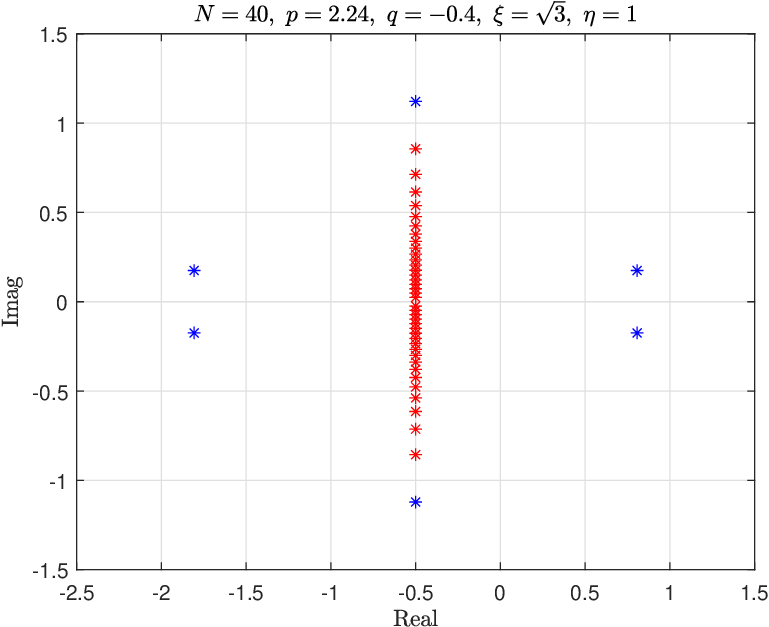}
  }
\caption{
Distribution of Bethe roots in the complex plane with parameters $N = 40$, $p = 2.24$, $q = -0.4$, $\xi = \sqrt3$, and $\eta = 1$. The blue points correspond to the paired-line roots.
}
\label{fig:cross9}
\end{figure}

Upon contact, a new paired-line root pair is formed.
Such paired-line root pairs are generated successively outward from the center, indicating that the central paired-line root pair that vanished during the first crossover has reappeared (Fig.~\ref{fig:cross9}).

\section{Supplementary Analysis of Surface Energy}
\label{app:F}
\setcounter{figure}{0}
\renewcommand{\thefigure}{F.\arabic{figure}}
In this section, we analyze the cases in Eq.~\eqref{eq:lab} with $l_\beta < 1$ and $a_\gamma < 1$.
We first consider the case $l_\beta<1$.
Numerical results show that such line roots exist only in the crossover region, as illustrated in Figs.~\ref{fig:cross2} and \ref{fig:cross3}.
Moreover, in this regime the line roots always appear in pairs.
These two roots are symmetric with respect to the line $\mathrm{Re}(u)=0$, namely, besides $\lambda^{(\mathrm l)}_1=\eta l_1-\tfrac{\eta}{2}$, there is also $\lambda^{(\mathrm l)}_2=\eta(1-l_1)-\tfrac{\eta}{2}$.
For simplicity, we restrict attention to the contribution of these two roots to the energy.
The corresponding root-density equation is then given by
\begin{align}
\delta \tilde{\rho}(\omega)
&=
-\frac{1}{2N}
\frac{
e^{-(1-l_1)|\omega|}
+
e^{-l_1|\omega|}
+
e^{-(l_1+1)|\omega|}
+
e^{-(2-l_1)|\omega|}
}{
1+e^{-|\omega|}}
\nonumber\\
&=
-\frac{1}{2N}
\frac{
(1+e^{-|\omega|})
\left(
e^{-l_1|\omega|}+e^{-(1-l_1)|\omega|}
\right)
}{
1+e^{-|\omega|}}
\nonumber\\
&=
-\frac{1}{2N}\left(
e^{-l_1|\omega|}+e^{-(1-l_1)|\omega|}
\right).
\end{align}
Then, the corresponding contribution to the energy is given by
\begin{align}
\delta E
&=
-\frac{4}{\left(l_1+\frac12\right)\left(l_1-\frac32\right)}
\nonumber\\
&=
-\frac{2\eta^2}{\left(\eta l_1-\frac\eta 2\right)\left(\eta l_1+\frac\eta 2\right)}
-\frac{2\eta^2}{\left[\eta(1-l_1)-\frac\eta 2\right]\left[\eta(1-l_1)+\frac\eta 2\right]}
\nonumber\\
&=
-\sum_{\beta=1}^{2}\frac{2\eta^2}{\lambda^{(\mathrm l)}_{\beta}(\lambda^{(\mathrm l)}_{\beta}+\eta)}.
\end{align}
This is consistent with the previous result.

Next, we consider the case $a_\gamma<1$.
For simplicity, we consider only a single conjugate pair of paired-line roots, $\lambda^{(\mathrm p)}_1=\eta a_1+i\eta b_1-\frac{\eta}{2}$ and $\lambda^{(\mathrm p)}_2=\eta a_1-i\eta b_1-\frac{\eta}{2}$, and focus on their contribution to the energy.
The corresponding root-density equation is then given by
\begin{align}
\delta \tilde{\rho}(\omega)
&=
-\frac{1}{N}
\frac{
\cos(\omega b_1)\Bigl(e^{-(1-a_1)|\omega|}+e^{-(a_1+1)|\omega|}\Bigr)
}{
1+e^{-|\omega|}}
\nonumber\\
&=
-\frac{1}{N}
\frac{
\cos(\omega b_1)e^{-(1-a_1)|\omega|}\left(1+e^{-2a_1|\omega|}\right)
}{
1+e^{-|\omega|}}.
\label{eq:a<1}
\end{align}
As shown in Appendix~\ref{app:E}, the paired-line roots are formed by the coalescence of line roots, and the real part of each paired-line root is equal to that of the line roots at the coalescence point.
Therefore, the paired-line roots with $a_\gamma<1$ can only originate from the line roots with $l_\beta<1$.
Since the line roots with $l_\beta<1$ exhibit the symmetric structure discussed above, one must have $a_\gamma=\tfrac{1}{2}$.
That is, if $a_\gamma<1$, then necessarily $a_\gamma=\tfrac{1}{2}$.
This is consistent with the numerical results, and thus Eq.~\eqref{eq:a<1} can be simplified as
\begin{align}
\delta \tilde{\rho}(\omega)
&=
-\frac{1}{N}
\cos(\omega b_1)e^{-\tfrac12|\omega|}.
\end{align}
Then, the corresponding contribution to the energy is given by
\begin{align}
\delta E
&=
\frac{4}{1+b_1^2}
\nonumber\\
&=
-\left[
\frac{2\eta^2}{(\eta a_1-i\eta b_1-\tfrac\eta 2)(\eta a_1-i\eta b_1+\tfrac\eta 2)}
+
\frac{2\eta^2}{(\eta a_1+i\eta b_1-\tfrac\eta 2)(\eta a_1+i\eta b_1+\tfrac\eta 2)}
\right]
\nonumber\\
&=
-\sum_{j=1}^{2}\frac{2\eta^2}{\lambda^{(\mathrm p)}_j(\lambda^{(\mathrm p)}_j+\eta)}
.
\end{align}
Similarly, this is consistent with the previous result.
When $a_{\gamma}=0$, $\lambda^{(\mathrm p)}_{\gamma}$ and $-\eta-\lambda^{(\mathrm p)}_{\gamma}$ form a complex-conjugate pair.
Since $\lambda^{(\mathrm p)}_{\gamma}$ lies beyond the edge of the regular-root distribution, it is pushed to infinity in the thermodynamic limit.
Therefore, the corresponding term can be omitted from the BAEs and does not contribute to the energy.

\clearpage


\begin{thebibliography}{99}

\bibitem{Bethe1931} H. Bethe, Zur Theorie der Metalle, Z. Phys. {\bf 71}, 205--226 (1931).

\bibitem{Korepin1993} V. E. Korepin, N. M. Bogoliubov, and A. G. Izergin, Quantum Inverse Scattering Method and Correlation Functions, Cambridge Univ. Press (1993).

\bibitem{Takahashi1999} M. Takahashi, Thermodynamics of One-Dimensional Solvable Models, Cambridge Univ. Press (1999).

\bibitem{Bax82} R. J. Baxter, Exactly Solved Models in Statistical Mechanics, Academic Press (1982).

\bibitem{Fad80} L.D. Faddeev, Quantum inverse scattering method. Sov. Sci. Rev. Math. C 1, 107 (1980).

\bibitem{Skl80} E.K. Sklyanin, L.A. Takhtajan, L.D. Faddeev, Qunatum inverse problem method. Theor. Math. Phys. 40, 688 (1980).

\bibitem{Sklyanin1988} E. K. Sklyanin, Boundary conditions for integrable quantum systems, J. Phys. A {\bf 21}, 2375--2390 (1988).

\bibitem{Cherednik1984} I. V. Cherednik, Factorizing particles on a half line and root systems, Theor. Math. Phys. {\bf 61}, 977--983 (1984).

\bibitem{Vega1993} H. J. de Vega and A. Gonzalez Ruiz, Boundary {K}-matrices for the six vertex and the n(2n-1)A$_{n-1}$ vertex models, J. Phys. A {\bf 26}, L519--L524 (1993).

\bibitem{Alcaraz1987} F. C. Alcaraz, M. N. Barber, M. T. Batchelor, R. J. Baxter, and G. R. W. Quispel, Surface exponents of the quantum XXZ, Ashkin-Teller and Potts models, J. Phys. A {\bf 20}, 6397--6409 (1987).

\bibitem{Mezincescu1991} L. Mezincescu and R. I. Nepomechie, Integrable open spin chains with nonsymmetric R-matrices, J. Phys. A {\bf 24}, L17--L23 (1991).

\bibitem{cao03} J. Cao, H. Q. Lin, K. J. Shi and Y. Wang, Exact solution of the XXZ spin chain with unparallel boundary fields, Nucl. Phys. B {\bf 663}, 487--519 (2003).
\bibitem{yun95} C. M. Yung and M. T. Batchelor, Exact solution of the spin-s XXZ chain with non-diagonal twists, Nucl. Phys. B {\bf 446}, 461--484 (1995).
\bibitem{nep021} R. I. Nepomechie, Solving the open XXZ spin chain with nondiagonal boundary terms at roots of unity, Nucl. Phys. B {\bf 662}, 615 (2002).
\bibitem{bas1} P. Baseilhac, The $q$-deformed analogue of the Onsager algebra: beyond the Bethe ansatz approach, Nucl. Phys. B {\bf 754}, 309--345 (2006).
\bibitem{bas2} P. Baseilhac and K. Koizumi, Exact spectrum of the XXZ open spin chain from the q-Onsager algebra representation theory, J. Stat. Mech. {\bf 2007}, P09006 (2007).
\bibitem{nie2-2} S. Niekamp, T. Wirth and H. Frahm, The XXZ model with anti-periodic twisted boundary conditions, J. Phys. A: Math. Theor. {\bf 42}, 195008 (2009).
\bibitem{nic13-1} G. Niccoli, Antiperiodic spin-1/2 XXZ quantum chains by separation of variables: complete spectrum and form factors, Nucl. Phys. B {\bf 870}, 397 (2013).
\bibitem{bel13-1} S. Belliard, Modified algebraic Bethe ansatz for XXZ chain on the segment-I: Triangular cases, Nucl. Phys. B {\bf 892}, 1 (2015).
\bibitem{bel13-2} S. Belliard and R. A. Pimenta, Modified algebraic Bethe ansatz for XXZ chain on the segment-II-general cases, Nucl. Phys. B {\bf 894}, 527 (2015).
\bibitem{cysw} J. Cao, W. L. Yang, K. Shi and Y. Wang, Off-diagonal Bethe ansatz and exact solution of a topological spin ring, Phys. Rev. Lett. {\bf 111}, 137201 (2013).
\bibitem{book} Y. Wang, W. L. Yang, J. Cao and K. Shi, {\it Off-diagonal Bethe Ansatz for Exactly Solvable Models} (Springer, Berlin, 2015).
\bibitem{yang69} C. N. Yang and C. P. Yang, Thermodynamics of a one-dimensional system of bosons with repulsive delta-function interaction, J. Math. Phys. {\bf 10}, 1115 (1969).
\bibitem{gau71} M. Gaudin, Thermodynamics of the Heisenberg-Ising Ring for $\Delta\geq 1$, Phys. Rev. Lett. {\bf 26}, 1301 (1971).
\bibitem{tak71} M. Takahashi, One-dimensional Heisenberg model at finite temperature, Prog. Theor. Phys. {\bf 46}, 401 (1971).
\bibitem{Frahm2026} H. Frahm, A. Kl\"umper, D. Wagner and X. Zhang, Non-linear integral equations for the XXX spin-$\tfrac{1}{2}$ quantum chain with non-diagonal boundary fields, SciPost Phys. {\bf 20}, 012 (2026).

\bibitem{qiao20} Y. Qiao, P. Sun, J. Cao, W. L. Yang, K. Shi and Y. Wang, Exact ground state and elementary excitations of a topological spin chain, Phys. Rev. B {\bf 102}, 085115 (2020).
\bibitem{qiao21} Y. Qiao, J. Cao, W. L. Yang, K. Shi and Y. Wang, Exact surface energy and helical spinons in the XXZ spin chain with arbitrary nondiagonal boundary fields, Phys. Rev. B {\bf 103}, L220401 (2021).


\bibitem{Cao2013} J. Cao, W.-L. Yang, K. Shi, and Y. Wang, Off-diagonal Bethe ansatz solution of the XXX spin chain with arbitrary boundary conditions, Nucl. Phys. B {\bf 875}, 152--165 (2013).

\bibitem{Nepomechie2013} R. I. Nepomechie, An inhomogeneous T-Q equation for the open XXX chain with general boundary terms: completeness and arbitrary spin, J. Phys. A {\bf 46}, 442002 (2013).

\bibitem{Cao2013c} J. Cao, W.-L. Yang, K. Shi, and Y. Wang, Off-diagonal Bethe ansatz solutions of the anisotropic spin-1/2 chains with arbitrary boundary fields, Nucl. Phys. B {\bf 877}, 152--175 (2013).

\bibitem{White1992} S. R. White, Density matrix formulation for quantum renormalization groups, Phys. Rev. Lett. {\bf 69}, 2863--2866 (1992).

\bibitem{Schollwock2005} U. Schollwöck, The density-matrix renormalization group, Rev. Mod. Phys. {\bf 77}, 259--315 (2005).

\bibitem{Schollwock2011} U. Schollwöck, The density-matrix renormalization group in the age of matrix product states, Ann. Phys. {\bf 326}, 96--192 (2011).

\bibitem{Verstraete2004} F. Verstraete, D. Porras, and J. I. Cirac, Density matrix renormalization group and periodic boundary conditions: A quantum information perspective, Phys. Rev. Lett. {\bf 93}, 227205 (2004).

\bibitem{McCulloch2007} I. P. McCulloch, From density-matrix renormalization group to matrix product states, J. Stat. Mech. {\bf 2007}, P10014 (2007).

\bibitem{Pollmann2010} F. Pollmann, A. M. Turner, E. Berg, and M. Oshikawa, Entanglement spectrum of a topological phase in one dimension, Phys. Rev. B {\bf 81}, 064439 (2010).

\bibitem{Murg2012} V. Murg, V. E. Korepin, and F. Verstraete, Algebraic Bethe ansatz and tensor networks, Phys. Rev. B {\bf 86}, 045125 (2012).

\bibitem{ITensor2022} M. Fishman, S. R. White, and E. M. Stoudenmire, The ITensor software library for tensor network calculations, SciPost Phys. Codebases {\bf 4}, 4 (2022).

\bibitem{Ahlfors1979} L. V. Ahlfors, Complex Analysis, McGraw-Hill (1979).

\bibitem{Dong2023} J.-S. Dong, P. Lu, P. Sun, Y. Qiao, J. Cao, K. Hao, and W.-L. Yang, Exact surface energy and elementary excitations of the XXX spin-1/2 chain with arbitrary non-diagonal boundary fields, Chin. Phys. B {\bf 32}, 017501 (2023).

\bibitem{nep022} R. I. Nepomechie and F. Ravanini, Completeness of the Bethe Ansatz solution of the open XXZ chain with nondiagonal boundary terms, J. Phys. A {\bf 36}, 11391 (2003).

\bibitem{Belliard2018} S. Belliard and A. Faribault, Ground state solutions of inhomogeneous Bethe equations, SciPost Phys. {\bf 4}, 030 (2018).

\bibitem{Xiang2023} T. Xiang, Density Matrix and Tensor Network Renormalization, Cambridge Univ. Press (2023).

\bibitem{Zhakenov2025} A. Zhakenov, P. Kattel, N. Andrei, Thermodynamics in a split Hilbert space: Quantum impurity at the edge of the Heisenberg chain, arXiv:2508.19334 (2025).

\bibitem{Gaudin1971} M. Gaudin, Boundary energy of a Bose gas in one dimension, Phys. Rev. A {\bf 4}, 386--394 (1971).
\bibitem{Wen2017} F. Wen, T. Yang, Z.-Y. Yang, J. Cao, K. Hao, and W.-L. Yang, Thermodynamic limit and boundary energy of the su(3) spin chain with non-diagonal boundary fields, Nucl. Phys. B {\bf 915}, 119--134 (2017).
\end{thebibliography}

\end{document}